\documentclass{aa}  

\usepackage[]{natbib}
\usepackage{lscape}
\usepackage{longtable}
\usepackage{float}
\usepackage{cuted}
\usepackage{sidecap}

\usepackage{color}
\newcommand{\om}[1]{{\color{blue} #1}}
\usepackage{caption}
\usepackage{multirow}
\usepackage{threeparttable}
\usepackage{nicefrac}
\usepackage{graphicx}
\usepackage{txfonts}
\usepackage{upgreek}

\usepackage{setspace}
\begin{document}

   \title{CHANG-ES XXI. Transport processes and the X-shaped magnetic field of NGC~4217: off-center superbubble structure}


  \author{Y. Stein
          \inst{1,2}
          \and
          R.-J. Dettmar
          \inst{2,3}
          \and
          R.~Beck
          \inst{4}
          \and
          J.~Irwin
          \inst{5}
          \and
          T.~Wiegert
          \inst{5}
          \and
          A.~Miskolczi 
           \inst{2}
           \and
          Q.D.~Wang
          \inst{6} 
          \and
          J.~English
          \inst{5}
          \and 
          R.~Henriksen
          \inst{5}
           \and 
          M.~Radica
          \inst{5}
          \and
          J.-T. Li 
          \inst{7}
           }

   \institute{Observatoire astronomique de Strasbourg, Universit\'e de Strasbourg, CNRS, UMR 7550, 11 rue de l'Universit\'e, \\ 
   	67000 Strasbourg, France; \email{yelena.stein@astro.unistra.fr}
   			\and
   				Ruhr-Universit\"at Bochum, Fakult\"at f\"ur Physik und Astronomie,	Astronomisches Institut (AIRUB), 
              Universit\"atsstrasse 150, 44801 Bochum, Germany
              \and
              Research Department: Plasmas with Complex Interactions, Ruhr-Universit\"at Bochum,
              Universit\"atsstrasse 150, 44801 Bochum, Germany
         			\and
         		 Max-Planck-Institut f\"ur Radioastronomie, Auf dem H\"ugel 69, 53121 Bonn, Germany
         				 \and
              Dept. of Physics, Engeneering Physics, \& Astronomy, Queen's University, Kingston, Ontario, Canada, K7L 3N6
              \and
              Department of Astronomy, University of Massachusetts, 710 North Pleasant Street, Amherst, MA 01003, USA  
              \and
              Department of Astronomy, University of Michigan, 311West Hall, 1085 S. University Ave, Ann Arbor, MI, 48109-1107,USA
              }

   \date{Received February 6 2020/Accepted 2 June 2020}

 
  \abstract
   {Radio continuum observations of edge-on spiral galaxies reveal the appearance of radio halos as well as the large-scale structure of their magnetic fields. Furthermore, with multiple frequency observations, it is possible to deduce the transport mechanisms of the cosmic ray electrons (CREs).}
   {In order to gain a better understanding of the influence of cosmic rays (CRs) and magnetic fields in the disk-halo interface of edge-on spiral galaxies, we investigate the radio continuum halo, the magnetic field, and the transport processes of the CRs of the edge-on spiral galaxy NGC~4217 using CHANG-ES radio data at two frequencies, 6\,GHz (C-band) and 1.5\,GHz (L-band), and supplemental LOFAR data of this galaxy at 150\,MHz. With additional X-ray Chandra data, we study the connection of radio features to the diffuse hot gas around NGC~4217.}
   {We investigate the total intensity (Stokes I) data in detail and determine the integrated spectral behavior. The radio scale heights of all three radio frequencies for NGC~4217 were extracted via exponential fits to the intensity profiles. From these, individual absolute flux densities of the disk and the halo were also calculated. Furthermore, we present magnetic field orientations from the polarization data using rotation measure synthesis (RM-synthesis), showing the large-scale ordered magnetic field of NGC~4217. After a  separation of thermal and nonthermal emission, we calculated the resolved magnetic field strength via the revised equipartition formula. Additionally, we modeled the transport processes of CREs into the halo with the 1D model \textsc{spinnaker}.}
   {NGC~4217 shows a large-scale X-shaped magnetic field structure, covering a major part of the galaxy with a mean total magnetic field strength in the disk of 9\,$\upmu$G. From the analysis of the rotation measure map at C-band, we found that the direction of the disk magnetic field is pointing inward. A helical outflow structure is furthermore present in the northwestern part of the galaxy, which is extended nearly 7\,kpc into the halo. More polarized emission is observed on the approaching side of the galaxy, indicating that Faraday depolarization has to be considered at C-band. With a simplified galaxy disk model, we are able to explain the finding of higher polarized intensity on the approaching side. We generalize the model to predict that roughly 75\% of edge-on spiral galaxies will show higher polarized intensity on the approaching side. Many loop and shell structures are found throughout the galaxy in total intensity at C-band. One structure, a symmetric off-center (to southwest of the disk) superbubble-like structure is prominent in total and polarized intensity, as well as in H$\upalpha$ and optical dust filaments. This is at a location where a second peak of total intensity (to the southwest of the disk) is observed, making this superbubble-like structure a possible result of a concentrated star formation region in the disk. The X-ray diffuse emission shows similarities to the polarized diffuse emission of NGC~4217. The flux density extension of the radio continuum halo increases toward lower frequencies. While the total flux density of the disk and halo are comparable at C-band, the contribution of the disk flux density decreases toward LOFAR to 18\% of the total flux density. Dumbbell-shaped structures are present at C-band and at the LOFAR frequency. Total intensity profiles at the two CHANG-ES bands and the LOFAR frequency show a clear two-component behavior and were fit best with a two-component exponential fit. The halo scale heights are 1.10~$\pm$~0.04\,kpc, 1.43~$\pm$~0.09\,kpc, and 1.55~$\pm$~0.04\,kpc in C-band, L-band, and 150\,MHz, respectively. The frequency dependence of these scale heights between C-band and L-band suggests advection to be the main transport process. The 1D CRE transport modeling shows that advection appears to be more important than diffusion.}
   {}

	
	\titlerunning{CHANG-ES XIV: Transport processes and the magnetic field of NGC~4217}
  \authorrunning{Stein et al.}
	
   \maketitle
%

\section{Introduction}
Studies of star forming galaxies in the local Universe provide information on evolutionary properties and the driving physical processes. 
The combined action of many supernova explosions and stellar winds of massive star (OB) associations in the disk can lead to the formation of “superbubbles \citep[e.g.,][]{maclowetal1988}, which are filled with hot gas and can last tens of millions of years \citep[e.g.,][]{elbadryetal2019}. The effects of superbubbles 
can be observed in the Milky way and external galaxies, where superbubbles that are breaking out of the disk cause gas and dust filaments to reach high above the disk into the halo, tracing chimney structures \citep[e.g.,][]{wang01}.

In addition to gas, cosmic rays (CRs) and magnetic fields play an important role in the disk-halo interaction of star forming galaxies, which leads to the formation of radio halos \citep[e.g.,][]{parker1992}. We note that, in this context, and in order to conform to previous papers in the CHANG-ES series, we use the word “halo” to refer to gas, dust, cosmic rays, and the magnetic field above and below the galaxy disk. This is not to be confused with stellar or dark matter halos. Specifically, as was defined in Irwin et al. 2012a, the halo refers to emission on larger scales, with a scale height of z > 1 kpc, while the disk–halo interface is at 0.2 < z < 1 kpc.

At radio frequencies, signatures of relativistic cosmic ray electrons (CREs) are observed since they gyrate around the magnetic field lines and radiate nonthermal synchrotron emission perpendicular to the field lines. This emission is linearly polarized and measured for example in radio continuum. Polarization from radio continuum observations of star-forming galaxies often show X-shaped magnetic field structures if observed edge-on \citep[][]{tullmannetal2000, krause2009}. One proposed mechanism to maintain these large-scale fields is the mean-field $\upalpha - \upomega$ dynamo \citep[see e.g.,][]{ruzmaikinetal1988, becketal1996, chamandy2016, henriksen2017, henriksenetal2018, beck2019}, where turbulent magnetic fields in spiral galaxies are amplified and ordered by the $\upalpha$-process (driven by e.g., supernova explosions and the Coriolis force), combined with the $\upomega$-process (shear motions due to differential rotation) to result in large-scale regular magnetic fields \citep[e.g.,][]{arshakianetal2009}.

Here, we analyze the edge-on spiral galaxy NGC~4217 using radio continuum data from Continuum HAlos in Nearby Galaxies - an Evla Survey (CHANG-ES), observed with the Karl G. Jansky Very Large Array (VLA) at C-band (6\,GHz) and L-band (1.5\,GHz). With these, and supplemental LOFAR data (150\,MHz), we investigate the radio halo, the polarization, the magnetic field structure as well as CR transport processes. We further use supplemental Chandra data to explore the relationship of the magnetic field properties to the diffuse hot gas around NGC~4217. 

NGC~4217 is a member of the Ursa Major cluster and is a possible companion of M101. Its distance is 20.6\,Mpc \citep{wiegertetal2015}. For basic galaxy parameters see Table~\ref{tab:bparameter}. Existing radio observations of NGC~4217 are from the VLA at 1.49\,GHz (20\,cm) by \citet{condon1987}, where the radio structure is washed out due to the large beam of 54". Furthermore, 6\,cm radio data from VLA/Effelsberg by \citet{hummeletal1991} exist, which show the structure of the radio disk with one intensity peak in the center and a smaller peak off-center in the southwest. In their map, the northeast of the galaxy appears to be thicker and more filamentary. The Hubble Space Telescope (HST) image \citep{thomsonetal2004} shows dusty filaments along the disk which reach 2\,kpc distance from the midplane. The authors also present a detection of two shell-like features extending 1\,kpc and a loop that reaches a height of 1.5\,kpc. These were probably driven out by stellar winds and supernovae. \citet{thomsonetal2004} claim a supernova (SN)-driven galactic fountain or chimney phenomenon as the responsible mechanism. They also suggest a possible but unknown influence of magnetic fields on the disk-halo interface. The HI data of \citet{verheijensancisi2001} have been reanalyzed by \citet{Allaertetal2015} and indicate a slight warp of the disk, which can be seen in the optical as well. Additionally, they suggest a coplanar ring-like structure outside the disk which is kinematically and spatially offset from the disk and shows a warp along the line of sight. They interpret this feature as the result of a recent minor merger.
\begin{center}
\begin{threeparttable}
\caption{Basic galaxy parameters.}  
\label{tab:bparameter}
\begin{tabular}{lc} 
\hline  \hline
Galaxy & NGC 4217  \\ 
\hline
Right Ascension	 & 12h15m50.9s\tnote{1} \\
Declination	 & +47d05m30.4s\tnote{1}\\
Distance (Mpc)					&  20.6\tnote{1}\\
Inclination ($^\circ$)	 	&	89\tnote{2}	\\
Position Angle ($^\circ$)								& 50\tnote{2} \\
Major Axis (arcmin)			& 5.2\tnote{3} \\
Minor Axis (arcmin)				& 1.5\tnote{3} \\
v$_{\text{sys}}$ (km s$^{-1}$) 		&	1027\tnote{3}\\
v$_{\text{rot}}$ (km s$^{-1}$) 		&	195\tnote{4}\\
SFR (M$_{\odot}$~yr$^{-1}$)		& 1.5\tnote{5}			\\
SFRD (10$^{-3}$\,M$_{\odot}$~yr$^{-1}$~kpc$^{-2}$)&	3.6\tnote{5}					\\			
Classification 			  				&	Sb\tnote{6}\\
\hline 
\end{tabular}
\begin{tablenotes}
\footnotesize
\item {\bf Reference:}
\item[1] \citet{wiegertetal2015} \item[2] determined in this work
\item[3] NASA/IPAC Extragalactic Database (NED, https://ned.ipac.caltech.edu)
\item[4] \citet{zaritzkyetal2014}
\item[5] Star formation rate (SFR) and star formation rate density (SFRD) from \citet{wiegertetal2015}
\item[6] \citet{irwinetal2012}
\end{tablenotes}
\end{threeparttable}
\normalsize
\end{center}

The paper is organized as follows: 
In Section~\ref{sec:data} we present the processing and imaging of the CHANG-ES data and introduce the LOFAR and Chandra data. In Section~\ref{sec:results} total intensity maps, the investigation of diverse structures throughout the galaxy, the scale height analysis, and the polarization including magnetic field orientations, as well as RM maps and their analysis are presented. The separation of the thermal and nonthermal emission for NGC 4217 is explained next and the nonthermal spectral index map as well as the magnetic field map derived with the equipartition formula are shown. Then we present the results of the 1D transport model \textsc{spinnaker}, applied to the total synchrotron intensity profile. Finally, the results regarding the main transport processes, the halo structures in the radio continuum and the polarized intensity are discussed in Section~\ref{sec:discussion} and summarized in Section~\ref{sec:summary}.

\section{Radio continuum, X-ray observations, and data analysis}
\label{sec:data}
\subsection{CHANG-ES}
\subsubsection{Data}
As part of the CHANG-ES survey \citep[][]{irwinetal2012}, observations in the radio continuum were obtained using B-, C-, and D-configuration at L-band (1.5\,GHz), and using C- and D-configuration at C-band (6\,GHz). The observation settings (32 spectral windows (spws) at 1.5\,GHz and 16 spws at 6\,GHz) allow us to apply RM-synthesis. Full polarization products (Stokes I, Q, U, and V) were obtained. The D-configuration data \citep{wiegertetal2015} as well as the B-configuration data \citep{irwinetal2019} are public\,\footnote{CHANG-ES data releases available at www.queensu.ca/changes} and the C-configuration data will be released soon as well (Walterbos et al. in prep).

We carried out the data reduction for the five data sets of NGC~4217 (see Table~\ref{tab:oparameter}) individually following the CHANG-ES calibration procedures \citep{irwinetal2013} and using the Common Astronomy Software Applications (CASA) package \citep[][]{mcmullinetal2007} with J1331+3030 (3C286) as the primary calibrator, which was also used as the bandpass and the polarization angle calibrator, J1219-4829 as the secondary calibrator and J1407+2827 as the zero polarization calibrator. The calibrated data from the different configurations were then combined for the two bands. With these combined data, imaging was performed for Stokes I and linear polarization. The Stokes I maps  were produced by cleaning with a robust zero weighting, at C-band with a uv-tapering of 12\,k$\lambda$. The linear polarization and magnetic field orientation maps are created via Stokes Q and Stokes U, which were imaged with a robust weighting of two in order to detect faint structures. The polarized intensity P is derived\,\footnote{A correction for the positive bias due to noise is not necessary because we consider only P values above three times the rms noise.} via P~=~$\sqrt{\text{Q}^2 + \text{U}^2}$ and the apparent orientation of the electric field vector $\Psi$ is calculated with $\Psi~=~1/2~\text{arctan}~(\text{U/Q})$. By definition, the direction of this vector is ambiguous as tan($\Psi$)~=~tan($\Psi \pm \, n \cdot 180^{\circ}$). Nevertheless, it gives a direct hint about the magnetic field orientation because the electric field vector and the magnetic field vector are perpendicular to each other. Additionally, we performed rotation measure synthesis (RM-synthesis), which is explained in the next section.

The achieved rms noise from the combined C-band data for the 12k$\uplambda$-uvtapered Stokes I is 5\,$\upmu$Jy/beam with a beam of 7.7"~$\times$~7.7". The resulting rms noise of the combined L-band data for Stokes I is 22\,$\upmu$Jy/beam with a beam of 12.3"~$\times$~13.4". These as well as smoothed images of Stokes I of both bands are shown to match the different resolutions of the polarization maps as well as the resolution of the maps from RM-synthesis. 

\begin{table}
\centering
\caption{Observation parameters.} 
\label{tab:oparameter}
\begin{tabular}{l c c c} 
\hline\hline   
Dataset	&Observing Date		& Time on Source\\
			 &                 		& (before flagging)\\ 
\hline
L-band B-configuration &  11. Aug 2012  & 2 hr\\
L-band C-configuration &  31. Mar 2012    & 30 min\\
L-band D-configuration &  18. Dec 2012   & 20 min\\
C-band C-configuration&  26. Feb 2012   & 3 hr\\
C-band D-configuration & 27. Dec 2012   & 40 min\\
\hline
\end{tabular} 
\end{table}

\subsubsection{RM-synthesis}
\label{sec:RMsynth}
 When a source emits polarized electromagnetic waves, these experience a wavelength-dependent rotation of the polarization plane when propagating through a magnetized plasma (e.g., the ISM of galaxies). 
RM-synthesis \citep{burn1966,brentjensdebruyn2005} is able to reproduce the intrinsic polarization angle of a source and to determine the amount of rotation of the polarization angle along the line of sight, which is dependent on the wavelength squared and the RM. 

For different regions at various distances $r_0$ from the observer, which emit polarized intensity and/or rotate the polarization angle, RM is replaced by the Faraday depth $\Phi$ \citep{burn1966}:
\begin{equation}
\Phi = 0.81 \int\limits_{r_0}^{0}\frac{n_e}{cm^{-3}} \ \frac{\text{B}_{\|}}{\mu G} \ \frac{\text{d}r}{\text{pc}}\ \text{rad} \ \text{m}^{-2} \, ,
\end{equation}
with the electron density $n_e$ and the magnetic field  B$_{\|}$ integrated along the line of sight (pathlength). By applying the method of RM-synthesis to wide-band multi-channel receiver data, it is possible to derive multiple sources along the line of sight \citep{brentjensdebruyn2005}. The basic idea \citep{burn1966} connects the complex polarized surface brightness in Faraday depth space via a Fourier transform with the complex polarized surface brightness in $\lambda^2$-space. We follow \citet{steinetal2019a} in applying this technique and present the important parameters derived via the formulas from \citet{brentjensdebruyn2005} in Table~\ref{tab:RMpar}.

The amount of rotation of the polarization angle $\Delta \chi$ within the bandwidth of the CHANG-ES data at C-band, ranging in frequency from 5\,GHz to 7\,GHz (wavelength range between $\lambda$~=~0.06\,m and $\lambda$~=~0.042\,m) and for a typical RM value of 100\,rad\,m$^{-2}$ at C-band can be calculated by:
\begin{align}
\Delta \chi &= \text{RM} \, (\lambda_1^2 - \lambda_2^2) = \text{RM} \, \Delta \lambda^2 \\ \nonumber
\Delta \chi_{\text{C-band}} &= 100\,\text{rad m}^{-2} \, (0.0036 - 0.00176)\,\text{m}^2 \\ 
                             &= 0.184\,\text{rad} = 10.5^{\circ}.
\label{eq:rmwinkelL}
\end{align}
At C-band, the rotation is not very strong and thus polarization maps at C-band with and without RM-synthesis look quite similar. With a similar RM value, this rotation is stronger at L-band and we need to apply RM-synthesis in order to get reliable magnetic field orientations. However, while the observed |RM| at C-band of NGC~4217 has values up to 200\,rad\,m$^{-2}$, at L-band the observed |RM| only goes up to 20\,rad\,m$^{-2}$.  Using the information in Table~\ref{tab:RMpar}, a typical RM of 10\,rad\,m$^{-2}$ would give a rotation across the CHANG-ES L-band of about 19 deg.

\begin{threeparttable}
\centering
\caption{RM-synthesis parameters derived via \citet{brentjensdebruyn2005}.}  
\label{tab:RMpar}
\begin{tabular}{l c c} \hline\hline
 \	                            																								& C-band     & L-band   \\ \hline
Bandwidth (GHz)		& 2         & 0.5     \\
$\upnu_{\text{min}}$ to $\upnu_{\text{max}}$ (GHz) & 5 to 7    & 1.2 to 1.75 [gap]\\
$\Delta \lambda^2$ = $\lambda_1^2 - \lambda_2^2$ (m$^2$) & 0.00184 	& 0.0336 	\\
$\delta \phi$ = $\frac{2 \sqrt{3}}{\Delta \lambda^2}$ (rad m$^{-2}$) & 1882  & 103    \\
\vspace{-0.35cm}\\
\hline
$\lambda_{min}$ (m)   	 																														& 0.042     & 0.17    \\
max$_{\text{scale}}$  = $\frac{\pi}{\lambda_{min}^2}$ (rad m$^{-2}$)  & 1781      & 109    \\ 
\vspace{-0.35cm}\\
\hline
spw-width $\delta$f (MHz)	& 125       & 16        \\
$\delta \lambda^2$ (m$^2$)												                   							 & 0.0001735 & 0.001447    \\
RM$_{\text{max}}$ = $\frac{\sqrt{3}}{\delta \lambda^2} $ (rad m$^{-2}$)											& 9983      & 1197       \\
\vspace{-0.35cm}\\
\hline
\end{tabular}
\begin{tablenotes}
\footnotesize
\item {\bf Explanations:}
\begin{itemize}
\item $\upnu_{\text{min}}$, $\upnu_{\text{max}}$: minimum and maximum CHANG-ES frequencies; $\Delta \lambda^2$: width of the $\lambda^2$ distribution; $\lambda_1^2$ and $\lambda_2^2$: boundary wavelengths of the band; $\delta \phi$ (rad\,m$^{-2}$): full width at half maximum (FWHM) of the resolution (RMSF) in Faraday depth ($\Phi$) space.
\item $\lambda_{min}$: shortest wavelength; max$_{\text{scale}}$ (rad\,m$^{-2}$): largest scale in $\Phi$ space to which the observation is sensitive.
\item $\delta \lambda^2$: spw-width squared; RM$_{\text{max}}$ (rad\,m$^{-2}$): maximum observable RM (resp. $\Phi$).
\end{itemize}
\end{tablenotes}
\end{threeparttable}
\normalsize
\ \\

We begin the RM-synthesis analysis by imaging Stokes Q and Stokes U for each spectral window with a robust two weighting using the CASA package. Since the resulting beam size (resolution) of the image is frequency dependent, all images are convolved to the largest beam in the frequency range. Then a cube is created from all images with the frequency as the third axis. This was done for the Stokes Q and U separately. From the cube we can produce maps of RM, the polarized intensity (PI), and the intrinsic polarization angle (PA), as well as their corresponding error maps. The RM map represents the fit peak position of each pixel, assuming there is only one component along the line of sight. Using the data in \citet{rm2014}, the Galactic foreground rotation measure in the direction to NGC~4217 was determined. This value (RM$_{\text{foreground}}$ = 9.4~$\pm$~1.2\,rad\,m$^{-2}$) was subtracted from the final RM maps. The results and the analysis on the RM maps are presented in Section~\ref{sec:magfield}.

\subsection{LOFAR}
 The LOFAR Two-meter Sky Survey (LoTSS) is a sensitive, high-resolution, low-frequency (120-168\,MHz) survey of the northern sky using LOFAR \citep{shimwelletal2017}. LoTSS data of NGC~4217 were obtained from data release 1 \citep[DR1,][]{shimwelletal2019}\,\footnote{LoTSS data are publicly available at: https://lofar-surveys.org}. In DR1, images with two resolutions of 6" and 20" are provided. We use the 6" image of NGC 4217 and applied Gaussian smoothing of 10" to show the halo structure and of 13.4" for the scale height analysis (Sec.~\ref{sec:scaleheights}). 

\subsection{Chandra}
To compare the radio data with X-ray emission from NGC~4217, we also present the Chandra ACIS-S observation of the galaxy (ID 4738). The observation was taken on Feb. 19, 2004 for an exposure time of 72.7\,ks. The global properties of the X-ray emission were presented by \citet{liwang2013a}, as part of the Chandra survey of galactic coronae around 53 nearby highly inclined disk galaxies. Here, we reprocessed the data in the same way, but with the latest software (CIAO version 4.11) and calibration files (CALDB  version 4.8.4.1), and produced diffuse (discrete source-excised) X-ray intensity images with a smoothing optimized for comparison with the radio data.

\section{Results and Discussion}
\label{sec:results}
\subsection{Radio continuum total intensity}

\begin{figure}
 \includegraphics[width=0.5\textwidth]{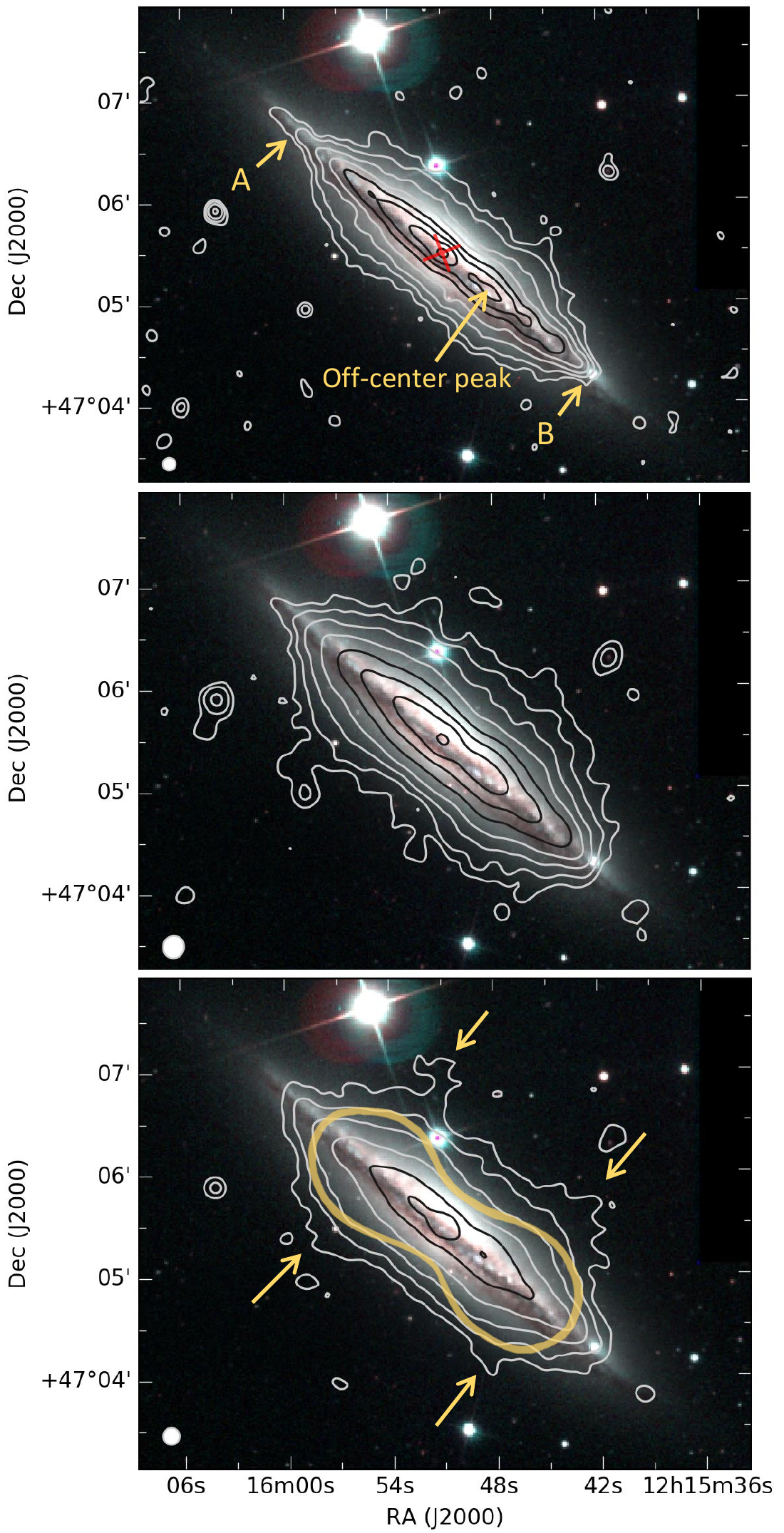}
	\caption{NGC~4217 total radio intensity contours on optical image made from the SDSS u, g, r-filters. Radio contours start at 3$\upsigma$ levels given below and increase in powers of 2 (up to 128). The beams are shown in the bottom left corner of each image. The robust parameter of clean was set to zero for all images. Top: combined C-band (C- and D-configuration), $uv$-tapering of 12\,k$\lambda$, rms noise $\upsigma$ of 5\,$\upmu$Jy/beam, the beam size is 7.7" $\times$ 7.7". A red cross indicates the center. The arrows point to feature "A", which is the finger-like elongation of the disk, and to feature "B", which shows a sharp cut-off of the radio map and to the off-center peak position in the southwest within the disk. Middle: combined L-band (B-, C-, and D-configurations), $\upsigma$ of 22\,$\upmu$Jy/beam, the beam size is 12.3" $\times$ 13.4". Bottom: LOFAR 150\,MHz, $\upsigma$ of 130\,$\upmu$Jy/beam, The image was smoothed to a beam size of 10" $\times$ 10". The arrows point to the extended radio emission, which appears to be a double-horn structure. The dumbbell-shape is overlayed in yellow. }
	\label{fig:N4217-Ccomb-uvtap}
	\label{fig:N4217-Lcomb}
	\label{fig:N4217-Lofar}
\end{figure}

The total power contours from Stokes~I are shown on the optical SDSS image. We used the formulas of Lupton 2005\,\footnote{www.sdss.org/dr12/algorithms/sdssUBVRITransform/\#Lupton2005} to compute RGB images from SDSS u, g, r-filters. In Figure~\ref{fig:N4217-Ccomb-uvtap} (top), the  Stokes I contour map of the combined C-band using a $uv$-tapered weighting of 12k$\uplambda$ is displayed. The radio flux density is extended into the halo of NGC~4217. The appearance of the radio halo to the southwest edge of the disk seems to be cut off sharply (marked with "B" in Fig.~\ref{fig:N4217-Ccomb-uvtap}), whereas the northeast edge of the disk is elongated with a finger-like structure (marked with "A" in Fig.~\ref{fig:N4217-Ccomb-uvtap}). In C-band, the halo is slightly dumbbell-shaped, where the vertical extent of contours toward the galaxies' center do not reach as far into the halo as at larger radii. The general shape is shown as the visual aid in yellow in the bottom panel. In the optical center of the galaxy (marked with a red cross in Fig.~\ref{fig:N4217-Ccomb-uvtap}), a small point source is detected in the total intensity maps. Based on the B-configuration L-band data \citep{irwinetal2019}, this point source is actually not point-like and therefore it was concluded that this galaxy does not host an AGN in the center. The main peak of the total intensity flux density (also at the position of this red cross) is close to the kinematic center of the galaxy and is elongated toward the northeast within the disk. A second intensity peak within the disk can be found to the southwest, which is labeled as "Off-center Peak" in Fig.~\ref{fig:N4217-Ccomb-uvtap}. The appearance of NGC~4217 in the map presented here are comparable to the one in the map of \citet{hummeletal1991}. 

In Figure~\ref{fig:N4217-Lcomb} (middle), the  Stokes I contours of the combined array configurations of L-band data of NGC~4217 are shown. The 3$\upsigma$ extent of the radio halo in L-band is up to 7\,kpc above and below the mid-plane. The radio halo is larger in comparison to the radio halo at C-band, but in general similar to the C-band map. The galaxy disk to the southwest also has a sharp cut-off and the galaxy disk to the north shows a finger-like extension. Overall, the radio halo in L-band seems to be elliptical or very slightly dumbbell-shaped. 

The LOFAR contours are shown in Fig.~\ref{fig:N4217-Lofar} (bottom). The major axis extent is similar to C- and L-band. Filamentary structures at the edge of the disk are reaching far out into the halo (marked with arrows). The halo above the disk is further extended than the L-band radio halo, while the halo below the disk is less extended than at L-band. The radio halo at the LOFAR frequency is dumbbell-shaped (as indicated in yellow) with clear signs of a double-horn structure (marked with arrows). 

\begin{figure*}[h!]
	\includegraphics[width=0.95\textwidth]{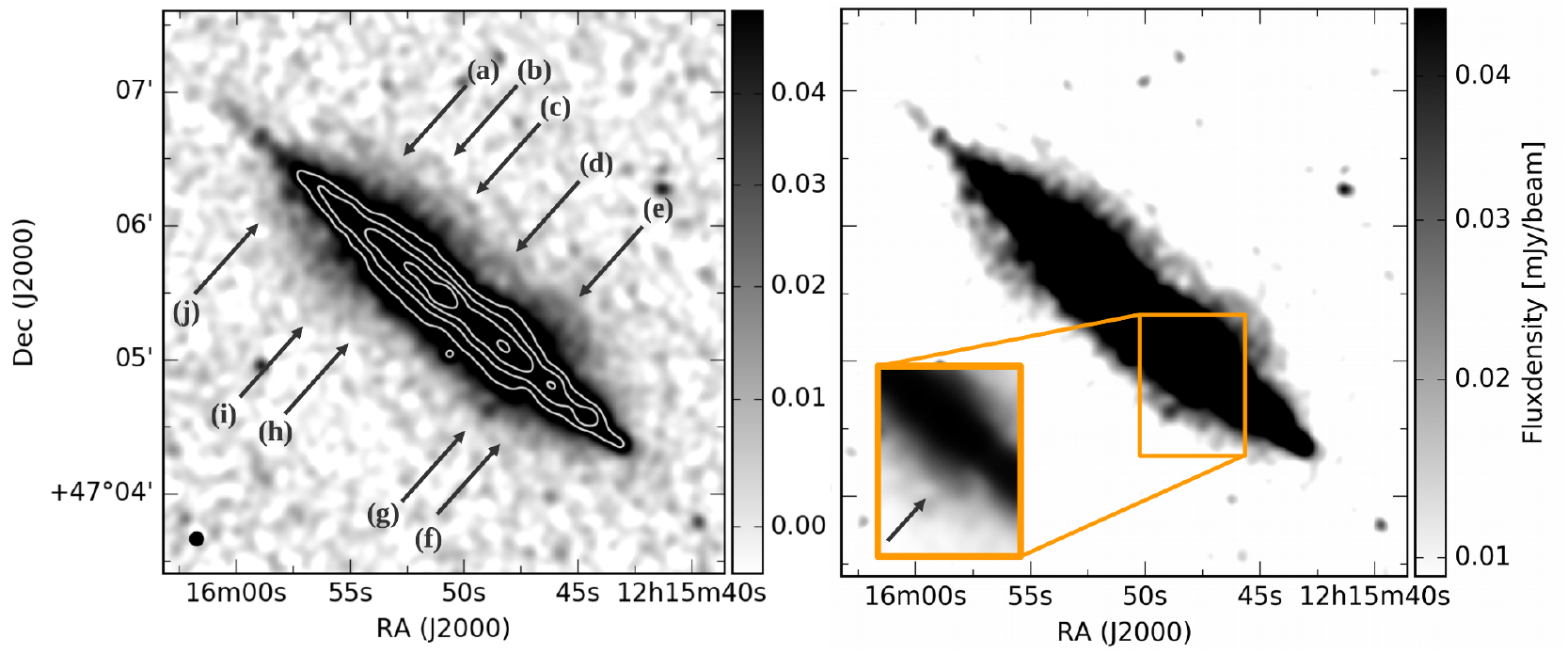}
	\caption{NGC~4217 total intensity image of C-band C-configuraion with logarithmic scaling. The robust parameter of  clean  was  set  to  zero and a $uv$-tapering of 16\,k$\uplambda$ was applied. Left: White contours at 0.09, 0.18, 0.35, 0.7\,mJy levels, with $\upsigma$ at 3\,$\upmu$Jy. The arrows point to the regions dicussed in the text and shown in Fig.~\ref{fig:N4217-galerie} in detail. The beam size of 5.8" x 5.9" is shown in the bottom left. Right: 
	Displays emission  above 3$\upsigma$. The insert shows hints of an outflow structure near location (g) and (f) at another gray scale stretching.	}
	\label{fig:N4217-bw}
\end{figure*}

\begin{figure*}[h!]
	\includegraphics[width=0.98\textwidth]{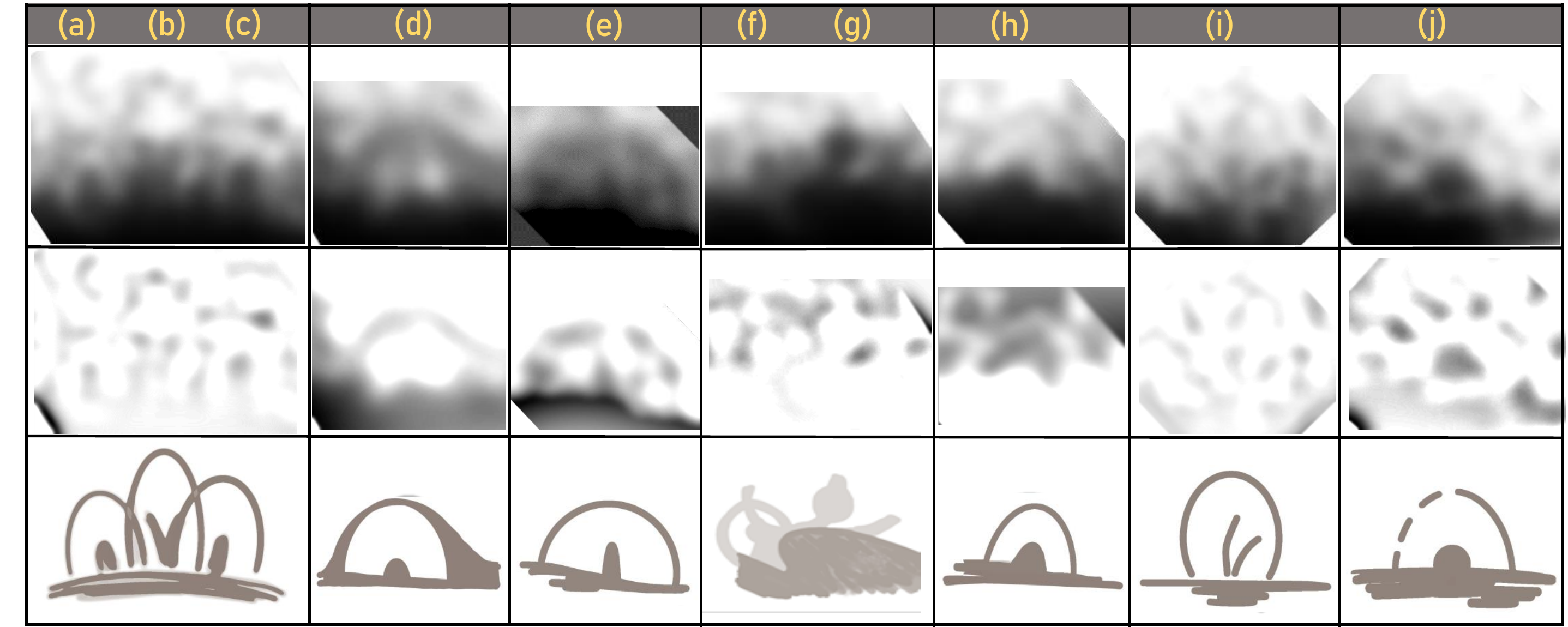}
	\caption{ NGC 4217 discussed radio structures. First row: Zoom-in of C-band image (Fig.~\ref{fig:N4217-bw}). Middle row: Edge-detection filter "difference of Gaussians" applied via GIMP. Bottom row: simplified cartoon made accordingly to the radio data.}
	\label{fig:N4217-galerie}
\end{figure*}

\begin{figure}
\centering
	\includegraphics[width=0.38\textwidth]{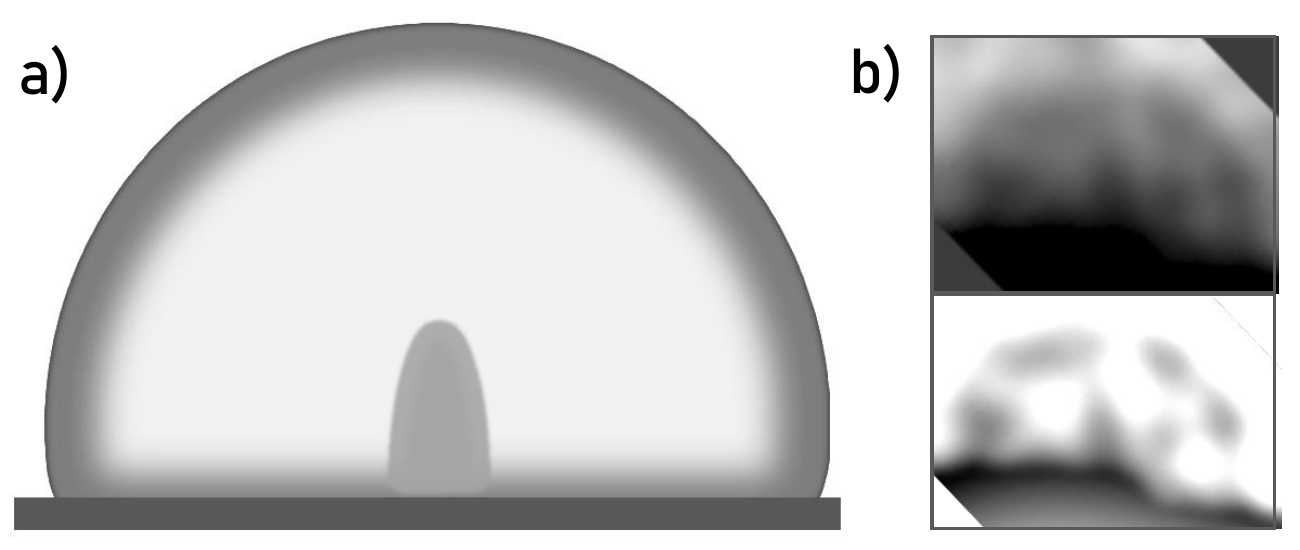}
	\caption{a) Simplified cartoon based on the radio observational morphologies of the loop- and bubble-like structures. b) Rotated superbubble-like structure (e). Top: zoom-in from Fig.~\ref{fig:N4217-bw}.  Bottom: same zoom-in with the edge-detection filter ``difference of Gaussians'' applied via GIMP.}
	\label{fig:bubble-model}
\end{figure}

\subsubsection{Filaments, loops, shells and superbubble-like structures in total intensity}
\label{sec:main_bubbles}

\begin{figure*}
\centering
	\includegraphics[width=0.84\textwidth]{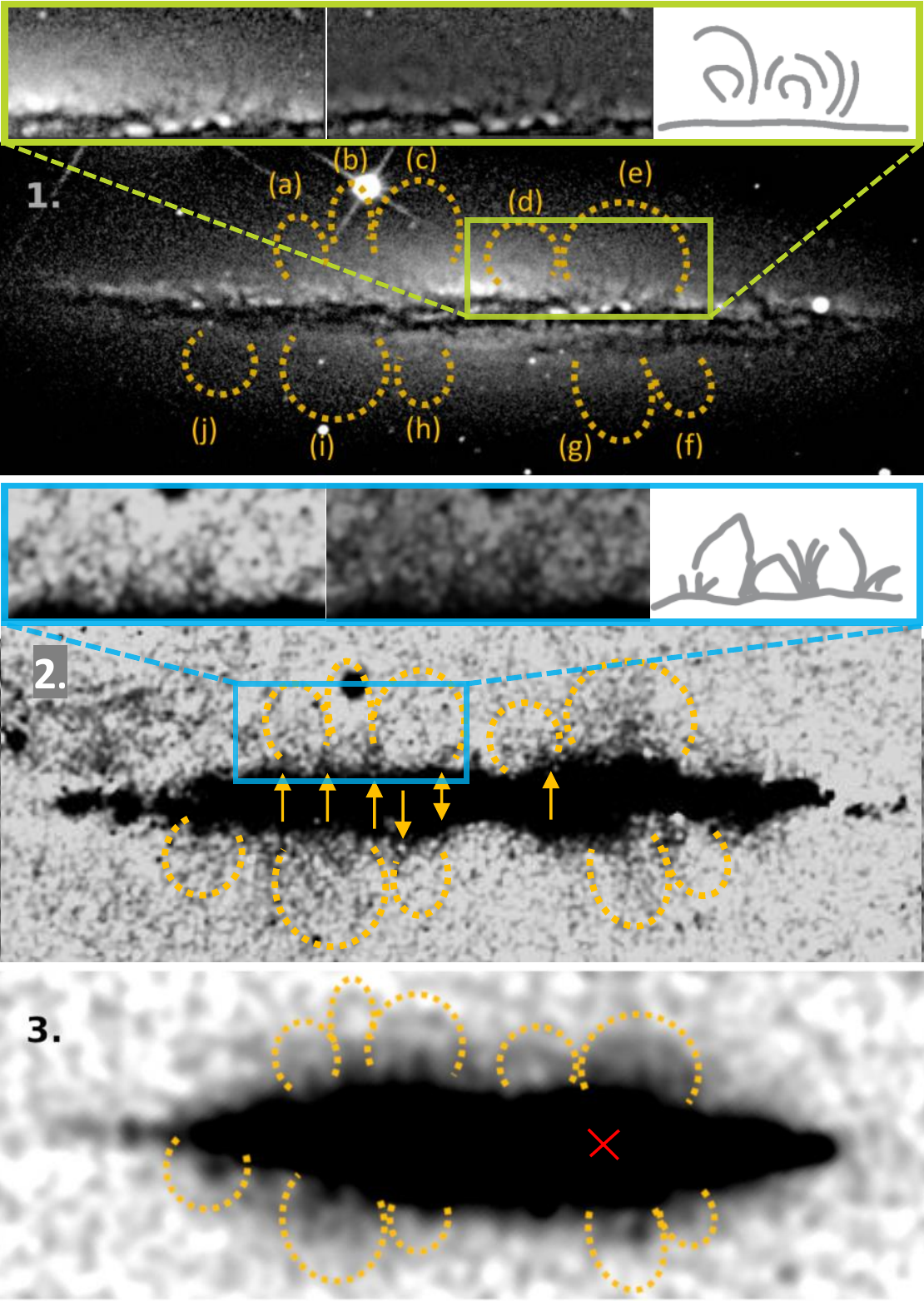}
	\caption{Rotated ($40\degr$) views of NGC~4217 all on the same scale with the same labels of the structures as in Figs. \ref{fig:N4217-bw} and \ref{fig:N4217-galerie}.\\
	{\bf 1.} Artificial optical B-band image created by combining the u and g filters from SDSS, sharpened by unsharp masked filter of GIMP. The green cutout shows loop structures from dusty filaments. Left: zoom-in, middle: wavelet decomposition applied via GIMP with the first 4 scales shown; right: simple cartoon representing the apparent structures.
	{\bf 2.} H$\upalpha$ image from \citet{rand1996}, using an additional edge-detection filter of GIMP. The blue cutout shows loops and extended structures. Left: zoom-in; middle: other gray scale stretch shown; right: simple cartoon representing the apparent structures. The arrows point to extended structures. {\bf 3.} Radio map from this work (Fig.~\ref{fig:N4217-bw}) indicating the investigated structures. The red cross marks the off-center peak position.}
	\label{fig:N4217-bubble_threeparts}
\end{figure*}

We investigate the structure of the C-band radio data in detail and show the total intensity of only the C-configuration in gray scale (Fig.~\ref{fig:N4217-bw}). In this figure, filamentary structures as well as loops, shells and at least one superbubble-like structure are visible in radio continuum. We visually inspected displays of the data that used a variety of so-called gray scale stretches, including a linear and a logarithmic scale for mapping the flux density. Judging from this, we show and discuss the most prominent structures here. These are marked in Fig.~\ref{fig:N4217-bw} with arrows. Most of them are above or around 3$\upsigma$. To show this, we present the figure with a 3$\upsigma$-level cutoff in the right of Fig.~\ref{fig:N4217-bw}. In Figure~\ref{fig:N4217-galerie} we show all the labeled structures in detail. The first row presents the zoom-in from the C-band image (Fig. 2). The middle row shows the same structure with an edge-detection filter "difference of Gaussians" applied via the image software GIMP\,\footnote{www.gimp.org}. In the bottom row, a simplified cartoon is presented, which illustrates our interpretation of the radio data. We consider structures (a), (c), (d) and (e) to be convincing loop structures (3$\upsigma$-detections). The other structures (b), (f), (g), (h), (i), (j) and (k) show either extended emission (like (f) and (g)) or loop structures mostly at a 2$\upsigma$-level, which we include for completeness sake (see below for description of individual structures). 

Generally, the halo below the disk is slightly more extended toward northeast (e.g., Fig. 1). The halo above the disk shows more prominent features. Looking carefully, we can discern a general behavior of most of the loops and bubble-like structures (see Fig.~\ref{fig:bubble-model}). The center of the structure has higher radio intensities, which seems to be finger-like or circular. This is surrounded by faint circular emission forming the loop or bubble-like structure.

Furthermore, we compare the structures in Fig.~\ref{fig:N4217-bubble_threeparts} to an optical artificial B-band image (combined from u and g filters of SDSS via the formulas of Lupton 2005$^3$), where sharpening was applied via the {\it unsharp masking} filter of GIMP and an H$\alpha$ image \citep{rand1996} with the {\it edge detection} filter applied via GIMP. We show these two images together with the C-configuration C-band radio map. This view is inspired by the comparison done in \citet{thomsonetal2004} with HST data, showing here all three images on the same scale. The yellow contour lines mark the location of the presented structures in the list below. The green and blue cutouts show a zoom-in to loop structures of the optical and H$\upalpha$ images with a cartoon made accordingly to the structures. The yellow arrows show extended features of the H$\upalpha$ image. Compare Appendix Fig.~\ref{fig:N4217-bband_halpha} to see the images without the marked locations. We include the complete set of radio halo features in the itemized description below since they all seem to have either footprints in the disk, extensions or halo features in H$\upalpha$ emission.

\begin{itemize}
 \item[(a)] This is a relatively small loop or bubble-like structure. In H$\upalpha$, extended emission is visible at the two locations where the radio structure connects to the disk (see yellow arrows). \\[-1em]
 
 \item[(b)] This loop or bubble-like structure reaches 4\,kpc out into the halo, which is one of the highest extents of the structures seen here. Although we note that the outer structure is roughly at the 2$\upsigma$ level only. However, at the two locations where the bubble connects to the disk, extended H$\upalpha$ emission is visible (see yellow arrows). \\[-1em]
 
\item[(c)] This is the second largest feature. H$\upalpha$ as well as the B-band image show corresponding features at that location. The H$\upalpha$ image seems to indicate a loop or bubble-like structure and extended features (see yellow arrows). The optical image shows dust filaments reaching into the halo. \citet{thomsonetal2004} found at that location two shell-like structures close to the disk.\\[-1em]

\item[(d)] This feature seems to be slightly rotated to the left side. In the optical B-band image, a loop seems to be present at this location, which was also found by \citet{thomsonetal2004}. We show the loop structure in the green cutout of panel 1. In H$\upalpha$, the right location where the structure connects to the disk is visible as extended emission (see yellow arrow).\\[-1em]

\item[(e)] This superbubble-like structure is the most prominent one and is present in H$\upalpha$ and optical as well. The extent of the radio emission into the halo is 3.3\,kpc. In H$\upalpha$ (Fig.~\ref{fig:N4217-bubble_threeparts}.2), diffuse emission is clearly visible, extending toward the halo. In the optical B-band image, the dust filaments indicate an outflow at this location. We present the green cutout in Fig.~5.1 to show the dust filaments forming arc-like structures. These filaments were also found by \citet{thomsonetal2004}. They detected ionized emission at the positions of the white spots within the disk, under loop (e), in the B-band image in Fig.~5.1. These are associated with prominent OB complexes. We find that these OB complexies coincide with the off-center total radio emission peak (position marked with a red cross in Fig. 5.3 and labeled in Fig. 1, top). This finding indicates that structure (e) is probably a superbubble.\\[-1em]

\item[(f)] This filamentary structure is at the 2$\upsigma$ level. It is not clear from the radio map whether the structures (f) and (g) have one origin or two. Nevertheless there is a prominent extension also in H$\upalpha$ at that location.\\[-1em]

\item[(g)] This structure consists of a couple of  extensions, possibly reminiscent of a fragmenting shell. Again, we include it since we see diffuse H$\upalpha$ emission in the halo at that location. \\[-1em]

\item[(h)] This bubble-like or loop feature is relatively small and faint with the outer loop being at a 2$\upsigma$ level. Nevertheless, extended H$\upalpha$ emission is seen at the two locations where the loop emanates from the disk (see yellow arrows).\\[-1em]

\item[(i)] This shell structure is also reminiscent of a fragmenting bubble. Its faintest structures reach far into the halo, with a 3$\upsigma$-level of intensity close to the disk. See also Fig.~\ref{fig:N4217-bw} for a better view of this structure in the radio. The H$\upalpha$ image in Fig. 5.2 shows ionized features within this shell structure, similar to superbubble (e).\\[-1em]

\item[(j)] A high intensity, relatively round dot-like feature extends into the radio halo at this location. The surrounding putative shell is only at the 1.5$\upsigma$ level. An H$\upalpha$ feature appears to emanate from the disk.
\end{itemize}


\subsubsection{Integrated spectral index}
\begin{figure}[h!]
\includegraphics[width=0.48\textwidth]{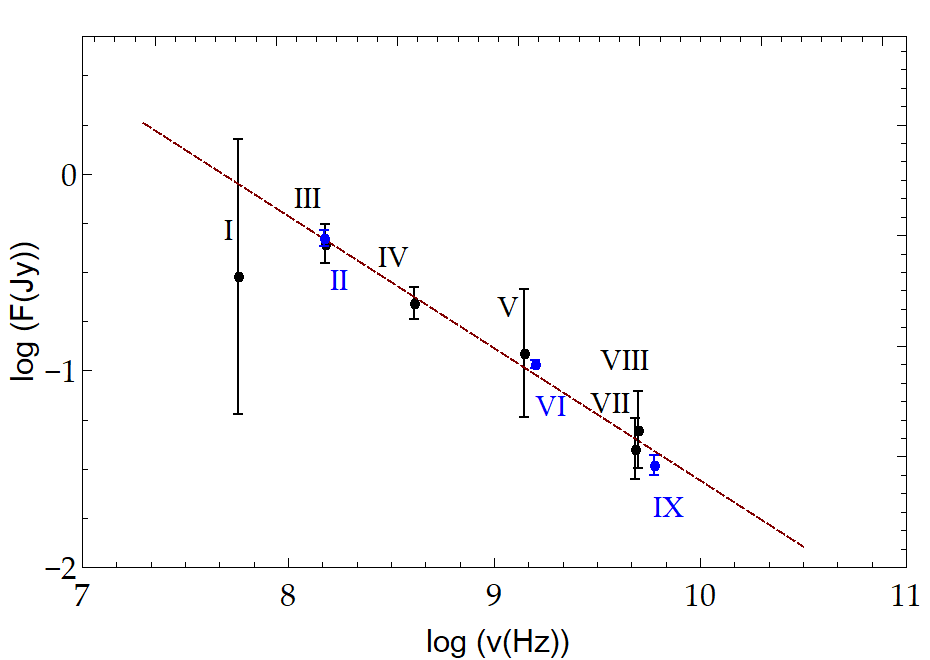}
\caption{NGC~4217 flux density measurements with a fit through all data points but the one at 57.5 MHz with a spectral slope of $\upalpha$~=~-0.67~$\pm$~0.05. References are given in Table~\ref{tab:fluxmeasuren4217}.} 
\label{fig:N4217_Fluxmeasure}
\end{figure}

Table~\ref{tab:fluxmeasuren4217} lists all flux density measurements from this galaxy in the literature with uncertainties as well as the flux density measurements from this work. The spectral index between the C-band and L-band data of CHANG-ES is $\upalpha$~=~$-0.85$~$\pm$~0.05. Figure~\ref{fig:N4217_Fluxmeasure} shows the integrated spectral behavior of NGC~4217 with a spectral slope of $\upalpha$~=~$-0.67$~$\pm$~0.05, which is a fit to all data points but the first one of \citet{israelmohoney1990}, since it has a large error. To further investigate the low-frequency spectral index, the LOFAR data (resolution of 10") are used to measure the total flux density of 470~$\pm$~40\,mJy at a frequency of 150\,MHz. This value is in accordance with 440~$\pm$~50\,mJy at 151\,MHz \citep{halesetal1988}. This indicates that there is no flattening of the spectral index toward low frequencies.

\begin{table}
\centering
\caption{Integrated flux densities of NGC~4217.}  
\label{tab:fluxmeasuren4217}
\begin{threeparttable}
\begin{tabular}{lcccc} 
\hline\hline
\#&Freq.&   Flux dens. 	& log(Freq.)& log(Flux dens.)\\
		& (GHz)		& 	(Jy)		& 	\	&	\	\\
\hline

I  & 0.0575  &0.30~$\pm$~0.20   &7.76	& -0.52~$\pm$~0.69 \\
\om{II} & 0.150  & 0.47~$\pm$~0.04 &   8.18   & -0.33~$\pm$~0.04\\
III & 0.151   &0.44~$\pm$~0.05   &8.18 & -0.36~$\pm$~0.09\\	
IV & 0.408  &0.22~$\pm$~0.02   &8.61  &-0.66~$\pm$~0.07\\	
V	  &1.40    &0.123~$\pm$~0.044  &9.15	&-0.91~$\pm$~0.32  \\
\om{VI}   &1.58     &0.108~$\pm$~0.005 & 9.20 & -0.96~$\pm$~ 0.02 \\	
VII     &4.85    & 0.040~$\pm$~0.007 &9.69	&-1.40~$\pm$~0.15	\\
VIII		&5.0		 & 0.050~$\pm$~0.011 &9.70	&-1.30~$\pm$~0.19	\\
\om{IX}	  &6.00    & 0.033~$\pm$~0.002 &9.78	&-1.48~$\pm$~0.05 \\	
\hline
\end{tabular}
\begin{tablenotes}
\footnotesize
\item {\bf References.}\\
I: Low-frequency radio continuum evidence for cool ionized gas in normal spiral galaxies \citep{israelmohoney1990}\\
II: LOTSS (this work)\\
III: The 6C survey of radio Sources \citep[6C,][]{halesetal1988}\\
IV: A new Bologna sky survey at 408 MHz \citep{ficarraetal1985}	\\
V: The NRAO VLA Sky Survey \citep[NVSS,][]{condonetal1998}\\
VI: CHANG-ES L-band (this work)\\
VII: The 87GB Catalog of radio sources \citep{gregorycondon1991}\\
VIII: The 5-GHz survey of bright sources \citep{Sramek1975}\\
IX: CHANG-ES  C-band (this work)
\end{tablenotes}
\end{threeparttable}
\end{table}
\normalsize

The integrated total flux density in L-band of all VLA array configurations is 108~$\pm$~5\,mJy with an effective frequency after flagging of 1.58\,GHz. The C-band (6\,GHz) flux density of the combined array configurations is 33~$\pm$~2\,mJy and in good agreement with the literature value of 40~$\pm$~7\,mJy at a frequency of 4.85\,GHz from \citet{gregorycondon1991}. The resulting spectral index from the latter two flux density measurements is $\upalpha$~=~$-0.91$~$\pm$~0.08. In comparison, the flux density measurement at 5\,GHz \citep{Sramek1975} is 50~$\pm$~11\,mJy, observed with the no longer operating 300-foot telescope in Green Bank (single-dish telescope). From this measurement, an indication of the total flux density at 6\,GHz is obtained and thus also whether missing flux density should be considered. If we assume the fit spectral index of $\upalpha$~=~$-0.67$~$\pm$~0.05 from Fig.~\ref{fig:N4217_Fluxmeasure}, a flux density of 44~$\pm$~10\,mJy is expected at 6\,GHz, which is not significantly different from the value measured in this paper.

VLA observations at C-band in C- and D-configuration are missing large-scale flux density for extended structures larger than about 240". As the radio extent of NGC~4217 at C-band is $\sim$200", our flux density is likely not influenced by missing spacings.

\subsubsection{Total intensity profiles - scale heights}
\label{sec:scaleheights}
Total intensity profiles of edge-on spiral galaxies can be best fit by exponential or Gaussian functions, which gives us scale heights of the thin and thick disk. These terms are used here to signify the disk (thin disk) and the halo (thick disk). 

\begin{figure}
	\centering
		\includegraphics[width=0.49\textwidth]{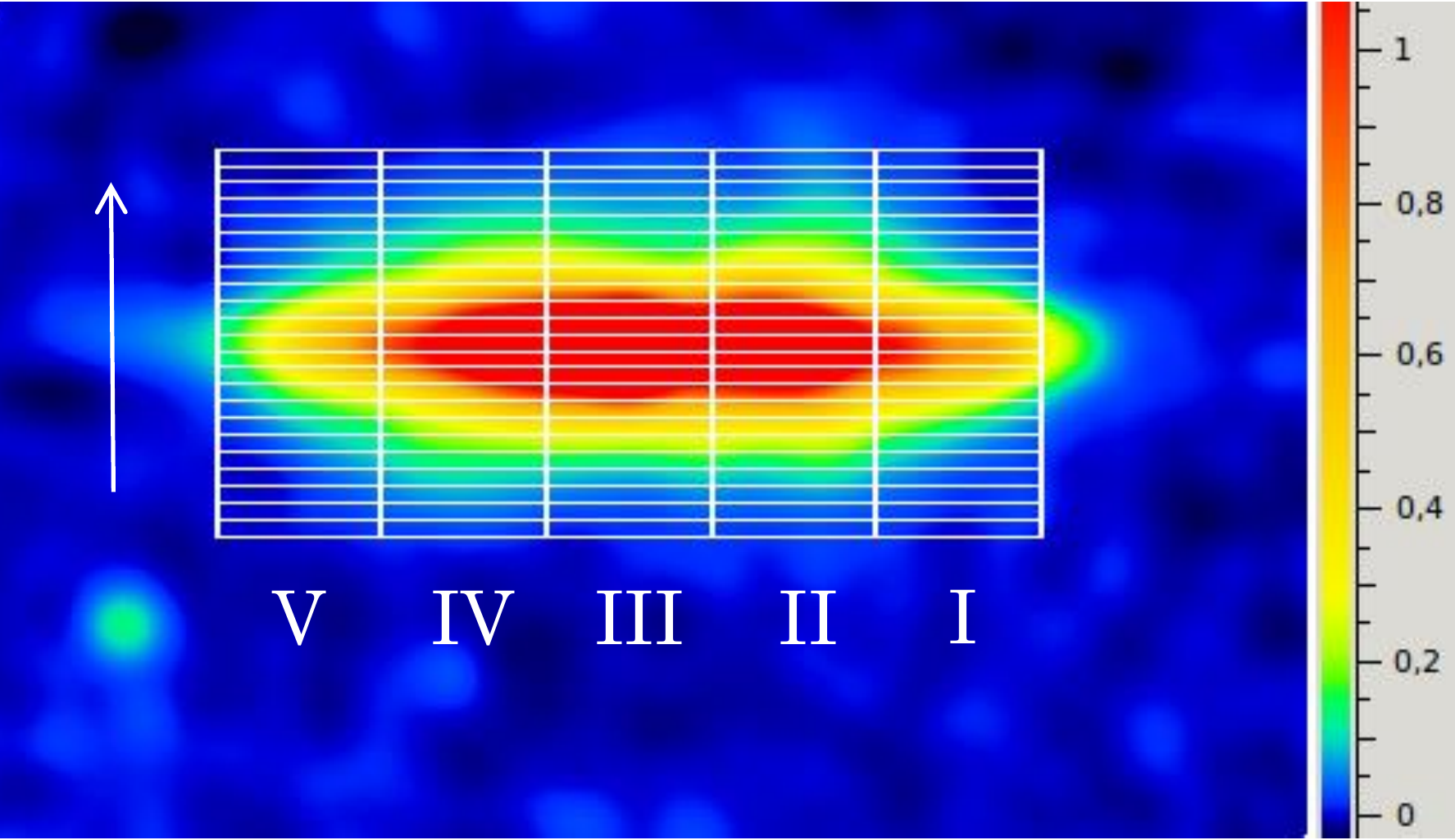}
\caption{NGC~4217 screenshot of the five strips fit with NOD3 (combined C-band). Fitting was done for all five vertical strips (labeled I - V), using a set of 20 points along the arrow corresponding to the mean value within each box. The box size is 5" x 35". The fitting results are presented in Figs. \ref{fig:N4217-Ccomb-nod3}-\ref{fig:N4217-Lofar-nod3}. }
	\label{fig:N4217_nod3_screenshot}
\end{figure}

\begin{SCfigure*}[][h!]
	\centering
		\includegraphics[width=0.70\textwidth]{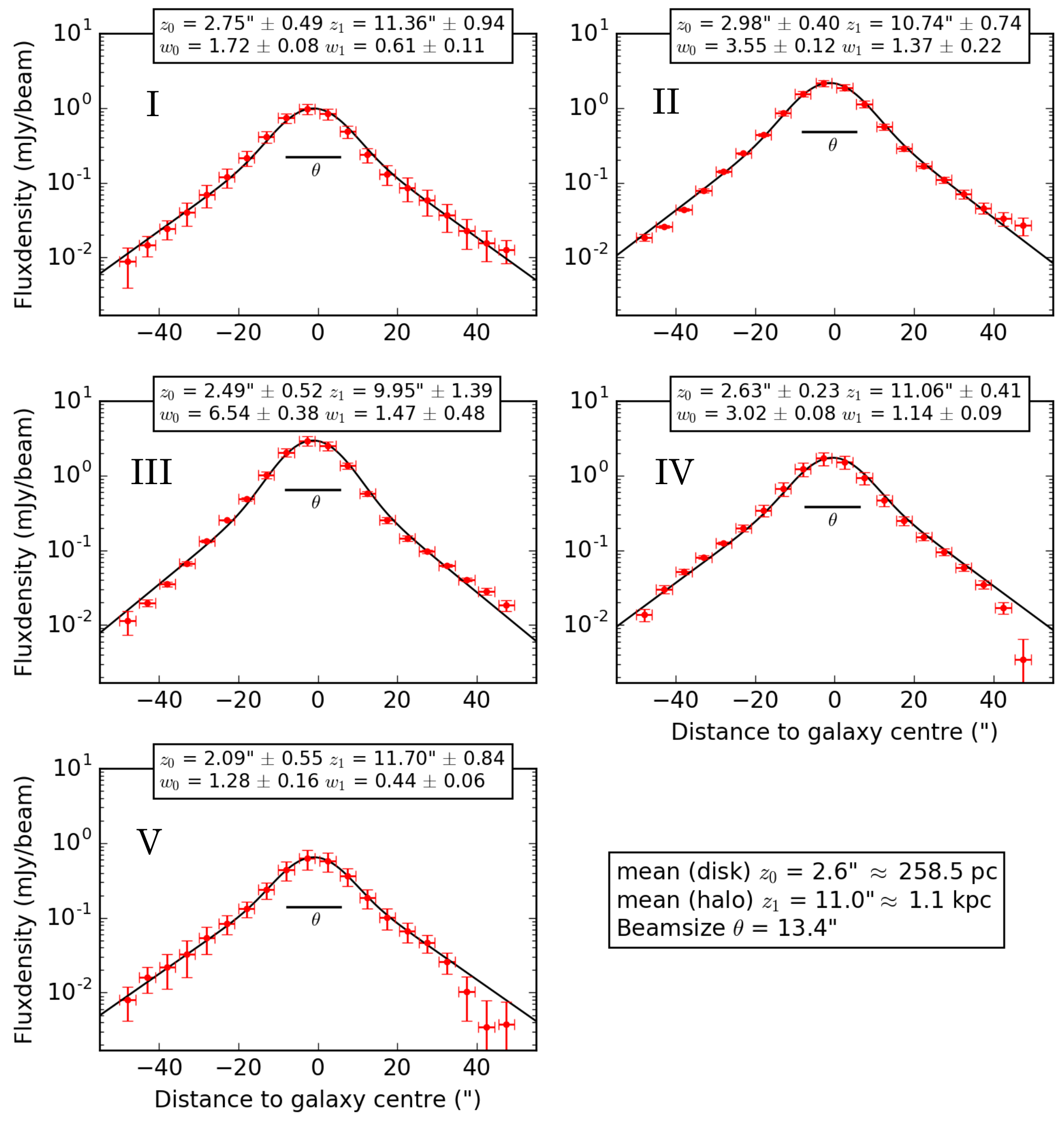}
	\caption{Strip fitting with NOD3 for five vertical strips of NGC~4217 (labeled according to Fig.~\ref{fig:N4217_nod3_screenshot}) on the combined and $uv$-tapered C-band data with a two-component exponential fit. The red dots represent the mean intensity of each box 
	(shown in Fig.~\ref{fig:N4217_nod3_screenshot}, the black line is the fit to the data done by NOD3. The black horizontal line shows the size of the beam $\theta$. The parameter of the two-component exponential fit function are given in the boxes on top of each strip: $z_0$ and $z_1$ are the scale heights of the disk and halo; $w_0$ and $w_1$ are the amplitudes of the disk and halo. The means of the five scale heights as well as the beam size are given in the box to the lower right.}
	\label{fig:N4217-Ccomb-nod3}
\end{SCfigure*}

\begin{table}[h!]
\caption{NGC~4217 NOD3 parameters.}  
\centering
\label{tab:N4217nod3para}
\begin{tabular}{lccc}
\hline \hline
Parameter 					&C-band	& L-band & LOFAR\\
\hline
Beam (")				&	13.4	& 13.4  &  13.4 \\
Inclination	($^\circ$)	&  89   	 &89	& 89							  \\
Position Angle ($^\circ$)&  50   	 &50	&	50						  \\	
rms	noise (mJy/beam)			& 0.004			 & 0.017&	0.09						\\
Box width (")			& 35   		&  35	& 35				  \\
Box height (")			& 5    		&5 		&	5					  \\
Number of boxes in X	&  5   		 &5		& 5						  \\
Number of boxes in Y	&  20   	& 20		&	20				  \\
\hline
\end{tabular}
\end{table}

\begin{SCfigure*}
	\centering
		\includegraphics[width=0.70\textwidth]{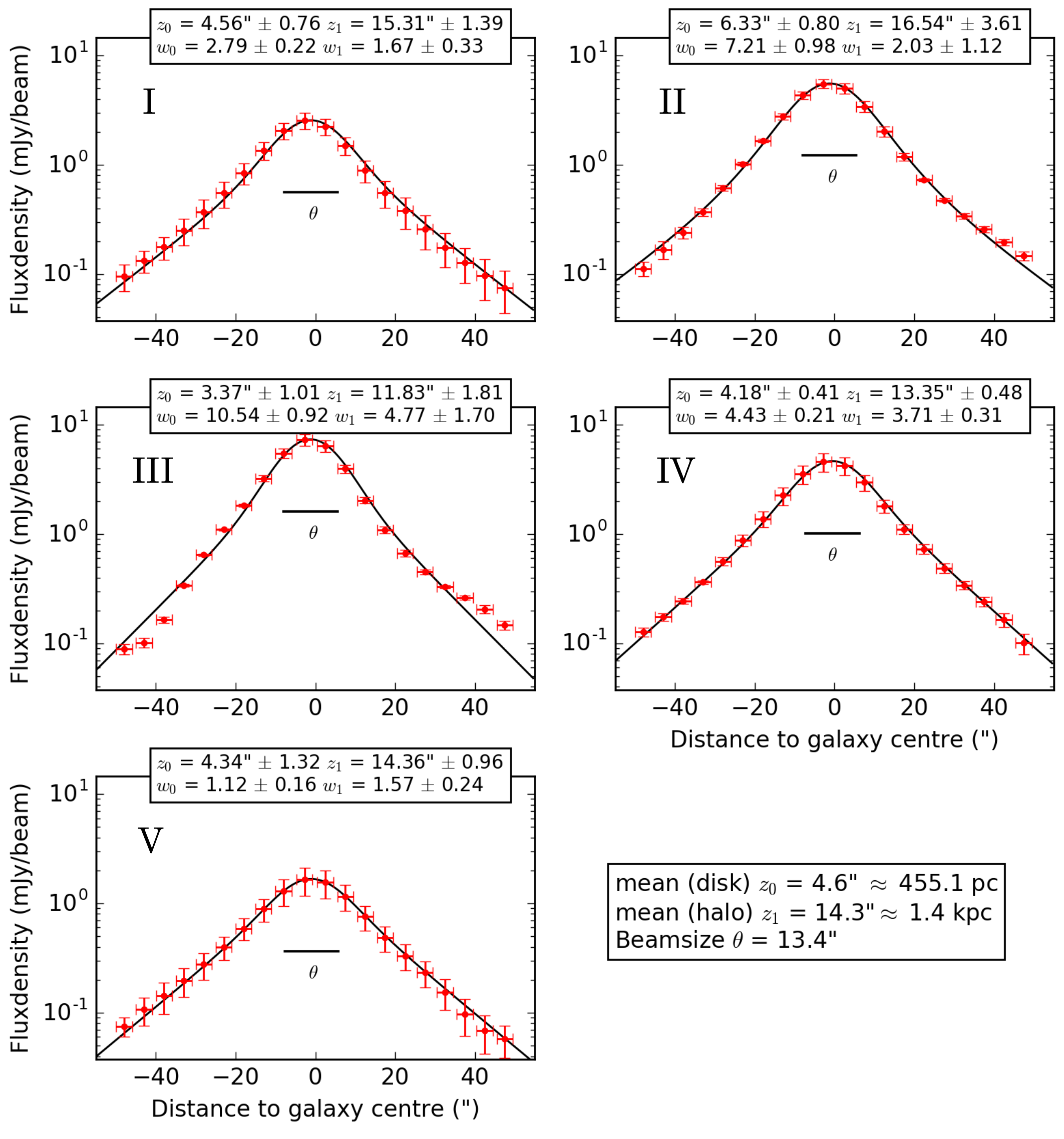}
	\caption[NGC~4217 strip fitting with NOD3 (combined L-band).]{Strip fitting with NOD3 for five strips of NGC~4217 on the combined L-band data with a two-component exponential fit. Designations are the same as for Fig.~\ref{fig:N4217-Ccomb-nod3}.}
	\label{fig:N4217-Lcomb-nod3}
\end{SCfigure*}


\begin{table*}[h!]
\centering
\begin{threeparttable}
 \captionof{table}{Properties of the radio emission of the disk and halo of NGC~4217 at three frequencies.}  
\label{tab:N4217albflux_guassian}
\label{tab:N4217scaleheights}
\begin{tabular}{lccccc|c}
\hline \hline
  \multicolumn{6}{c}{Exponential fits} & \multicolumn{1}{|c}{Integrated flux density}\\  
     & $\bar{w}$ & $\bar{z}$  & $\bar{z}$ &  Flux density & Flux density & disk + halo \\
     &(mJy/beam)       & (")                &     (kpc)   &        (mJy)           &       disk+halo (mJy)          &  (mJy)         \\
\hline
C-band &        &     & & & &  \\
Disk  & 3.2~$\pm$~0.1 &  2.6 $\pm$ 0.2 & 0.26 $\pm$ 0.02&14.2 $\pm$ 1.2 & \multirow{2}{*}{33.0 $\pm$ 2.3} & \multirow{2}{*}{33 $\pm$ 2} \\
Halo & 1.0~$\pm$~0.1 & 11.0 $\pm$ 0.4 & 1.10 $\pm$ 0.04 & 18.8 $\pm$ 2.0 & &\\
Ratio & 3.2~$\pm$~0.3  & 0.24~$\pm$~0.02 $^1$                   &&  0.75~$\pm$~0.10  & &   \\
\hline
L-band  &        &   & & &  & \\
Disk  & 5.2~$\pm$~0.2 & 4.6 $\pm$ 0.4  & 0.46 $\pm$ 0.04 & 41.4 $\pm$ 3.9 &\multirow{2}{*}{108.1 $\pm$ 7.3} &  \multirow{2}{*}{108 $\pm$ 5} \\
Halo & 2.7~$\pm$~0.2 & 14.3 $\pm$ 0.8   & 1.43 $\pm$ 0.09  & 66.7 $\pm$ 6.2 & &   \\
Ratio &  1.9~$\pm$~0.2 & 0.32~$\pm$~0.03 &     & 0.62~$\pm$~0.08 & &  \\
\hline
LOFAR  &        &     & & & &  \\
Disk  & 10.3~$\pm$~1.1 & 4.9 $^2$ $\pm$ 1.4 &  0.49 $\pm$ 0.15 & 87.3 $\pm$ 8.4 &\multirow{2}{*}{481.5 $\pm$ 20} & \multirow{2}{*}{470 $\pm$ 40} \\
Halo &  14.7~$\pm$~1.1 & 15.5 $^3$ $\pm$ 0.36   & 1.55 $\pm$ 0.04 &  394.2 $\pm$ 18.7 & & \\
Ratio &  0.70~$\pm$~0.09      & 0.32~$\pm$~0.09  && 0.22~$\pm$~0.07  & & \\
\hline
\end{tabular}
\begin{tablenotes}
\item \textbf{Notes.} $\bar{w}$ as the mean amplitudes and $\bar{z}$ the mean scale heights of the exponential fits. The column ``flux density'' lists the integrated flux densities of the disk and halo calculated from the exponential fits. The last column displays the integrated flux densities from Table~\ref{tab:fluxmeasuren4217}. \\
$^1$ Lower limit if only the synchrotron emission is considered. $^2$ Strip no.\,2 not included. $^3$ Strip no.\,1 not included.
\end{tablenotes}
\end{threeparttable}
\end{table*}
\normalsize

\begin{SCfigure*}
	\centering
		\includegraphics[width=0.70\textwidth]{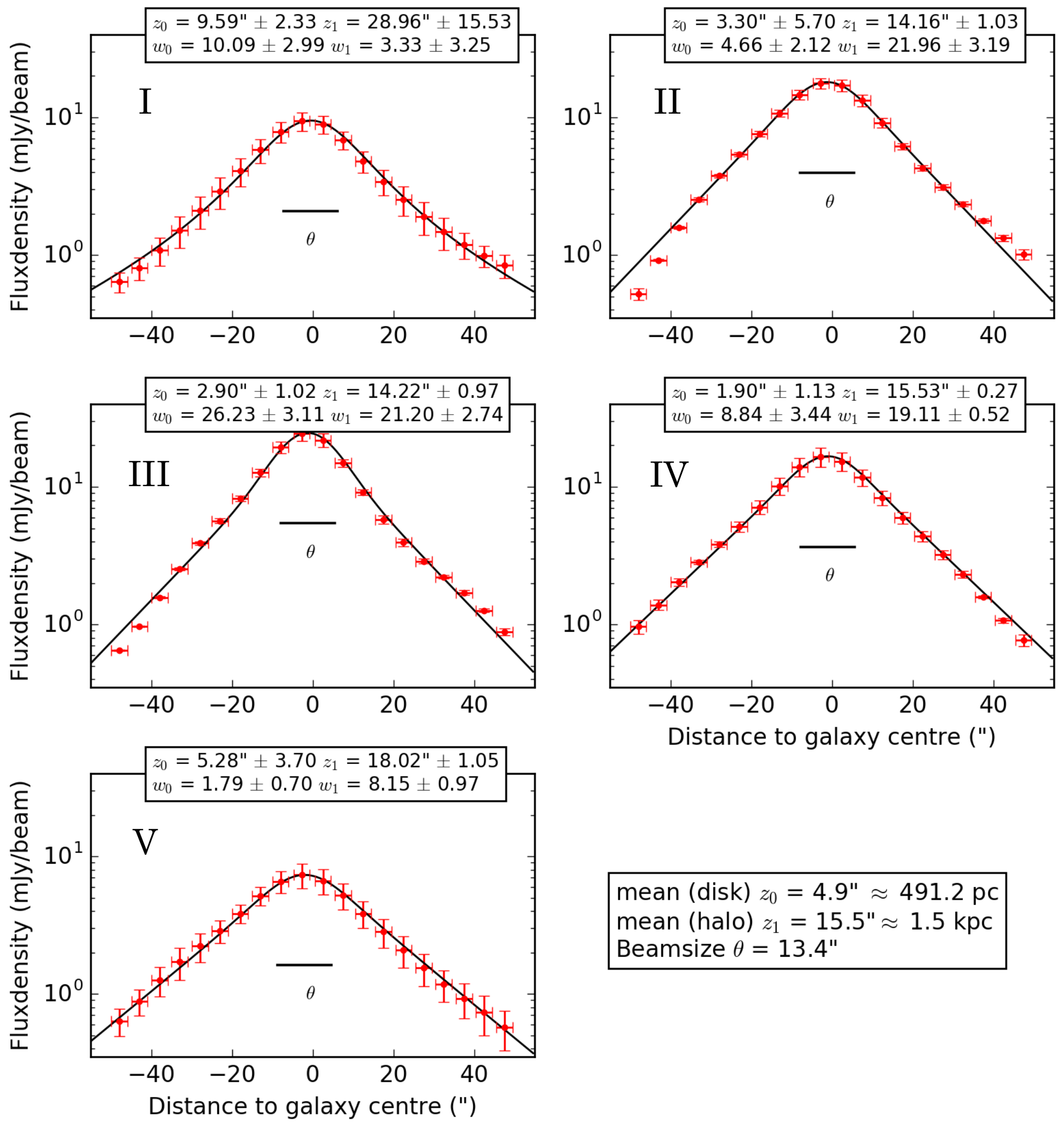}
	\caption[NGC~4217 strip fitting with NOD3 (LOFAR).]{Strip fitting with NOD3 for five strips of NGC~4217 on LOFAR data with a two-component exponential fit. Designations are the same as for Fig.~\ref{fig:N4217-Ccomb-nod3}. The mean of $z_0$ was calculated without strip two, the mean of $z_1$ was calculated without strip one.}
	\label{fig:N4217-Lofar-nod3}
\end{SCfigure*}

We follow \citet{dumke1995} to describe the observed intensity profile as the convolution of a Gaussian beam profile with the true intensity profile of the galaxy, which could be either exponential or Gaussian. We use the NOD3 software package \citep{mulleretal2017} with the implemented "BoxModels" tool to fit two-component exponential functions, determined by the amplitudes of the first and second component, w$_0$ and w$_1$, and the scale heights of the first and second component, z$_0$ and z$_1$. We apply the fit to the intensity profiles of the combined configuration maps of C-band and L-band as well as to the LOFAR map, which were smoothed to the same beam size of 13.4" x 13.4" in order to compare the results. The scale height analysis was carried out in an analogous way to \citet{krauseetal2018} using five strips perpendicular to the major axis (see Fig.~\ref{fig:N4217_nod3_screenshot}). Each of the 20 boxes in each vertical strip in Fig.~\ref{fig:N4217_nod3_screenshot} gives one intensity measurement corresponding to the mean inside the box. The error is calculated by the standard deviation. The box height was chosen to be approximately half of the beam, which results in a height of 6" for the three bands. The box width was set to 35", thus the five strips cover the entire galaxy NGC~4217. The total intensity profiles of NGC~4217 show a well defined two-component exponential disk which is represented by the thin disk and the thick disk, which we refer to as the disk and the halo, respectively, as described in Section 1.

The parameters used in NOD3 are summarized in Table~\ref{tab:N4217nod3para}. Figure~\ref{fig:N4217-Ccomb-nod3} displays the intensity profiles of the five strips and the fit two-component exponential at C-band from NOD3. The intensity profiles at L-band are shown in Fig.~\ref{fig:N4217-Lcomb-nod3} and the profiles at the LOFAR frequency are shown in Fig.~\ref{fig:N4217-Lofar-nod3}. The order of the strips in these figures are labeled corresponding to Fig.~\ref{fig:N4217_nod3_screenshot}. Negative distances to the midplane of the galaxy are below the midplane, positive above the midplane.

The resulting scale heights of the fitting are presented in Table~\ref{tab:N4217scaleheights}. The scale height analysis from NOD3 shows a disk component of 260~$\pm$~20\,pc (2.6~$\pm$~0.2") at C-band, 460~$\pm$~40\,pc (4.6~$\pm$~0.4") at L-band, and 490~$\pm$~150\,pc (4.9~$\pm$~1.4") at 150\,MHz. The halo component is 1.10~$\pm$~0.04\,kpc (11.0~$\pm$~0.4") at C-band, 1.43~$\pm$~0.09\,kpc (14.3~$\pm$~0.8") at L-band, and 1.55~$\pm$~0.04\,kpc (15.5~$\pm$~0.36") at 150\,MHz. \citet{hummeletal1991} also found a clear two-component structure from VLA C-band observations of 5.3" for the thin disk (disk) and 12.5" for the thick disk (halo). These values are slightly higher in comparison to the scale heights found here. \citet{krauseetal2018} found the halo scale heights in NGC~4217 to be 1.08~$\pm$~0.05\,kpc at C-band and 1.06~$\pm$~0.09\,kpc at L-band. While the C-band value is comparable, their L-band value is smaller than our result at the 2.8 sigma level. One reason might be that we use the combined array configurations to fit the data, while \citet{krauseetal2018} only included C-band D-configuration and L-band C-configuration.

\subsubsection{Integrated flux density ratios of the disk and halo}

Following \citet{steinetal2019b}, we determine the absolute flux density ratios between disk and halo for the three frequencies from the scale height exponential fit results. We use the mean values for the scale heights ($\bar{z}$) and the amplitudes ($\bar{w}$) at each frequency. All parameters are listed in Table~\ref{tab:N4217scaleheights}. 

The exponential distribution of an intensity profile of a galaxy is:
\begin{align}
w(z) = w_0 \ \text{exp}\,(-\,z/z_0).    
\end{align}
We compute the integral of the above exponential (taken twice between zero and infinity):
\begin{align}
W (\text{mJy $\cdot$ arcsec /\,beam}) = 2 \, w_0 \, z_0.    
\end{align}
Here, $w_0$ is in mJy/beam and $z_0$ in arcsec. To get the flux density in the correct unit of mJy, we firstly need to divide by the beam area in arcsec ($b$=203.5\,arcsec$^2$). To determine the total flux density $F$ of the disk and the halo, we multiply with the extent of the major axis, $e$ = 174" at C-band, $e$ =  176" at L-band, and $e$ =  176" at the LOFAR frequency by again using the mean values of the amplitude and scale height: 
\begin{align}
F (\text{mJy}) = 2 \,  \bar{w} \, \bar{z} \, e\, /\, b.   
\end{align}

The results are presented in Table~\ref{tab:N4217scaleheights}. Adding the flux densities of the disk and halo, these values agree with the total flux densities given in Table~\ref{tab:fluxmeasuren4217} within the uncertainties. While at C-band the disk contributes nearly as much as the halo to the total flux density, at L-band the fraction of the disk contribution decreases, and at the LOFAR frequency the disk only contributes 18\% to the total flux density. A comparison to the results of NGC~4013 and NGC~4666 from \citet{steinetal2019b} is given in the Discussion (Sec.~\ref{sec:discussion_transport}).

The integrated flux densities in Table~\ref{tab:N4217scaleheights} allow us to compute the spectral indices $\alpha$ for the disk and halo of NGC~4217 separately. For the disk we obtain $\alpha=-0.32\pm0.06$ between LOFAR 150\,MHz and L-band and $\alpha=-0.77\pm0.09$ between L- and C-band, and for the halo  $\alpha=-0.77\pm0.05$ between LOFAR 150\,MHz and L-band and $\alpha=-0.91\pm0.13$ between L- and C-band.

\subsection{Magnetic field}
\label{sec:magfield}

With the linear polarization data (Stokes Q and U) of NGC~4217, it is possible to constrain the magnetic field. We do this first in the classical way by using the Stokes Q and U maps to calculate the polarized intensity and magnetic field orientations. The latter is determined by rotating the polarization angles, which are the electric field angles, by 90 degrees. Secondly, using RM-synthesis, where the magnetic field orientations are a direct outcome. We furthermore apply RM-synthesis to determine the RM map, which is dependent on the magnetic field along the line of sight.

\begin{figure*}[h!]
	\centering
		\includegraphics[width=1.0\textwidth]{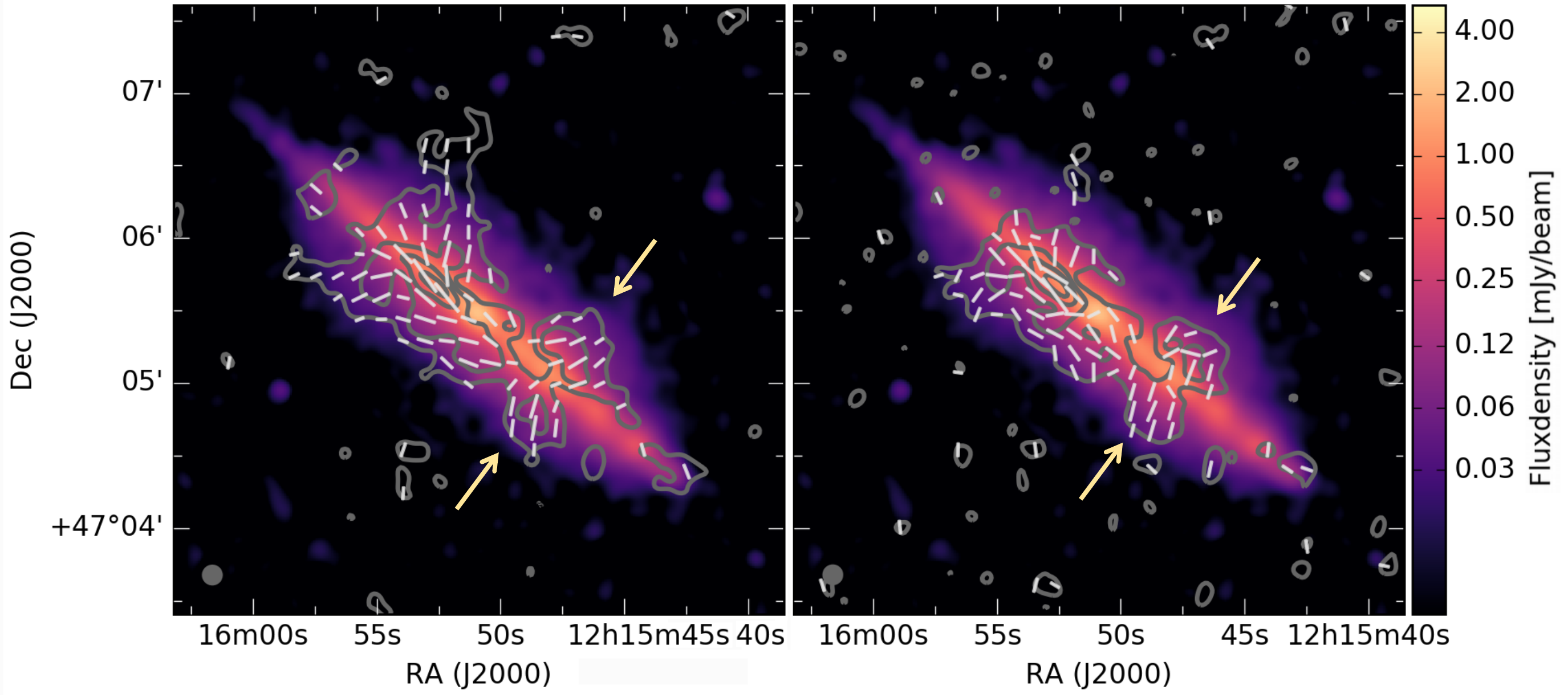}
	\caption{Total intensity image of NGC~4217 at C-band (same as Fig.~\ref{fig:N4217-Ccomb-uvtap}, top). Gray polarized intensity contours  at 3, 6, 9, 12, 15 $\upsigma$ levels. The robust weighting parameter of clean was set to two and then smoothed to a resulting beam size of 7.7" $\times$ 7.7", shown in the bottom left. The arrows mark the super-bubble structure, visible here in polarized emission. Left: Magnetic field {\bf derived from Stokes Q and U maps} with a $\upsigma$ of 4.8\,$\upmu$Jy/beam. The apparent magnetic field orientations, not corrected for Faraday rotation, are shown in white and clipped below 3$\upsigma$. Right: Magnetic field {\bf derived via RM-synthesis} with a $\upsigma$ of 4.7\,$\upmu$Jy/beam. The intrinsic magnetic field orientations, corrected for Faraday rotation, are shown in white and clipped below 3$\upsigma$.}
	\label{fig:N4217_Ccomb_pol7}
	\label{fig:N4217_Ccomb_pol_synth7}
\end{figure*}
\begin{figure*}[h!]
	\centering
		\includegraphics[width=1.0\textwidth]{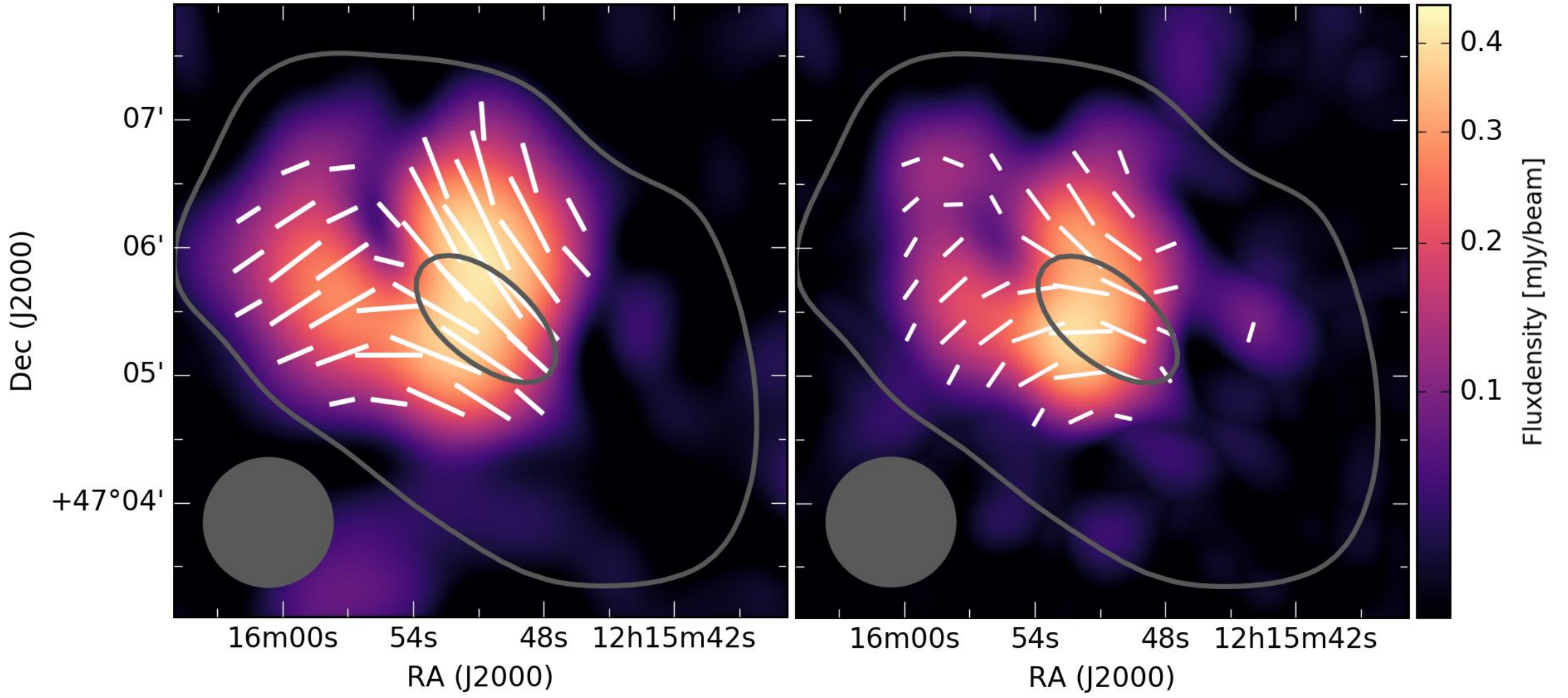}
	\caption{Polarized intensity image of NGC~4217 at L-band. The robust weighting parameter of clean was set to two, smoothed to a resulting beam size of 60" $\times$ 60", shown in the bottom left. Gray total intensity contours from D-configuration L-band smoothed to 60", contours at 3 and 270 $\upsigma$ levels with a $\upsigma$ of 136.7\,$\upmu$Jy/beam. Left: {\bf From Stokes Q and U} with a $\upsigma$ of 35\,$\upmu$Jy/beam. The apparent magnetic field orientations, not corrected for Faraday rotation, are shown in white and clipped below 3$\upsigma$. Right: {\bf From RM-synthesis} with a $\upsigma$ of 32\,$\upmu$Jy/beam. The intrinsic magnetic field orientations corrected for Faraday rotation, are shown in white and clipped below 3$\upsigma$.}
	\label{fig:N4217_DLcomb_synth}
	\label{fig:N4217_DLcomb_pol}
\end{figure*}

\subsubsection{Magnetic field in the sky plane - polarization}
In Figure~\ref{fig:N4217_Ccomb_pol7} on the left, the C-band polarized intensity contours using the classical approach are shown overlaid on the total intensity color image (from Fig.~\ref{fig:N4217-Ccomb-uvtap}). The apparent magnetic field orientations are shown in white. The polarized emission covers many parts of the disk and the halo. The furthest extents are observable in the northeast of the galaxy on both sides of the halo, with the central parts of the galaxy being less extended in polarized emission. The emission on the southwestern side of the galaxy is less prominent in comparison to the emission on northeastern halo. 

The peak of the polarized intensity is toward the northeast within the disk, which is the approaching side of NGC~4217 \citep{verheijensancisi2001}. In the midplane, magnetic field orientations are parallel to the plane of the disk. 

Toward larger radii, the field is more perpendicular to the plane of the disk. The resulting X-shaped magnetic field of this galaxy is compelling. At the end of each side of the disk, there is a patch of polarized emission with plane-parallel magnetic field orientations.

In Figure~\ref{fig:N4217_Ccomb_pol_synth7} (right panel), the polarized emission contours obtained via RM-synthesis are shown on top of the total intensity color image (from Fig.~\ref{fig:N4217-Ccomb-uvtap}). The general appearance of the polarized emission is comparable to Fig.~\ref{fig:N4217_Ccomb_pol7} (left panel), but fewer faint emission structures are recovered. The magnetic field orientations are shown in white. The difference between the magnetic field orientations in the left and right panels is due to Faraday rotation; for $|RM|\lesssim200$\,rad\,m$^{-2}$ (see Section~\ref{sec:RMsynth}), the rotation at C-band should be $\lesssim 30\degr$, while the rotation at L-band for $|RM|\lesssim20$\,rad\,m$^{-2}$ should be $\lesssim 40\degr$, which is indeed recognized in the figures.
 
In both figures, a symmetric bubble-like structure is detected above and below the location of the disk in the southwest, where the off-center second peak of total intensity is found (indicated with arrows in both figures). This superbubble-like structure is found in total intensity as well (feature (e) and (g)/(f) in Fig.~\ref{fig:N4217-bw}). With a diameter of the polarization structure of $\sim$3\,kpc on each side of the halo, it is only slightly smaller than the corresponding extend in total intensity. This structures could be explained by strong star formation in the disk, which leads to several tens of SN explosions and stellar winds, consequently forming a superbubble structure. We find it very interesting that the structure is highly symmetric on both sides of the halo, which implies a SN induced blown-out centrally coming from the disk. We investigate further connections between the polarized emission and the loop and bubble-like structures of total intensity in the Discussion (Sec.~\ref{sec:discussion_loops}).

In summary, in the combined C-band data, nearly the same polarization structures are recovered both with and without RM-synthesis. The faintest polarization structures reaching far into the halo are better recovered with the classic method deriving the polarization.

In the combined L-band data, no significant linear polarized emission could be found with or without RM-synthesis. Thus, the L-band data of all array configurations were analyzed separately. Of the three different array configurations at L-band, only the D-configuration data exhibits polarized emission, located in the approaching half and the center of the galaxy. With the D-configuration having the largest beam of the array configurations, this argues against the possibility that the low polarization is due to beam depolarization. RM-synthesis was not able to reconstruct the polarized emission in the combined L-band due to the small amount of data from the D-configuration in comparison to the other two array configurations (shorter integration time) and thus weak influence on the reconstructed image of polarized intensity. Consequently, we analyzed the D-configuration data separately. The results are shown in Fig.~\ref{fig:N4217_DLcomb_pol} (left) for the classic method deriving the polarization and in Fig.~\ref{fig:N4217_DLcomb_synth} (right) using RM-synthesis. The magnetic field orientations of both figures have comparable orientations to the C-band magnetic field orientations, that is a large-scale X-shaped regular magnetic field structure, at least for the approaching side and the central part of the galaxy. The intensity of the polarization seems to be higher  without RM-synthesis, though some faint emission on the receding side was recovered by RM-synthesis. Major parts of the receding side (southwest) seem to be depolarized in L-band. Potential reasons are investigated in the Discussion (Sec.~\ref{sec:discussion_pol}). 

In summary, with the CHANG-ES data, a large-scale X-shape structure of the magnetic field is revealed, particularly visible at C-band. RM-synthesis produces less strong polarized signals at both CHANG-ES bands and was not able to recover the faint emission equally well at C-band, compared to the classical Stokes Q and U derived polarization. The reason for this is that Stokes Q and U images are produced for each spectral window separately in RM synthesis (see Sec.~\ref{sec:RMsynth}), rather than across the entire bandwidth, causing the signal to noise to decrease and the faint emission is harder to detect.

\subsubsection{Large-scale magnetic field at C-band}

\begin{figure}
	\centering
		\includegraphics[width=0.4\textwidth]{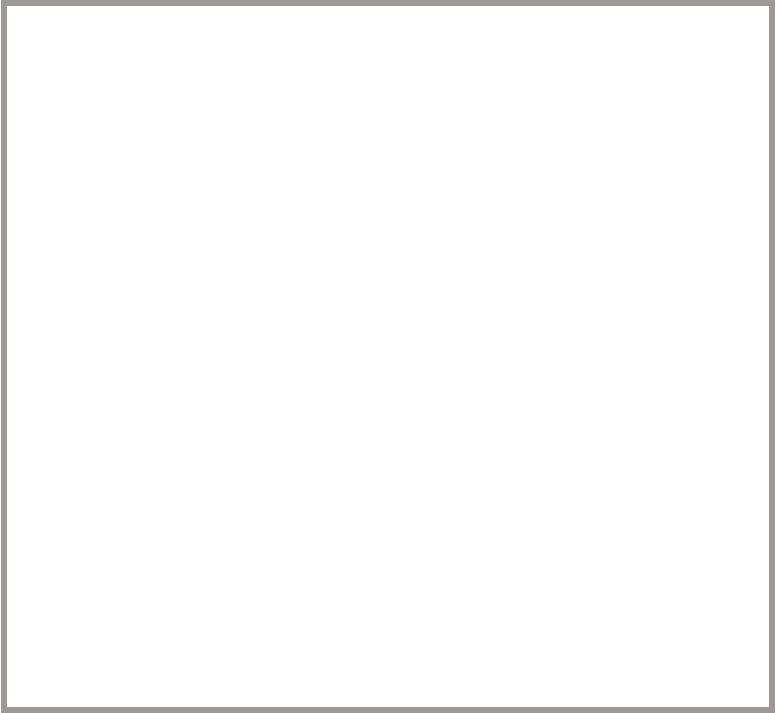}
	\caption{NGC~4217 magnetic field orientations represented by flow lines in green on optical SDSS image as well as the H$\upalpha$ image of \citet{rand1996}. The magnetic field flow lines are produced from the same data as shown in Fig.~\ref{fig:N4217_Ccomb_pol18}.}
	\label{fig:N4217_Ccomb_schoen_LIC2}
	\label{fig:N4217_Ccomb_schoen}
\end{figure}

\begin{figure}
	\centering
		\includegraphics[width=0.50\textwidth]{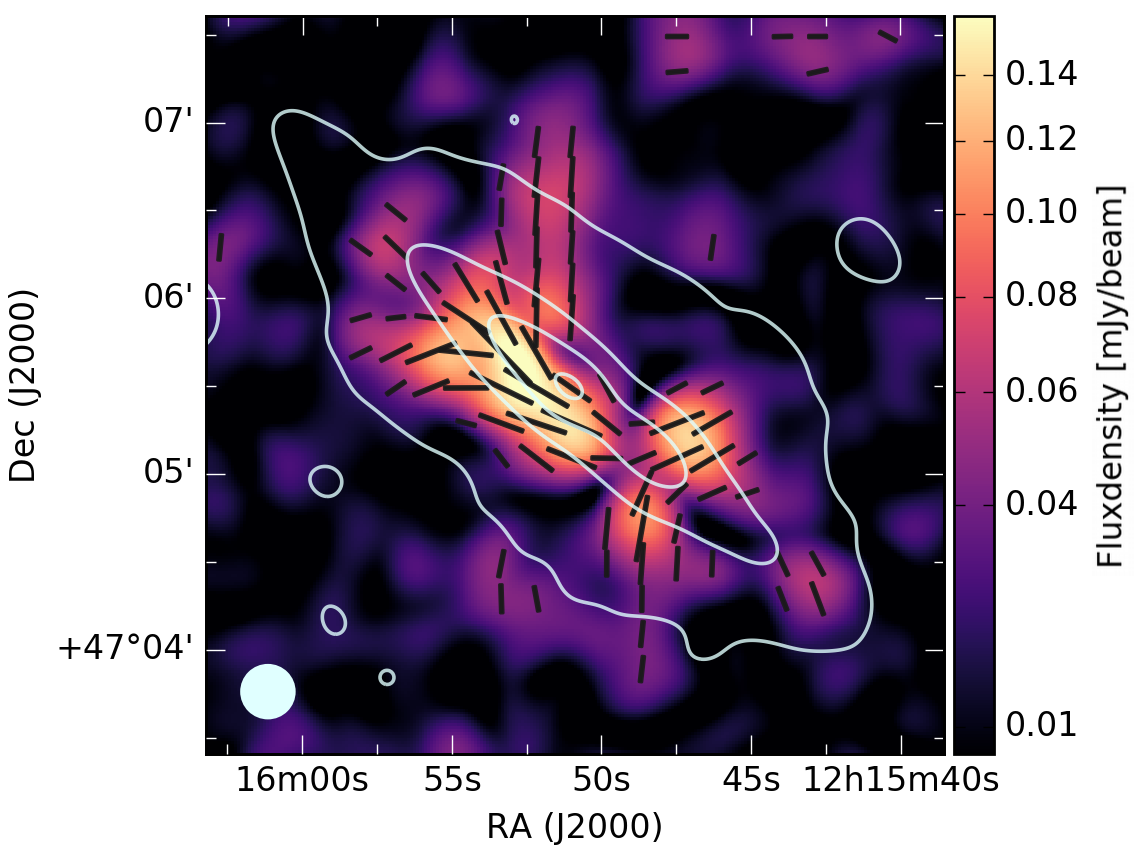}
	\caption{Polarized intensity image of NGC~4217 from Stokes Q and U at C-band with a $\upsigma$ of 13.5\,$\upmu$Jy/beam. The robust parameter of clean was set to two, smoothed to a resulting beam size of 18" $\times$ 18", shown in the bottom left. Total intensity (Fig.~\ref{fig:N4217-Ccomb-uvtap} smoothed to 18") contours at 3, 120, 360, 690 $\upsigma$ levels with a $\upsigma$ of 7.0\,$\upmu$Jy/beam. The apparent magnetic field orientations, not corrected for Faraday rotation, are shown in black and clipped below 3$\upsigma$.}
	\label{fig:N4217_Ccomb_pol18}
\end{figure}

\begin{figure*}[h!]
	\centering
		\includegraphics[width=\hsize]{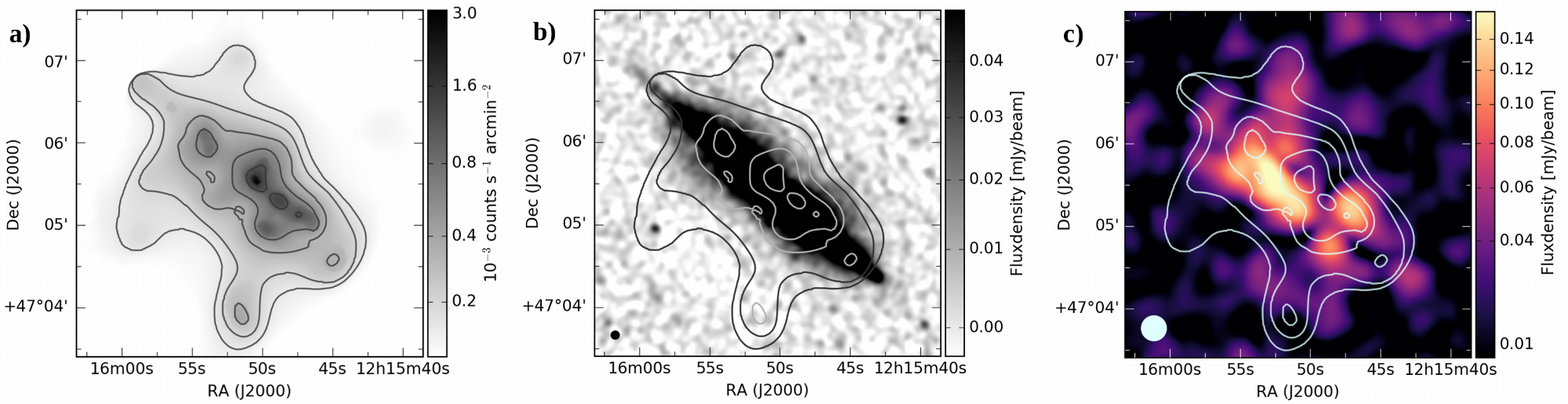}	
		\caption{ NGC~4217 diffuse X-ray emission: {\bf a)} Chandra diffuse X-ray intensity image in the 0.3-1.5 keV band. The contours are at 1.4, 1.6, 2.0, 2.6 and 3.4  $\times 10^{-3} {\rm~counts~s^{-1}~arcmin^{-2}}$ (or about 3, 4, 6,  9, and 13  $\sigma$ above the local background). The image is adaptively smoothed after detected point-like sources are excised. {\bf b)} and {\bf c)} The contours from a) are overlaid on the C-configuration C-band radio intensity image (Fig.~\ref{fig:N4217-bw}, beam size 5.8" x 5.9") and the polarized intensity image (Fig.~\ref{fig:N4217_Ccomb_pol18}, beam size 18").}
	\label{fig:N4217_chandra}
\end{figure*}
To show the large-scale structure of the magnetic field, a polarization intensity map at C-band is produced at a lower resolution of 18", shown in Fig.~\ref{fig:N4217_Ccomb_schoen} and Fig.~\ref{fig:N4217_Ccomb_pol18}. In the central part of the disk, a plane-parallel magnetic field is visible. The polarized intensity is more extended below the disk than above the disk in the central region. In these figures, the large-scale X-shaped magnetic field structure of NGC~4217 is present and even more extended than in the higher-resolution maps. The extent of the X-shaped structure is larger on the northern side of the disk. On the southern side of the disk, where the symmetric superbubble-like structure is located, the magnetic field shows orientations almost perpendicular to the major axis of the disk. Furthermore, the polarized emission extends beyond the total intensity 3$\upsigma$-contour, because we produced Stokes Q and U images with the robust weighting parameter set to two, recovering faint extended emission in polarization\,\footnote{For total power, the robust weighting parameter was set to zero.}

In Figure~\ref{fig:N4217_Ccomb_schoen}, the C-band polarization data with the magnetic field orientations of NGC~4217, are visualized as "flow lines" created by the Line Integral Convolution (LIC) algorithm\,\footnote{modified from\\https://scipy-cookbook.readthedocs.io/items/LineIntegralConvolution.html}. For this, the polarized intensity map as well as the polarization angle map are used as inputs to create a flow map, which connects the magnetic field orientation at each pixel. The flow map is thus similar to plotting the orientations for every pixel. The corresponding data are the same as presented in Fig.~\ref{fig:N4217_Ccomb_pol18}. For the optical composite image, SDSS data (u, g, r filters) as well as the H$\upalpha$ map from \cite{rand1996} were combined in GIMP.

This image visualizes the impressive extent of large-scale magnetic fields; they are maintained throughout the entire galaxy reaching far into the halo.

\subsubsection{Connection to X-ray-emitting hot gas}

Figure~\ref{fig:N4217_chandra} presents the diffuse soft X-ray emission from NGC~4217 (panel a). The image is adaptively smoothed to achieve a non-X-ray background-subtracted signal-to-noise ratio greater $\sim 4$. The detected sources are excised. The X-ray contours encompass the major radio features seen in either the total intensity (panel b) or the polarized component (panel c). Radially (extent of the galactic disk) the soft X-ray emission extends similar to the radio emission, vertically the X-ray is farther extended. In general, the X-ray emission of NGC~4217 seems to better correlate with polarized intensity than to the total radio intensity, as also observed for example in NGC~4666 \citep{steinetal2019a} or NGC~4013 \citep{steinetal2019a}. This may be explained by recent blown-out superbubbles of hot gas surrounded by supershells of enhanced CRE and aligned magnetic fields. We consider the outer X-ray contour to the east as a possible contamination by a background source (see panel b), where there is a point source at that location of the contour in radio or statistical fluctuations. Looking at panel (c) most extended features in polarized emission have a counterpart in X-ray. There is a general similarity between the X-ray and radio emission, although detailed one-to-one correspondence is not apparent -- a case similar to what is seen in a few other edge-on galaxies (e.g., \citealt{wang01,wang03}). This highlights the need to further explore the physical relationship between diffuse hot gas and cosmic-ray/magnetic field and their role in driving the galactic disk/halo interaction.


\subsubsection{Line of sight magnetic field -- rotation measures}

The RM map, produced via RM synthesis (Section~\ref{sec:RMsynth}), represents the fit peak position along the RM axis of the polarization cube for each pixel. It is dependent on the mean magnetic field component along the line of sight, where the RM value is positive for a field pointing toward the observer and negative for a field pointing away from the observer. 

\begin{figure}
	\centering
		\includegraphics[width=0.5\textwidth]{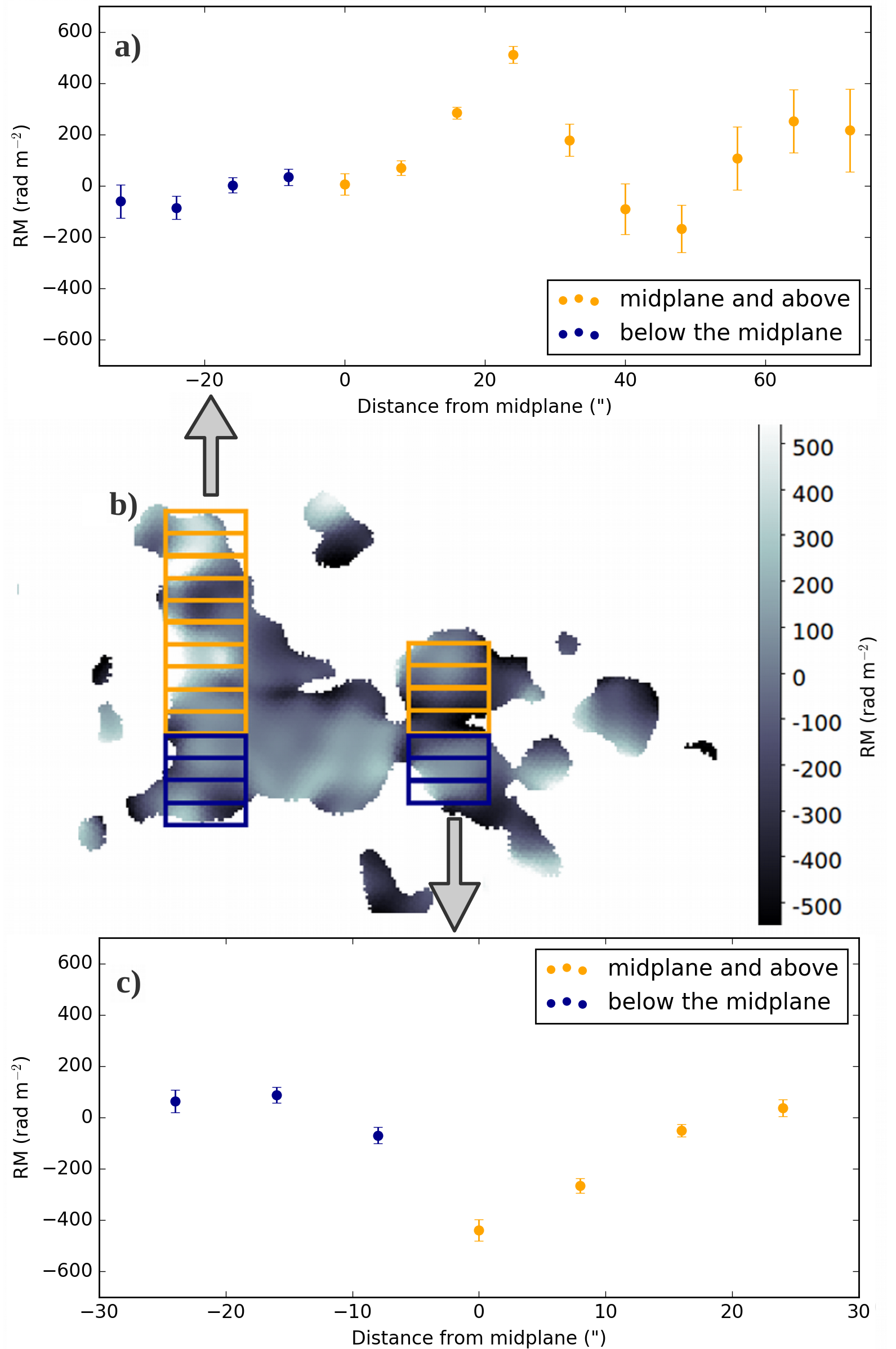}
    \caption{RM map of NGC 4217 at C-band rotated by $40\,\degr$ (panel b), beam size of 13", cut to a level of 3$\upsigma$ of the polarized intensity map. The two strips to produce the vertical profiles are shown on the map with the left one showing the helical-like structure in panel a) and the right one showing the bubble-like structure in panel c). The box size is 18" x 8". }
	\label{fig:N4217_Ccomb_RM}
\end{figure}

The RM map at C-band is shown in Fig.~\ref{fig:N4217_Ccomb_RM}, panel b) and Fig.~\ref{fig:N4217_Ccomb_RM_disk}, panel a). In both figures, the galaxy was rotated by 40\,deg. The values are mostly between -400 and 400\,rad\,m$^{-2}$ and are shown in the two figures with different color maps. Two vertical profiles as well as the disk profile were investigated in detail. To do so, mean Stokes Q and U values were produced for each box (shown in both figures) at each spectral window, then a corresponding Q and U cube was created and finally RM-synthesis applied on these cubes. The resulting profiles are presented here. 

In the rotated view of the galaxy, the left part shows a helical-like structure above the halo. An RM profile was produced for this region shown in Fig.~\ref{fig:N4217_Ccomb_RM}, panel a). Here, the helica structure is visible with a distance between the two peak positions of $\sim$~4\,kpc and a decreasing amplitude from the first to the second peak. This could be caused by a smaller pitch angle of the helix or smaller number density of thermal electrons at a larger distance from the plane of the galaxy.

\begin{figure}
	\centering
		\includegraphics[width=0.5\textwidth]{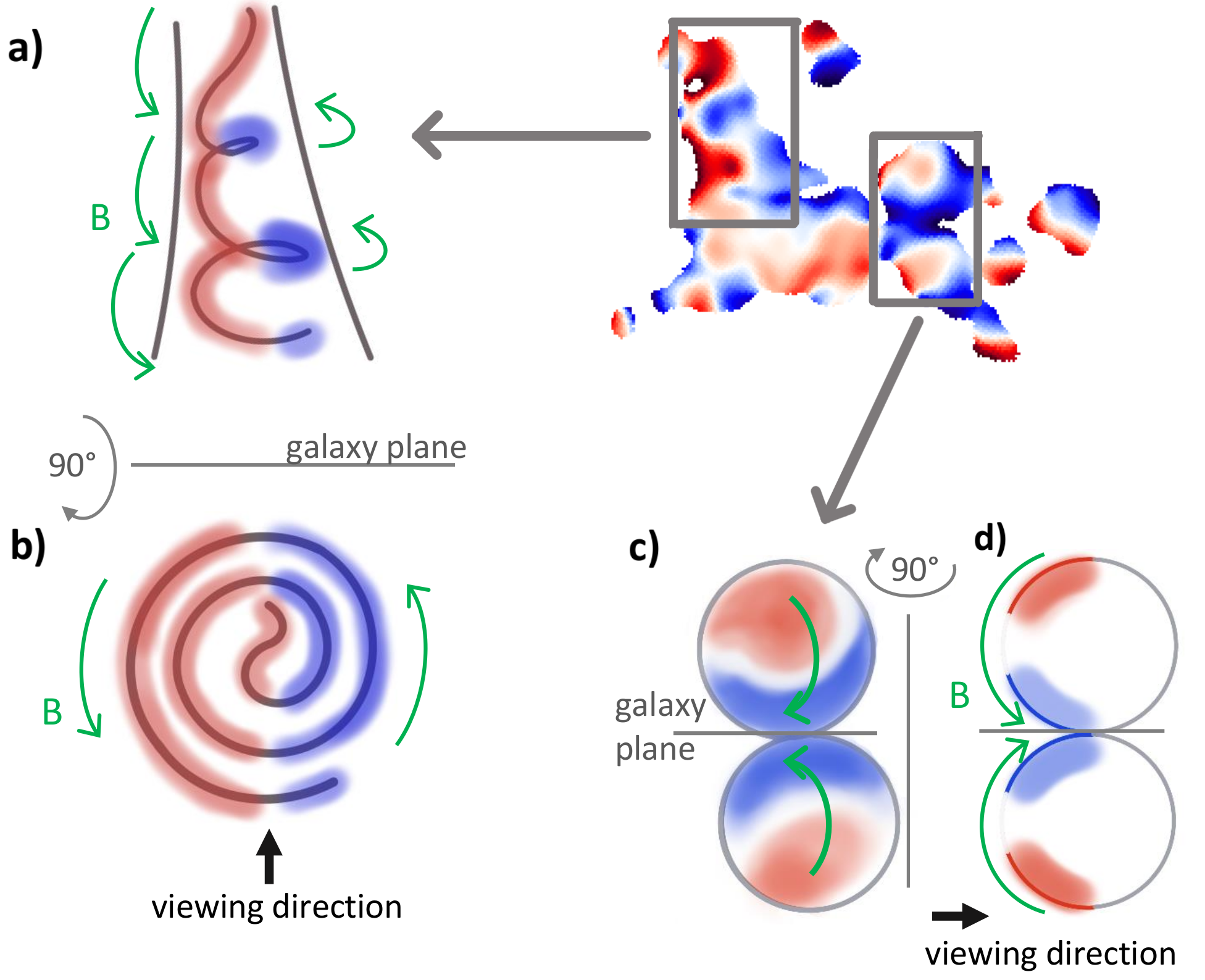}
    \caption{Simplified cartoon of a possible configuration of the magnetic field of the helical-like and bubble-like structures in NGC~4217 (marked with boxes in the RM map in the upper right). Red shaded regions show positive RM values where the field points toward the observer, blue shaded areas show negative RM values where the field points away from the observer. The colors are chosen to match the colors of the RM to the upper right and Fig.~\ref{fig:N4217_Ccomb_RM_disk}. The green arrows mark the magnetic field direction. a) The helical-like structure from the observer's perspective, where the left parts point toward the observer and the right parts point away from the observer; b) a 90$^\circ$ rotated (around the horizontal line) version from a point of view from above, where the projected magnetic field follows a spiral pattern. c) RM of the bubble-like structures from the observer's perspective; d)  a 90$^\circ$ rotated  (around the vertical line) model version of the bubbles, where the magnetic field points to the observer on the outer sides of the bubbles and away from the observer on the sides closer to the disk.
    The straight lines in panels a), c), and d) indicate the galaxy plane.}
	\label{fig:helix-bubble-skizze}
\end{figure}

\begin{figure}
	\centering
		\includegraphics[width=0.5\textwidth]{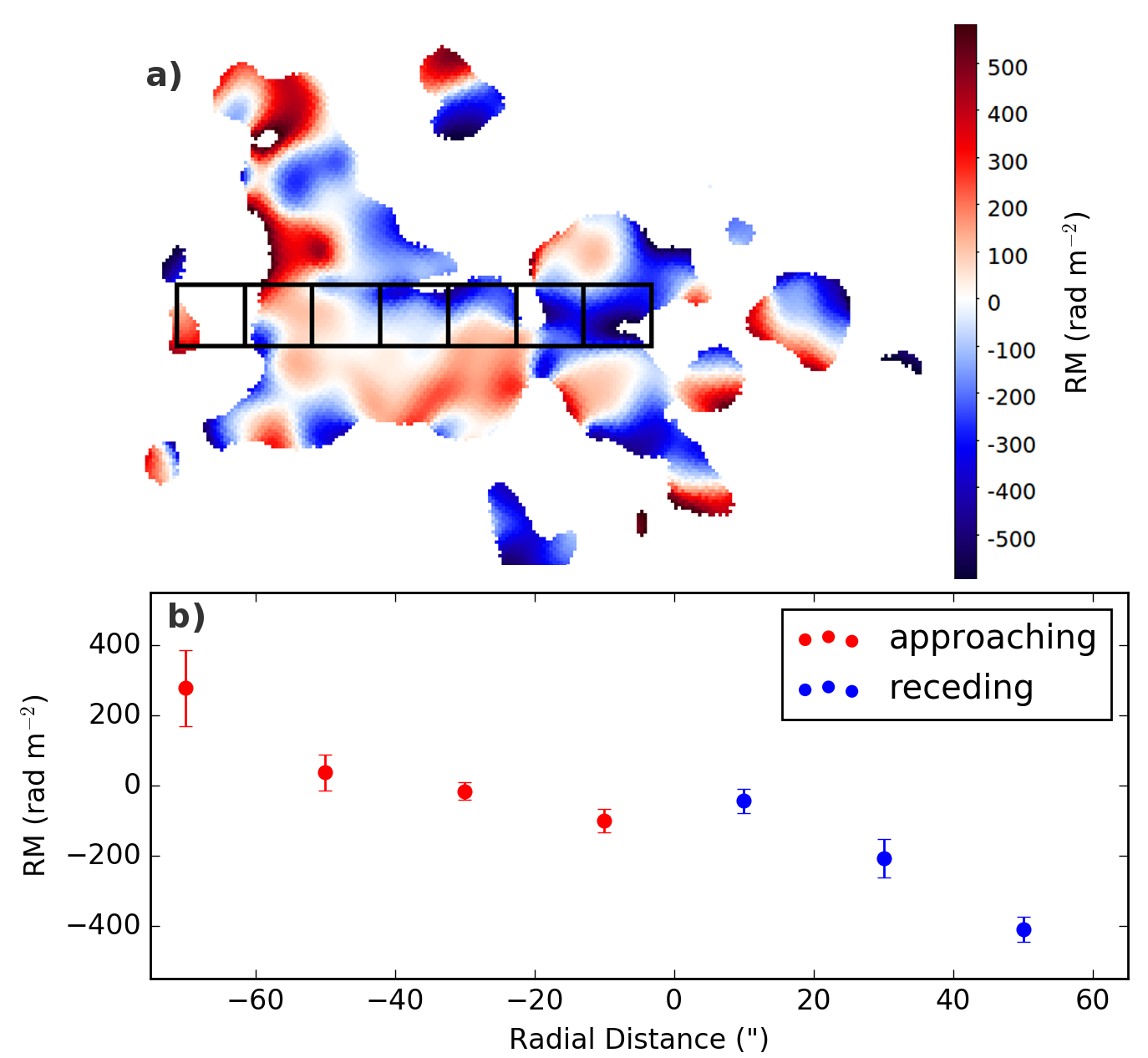}
    \caption{RM map of NGC 4217 at C-band (same as Fig.~\ref{fig:N4217_Ccomb_RM}, panel b) with box regions used for the disk profile (panel b). The box size is 20" x 18".}
	\label{fig:N4217_Ccomb_RM_disk}
\end{figure}

In Fig.~\ref{fig:helix-bubble-skizze}, we show the RM map to the upper right, where the helical-like structure of the galaxy is marked with the box on the left side. A simplified cartoon of a possible helical magnetic field configuration consistent with the observations is presented. The colors are chosen to match the colors of the RM map.

In Fig.~\ref{fig:helix-bubble-skizze}a) the helical magnetic field is presented from the observer's point of view. The left magnetic field structures of the helix point toward us, structures to the right are pointing away from us. To get a better understanding of the three dimensional shape of the helix, we present in panel b) a version of a view seen from above. With the observers viewing direction from below, the left parts point toward the observer and the right parts point away from the observer.

Rising helical magnetic field structures have been expected from theoretical work to be roughly on scales of 0.2 to 0.4 galactic radii \citep[e.g.,][]{henriksenirwin2016, woodfindenetal2019}, which in the case of NGC~4217 corresponds to $\sim$3 to 6\,kpc, consistent with our observations. Additionally, this scale is roughly the same as for the field reversals between the ``giant magnetic ropes'' (GMRs) in the halo of NGC~4631 \citep{moraetal2019a}, as well as for the disk magnetic field reversals of NGC~4666 \citep{steinetal2019a}, indicating that there may be a general scale of magnetic reversals around that value.

The second RM profile was plotted vertically through the bubble-like structure shown in Fig.~\ref{fig:N4217_Ccomb_RM}, panel c). The bubble-like structure toward the southwest of the galaxy is clearly seen in the RM map (panel b). Both bubbles show a pattern of negative and positive values. This pattern is probably slightly shifted toward negative RM values, because the line of sight magnetic disk field at that location is negative, which adds up with the magnetic field of the bubbles. However, the general profile of panel c) is typical for a real bubble structure in RM because the field is wrapped around the bubble. A cartoon of this is shown in Fig.~\ref{fig:helix-bubble-skizze}. The region of the bubbles is marked with a box. Panel c) shows a simplified presentation of the observations, where in the outer parts of the bubble-like structures the magnetic field points toward the observer, whereas the magnetic field points away from the observer on the parts of both bubbles that are closer to the disk. To better understand a possible magnetic field configuration, we show a view rotated by 90$^\circ$ in panel d). The magnetic field lines (in green) follow the shape of the bubbles, but due to Faraday depolarization we see mainly the front (left) side of the bubbles.


\begin{figure}
	\centering
		\includegraphics[width=0.28\textwidth]{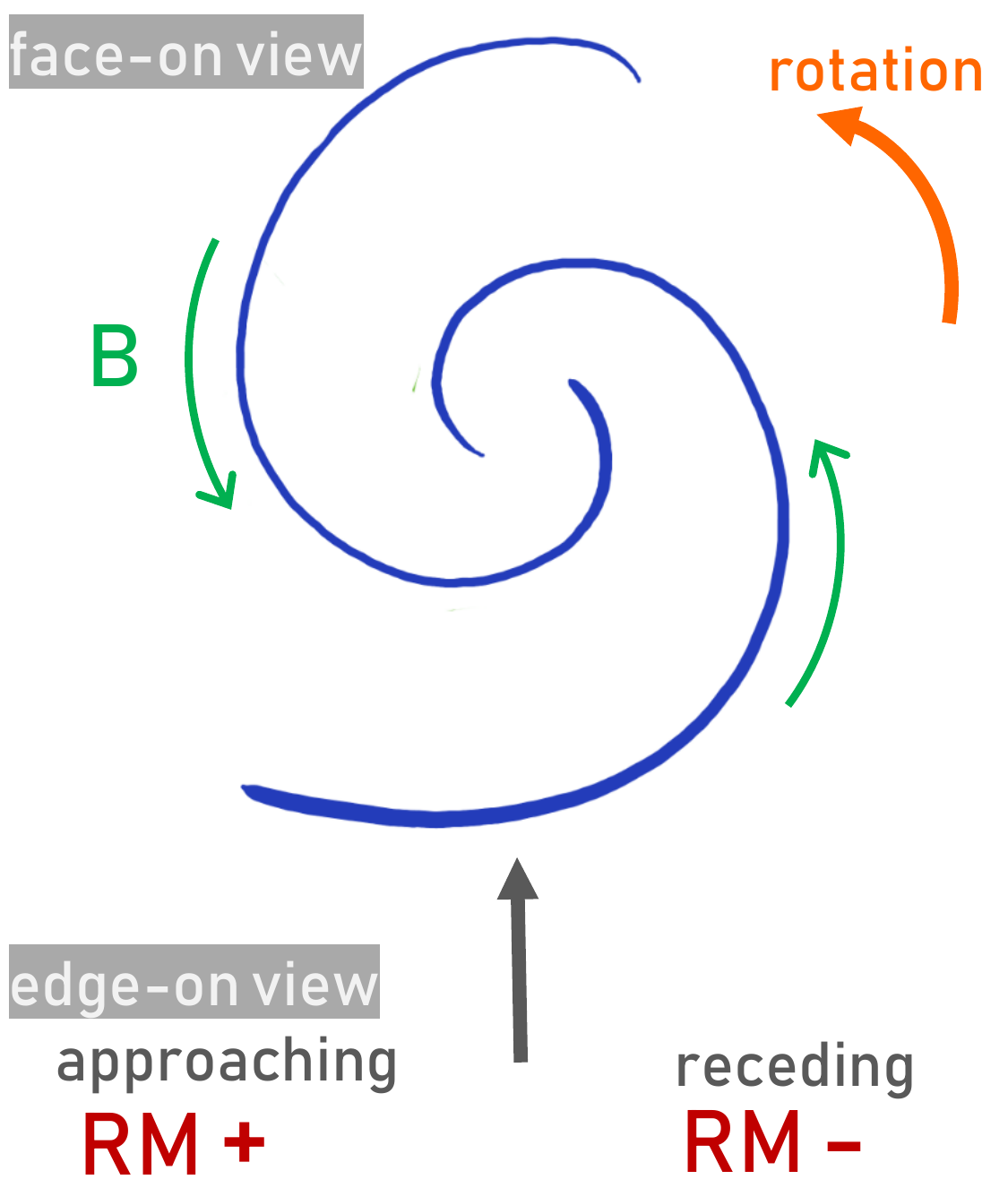}
    \caption{Top: Face-on view of a spiral galaxy with a spiral magnetic field (green arrows) pointing inward. Bottom: In edge-on view the magnetic field direction leads to positive RM values on the approaching side and to negative RM values on the receding side.}
	\label{fig:N4217_rm_mag_model}
\end{figure}

Information about the direction of the disk magnetic field of spiral galaxies can be drawn from the comparison of the rotation curve, that is the approaching or the receding side of the disk, and the corresponding RM values \citep{krausebeck1998}. Thus we use the RM profile along the disk (see Fig.~\ref{fig:N4217_Ccomb_RM_disk}) to investigate the direction of the magnetic field further. The resulting disk profile is symmetric around the center of the galaxy. The left part (approaching) of the disk shows positive RM values and the right part (receding) negative RM values. Figure~\ref{fig:N4217_rm_mag_model} illustrates this for a spiral galaxy with two trailing spiral arms. The approaching side leads to positive RM values in the edge-on view if the B$_{\parallel}$ points toward the observer. This indicates that the vectors of the magnetic field point inward along the spiral arms. Thus the spiral field in the disk has a radial component that is pointing inward.

\subsection{Separation of the thermal and nonthermal emission}

We performed a separation of the thermal and nonthermal emission to derive the synchrotron (nonthermal) maps of C-band and L-band, following \citet{vargasetal2018} and \citet{steinetal2019a}. As a tracer of the thermal emission, the H$\alpha$ emission from the H$\alpha$ map of \citet{rand1996} was used. The [NII]-corrected H$\alpha$ flux density ($L_{\text{H}\upalpha, \text{obs}}$) was then absorption corrected via Infrared WISE data at 22\,$\upmu$m ($\upnu L_{\nu}(22\upmu$m)) with the new calibration factor of 0.042 found by \citet{vargasetal2018}:
\begin{equation}
L_{\text{H}\upalpha,\text{corr}}[\text{erg s}^{-1}] = L_{\text{H}\upalpha, \text{obs}}[\text{erg s}^{-1}] + 0.042 \cdot \upnu L_{\upnu}(22\upmu\text{m})\,[\text{erg s}^{-1}].
\end{equation}
The thermal emission was calculated by using the corrected H$\upalpha$ flux density (L$_{H\upalpha, \text{corr}}$):
\begin{align}
F_{\text{thermal}} & [\text{erg s}^{-1} \text{Hz}^{-1} \text{pix}^{-1}] = \\ \nonumber
				& 1.167 \times 10^{-14} \cdot \left(\frac{\text{T}_\text{e}}{10^4 K}\right)^{0.45} \cdot \left(\frac{\upnu}{GHz}\right)^{-0.1} \cdot L_{\text{H}\upalpha, \text{corr}}\, [\text{erg s}^{-1}].
\label{eq:murphy}
\end{align}
An electron temperature of T$_\text{e}$ = 10,000\,K is assumed. $\upnu$ is the central frequency (6\,GHz and 1.5\,GHz for C- and L-band, respectively). The derived thermal map was subtracted from the radio map to get the nonthermal map, which represents the synchrotron emission alone. The nonthermal maps of the CHANG-ES frequencies have a beam size of 13.4"~$\times$~13.4". The separation of the thermal and nonthermal emission was not undertaken for the LOFAR data as the expected thermal contribution at that frequency is small. We refer to \citet{vargasetal2018} for more details on the separation of the thermal and nonthermal emission in edge-on spiral galaxies.


\subsubsection{Nonthermal fractions}
\begin{figure}[h]
	\centering
		\includegraphics[width=0.5\textwidth]{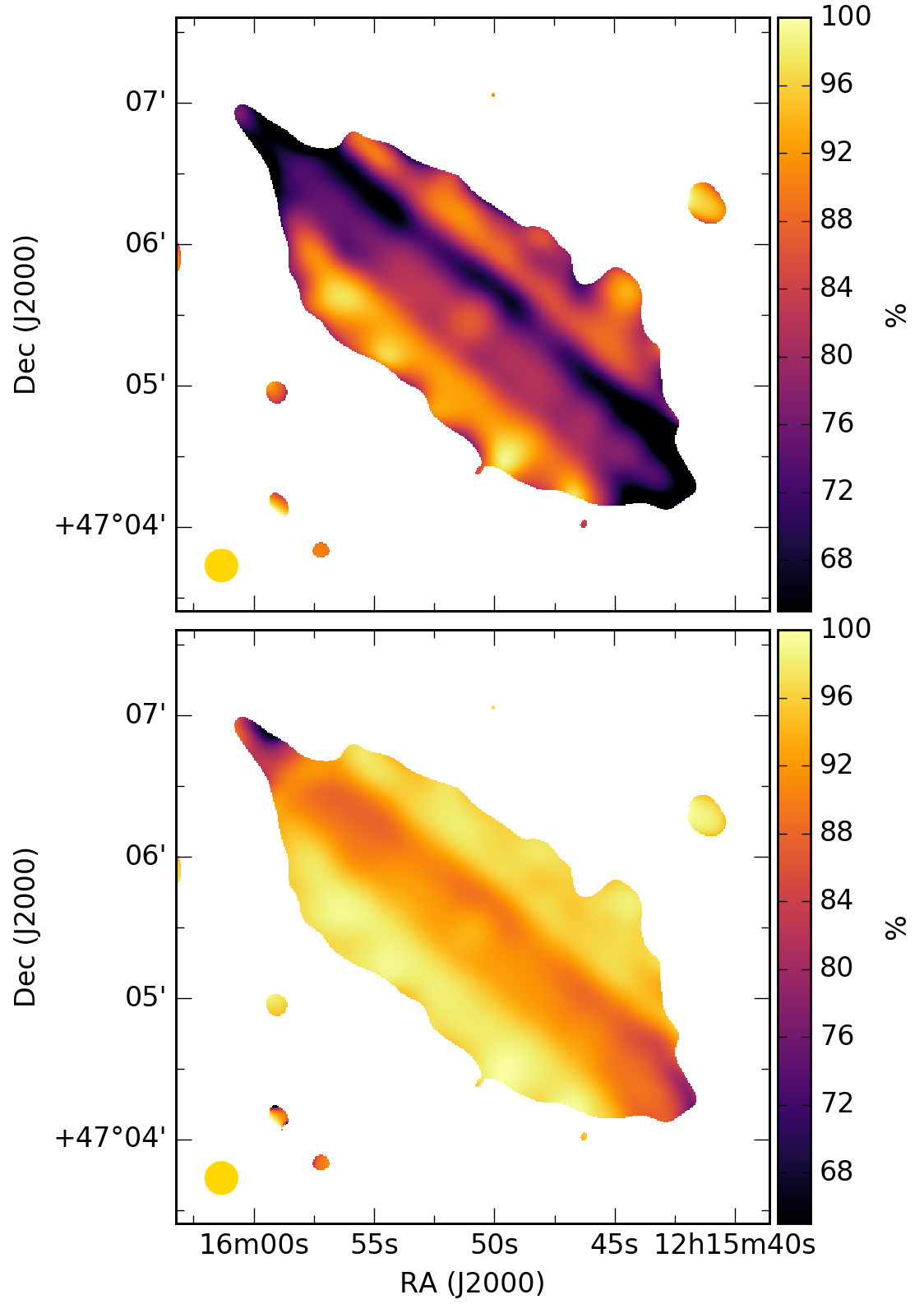}
	\caption{Nonthermal fraction of the total radio emission of NGC~4217 in percent. Top: C-band, Bottom: L-band, the beam size is 13.4".}
	\label{fig:N4217_nonthermfrac}
\end{figure}

The nonthermal fraction represents how much percent of the total radio emission is produced by synchrotron radiation. The remaining part originates from thermal emission. The nonthermal fractions in percent are shown for C-band and L-band in Fig.~\ref{fig:N4217_nonthermfrac}. The general trend looks similar for both images with different absolute values. The mean disk values are 76\% and 91\% nonthermal emission for C-band and L-band, respectively. These values agree well with the predictions by \citet{Condon1992}. The halo fraction of nonthermal emission is higher, reaching a mean halo value of 87\% and 97\% for C-band and L-band, respectively. Therefore, in L-band there is nearly no thermal radio emission in the halo. 

\subsubsection{Nonthermal spectral Index}

\begin{figure}
	\centering
		\includegraphics[width=0.5\textwidth]{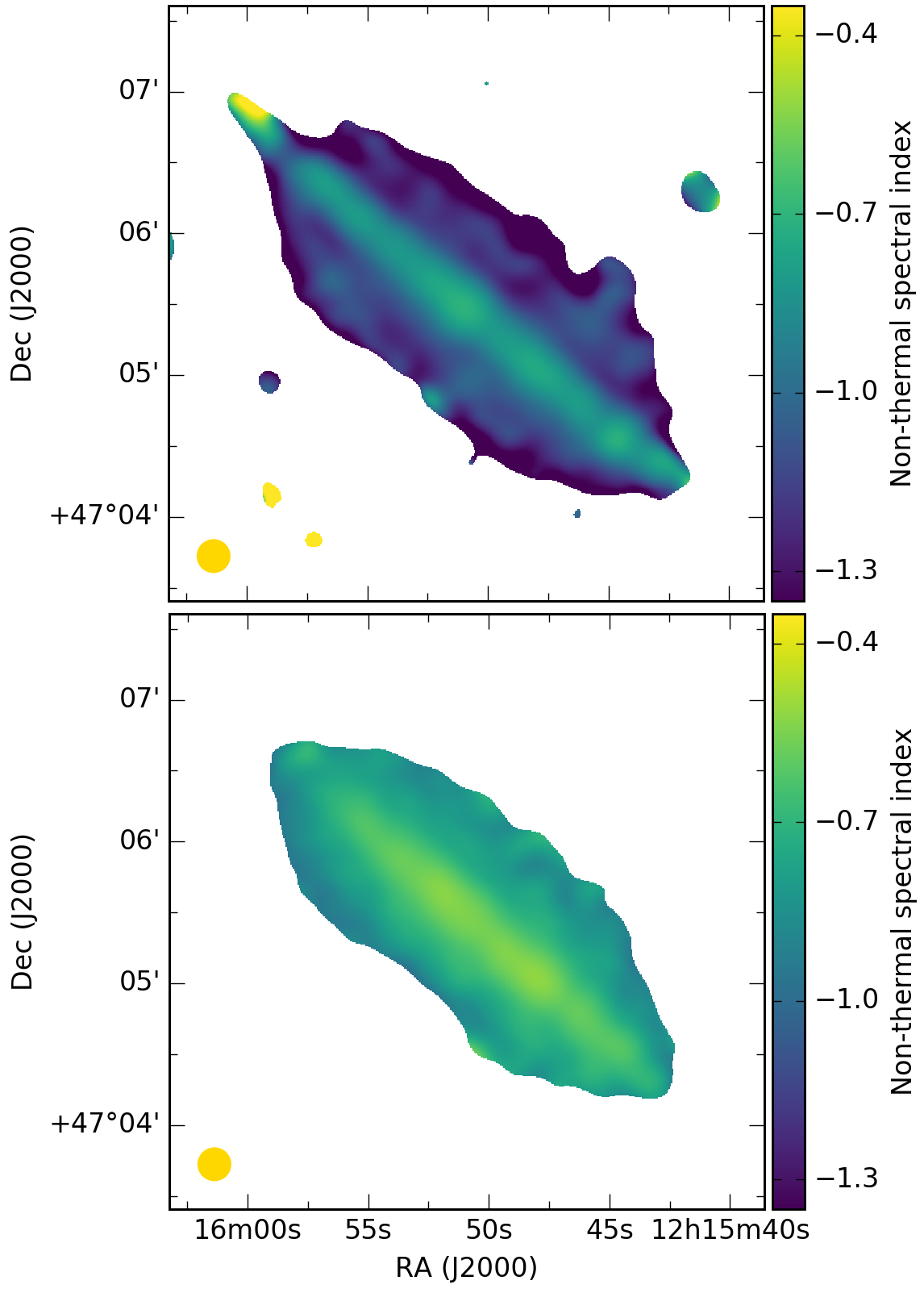}
	\caption{NGC~4217 nonthermal spectral index maps. Top: C-band/L-band. Bottom: L-band/LOFAR. The beam size is 13.4".}
	\label{fig:N4217_nonthermal_SpI}
\end{figure}

The nonthermal spectral index maps are presented in Fig.~\ref{fig:N4217_nonthermal_SpI}. We calculated the nonthermal spectral index maps between (a) the nonthermal maps of C- and L-band and (b) the nonthermal map of L-band and the LOFAR map. As discussed above, we assume the thermal emission at LOFAR frequencies to be negligible.

The nonthermal spectral index maps were derived using the following equation:
\begin{equation}
\upalpha_{\text{nt}} = \frac{\text{log} I_{nth}(\upnu_1) - \text{log} I_{nth}(\upnu_2)}{\text{log} \upnu_1 - \text{log} \upnu_2} \, ,
\label{eq:spi}
\end{equation}  
where $\upnu_1$ and $\upnu_2$ are the central frequencies of two different bands. 
The error map was derived as described in \citet{steinetal2019a}. The errors are between 0.05 and 0.3 toward the edges. 

The spectral index values are clearly separated between the disk and the halo. In both of the nonthermal spectral index maps, a clear two-component structure, dividing disk and halo, is visible, which was also found in the total intensity profiles of the galaxy. The clear distinction between the disk and the halo in the map is, however, less prominent in the nonthermal spectral index map between L-band and LOFAR in comparison to the spectral index map between C-band and L-band.

(a) C-band/L-band: The mean spectral index of the disk is $\upalpha$~=~$-$0.70~$\pm$~0.03, whereas the halo spectral index is $\upalpha$~=~$-$0.90~$\pm$~0.05. There is a region to the northeast of the galaxy, which shows a flat spectral index of $\upalpha$~=~$-$0.2~$\pm$~0.2. As indicated by the large uncertainty, this could be an artifact. Nevertheless, the H$\upalpha$ map displays that this region coincides with a star forming region in the northeast, which would explain the flat spectral index. The mean over the whole galaxy is $\upalpha$~=~$-$0.8~$\pm$~0.05. The steepening of the spectral index from disk toward the halo is expected as we trace aging cosmic-ray electrons in the halo, whereas in the disk we find younger particles.

(b) L-band/LOFAR: The mean spectral index of the disk is $\upalpha$~=~$-$0.60~$\pm$~0.03. The halo spectral index is $\upalpha$~=~$-$0.77~$\pm$~0.05.
Generally, the spectral index flattens slightly toward the LOFAR frequency.


\subsection{Depolarization}

Wavelength-dependent Faraday depolarization (DP) can be derived as the ratio of relative polarized intensities at two wavelengths \citep[e.g.,][]{fletcheretal2004}:
\begin{equation}
DP_{L/C} = \frac{P_\mathrm{L-band}}{P_\mathrm{C-band}},
\end{equation}
with the nonthermal degree of polarization at a certain wavelength P, defined as the ratio of polarized flux density to nonthermal flux density. 

\begin{figure}[h]
	\centering
		\includegraphics[width=0.5\textwidth]{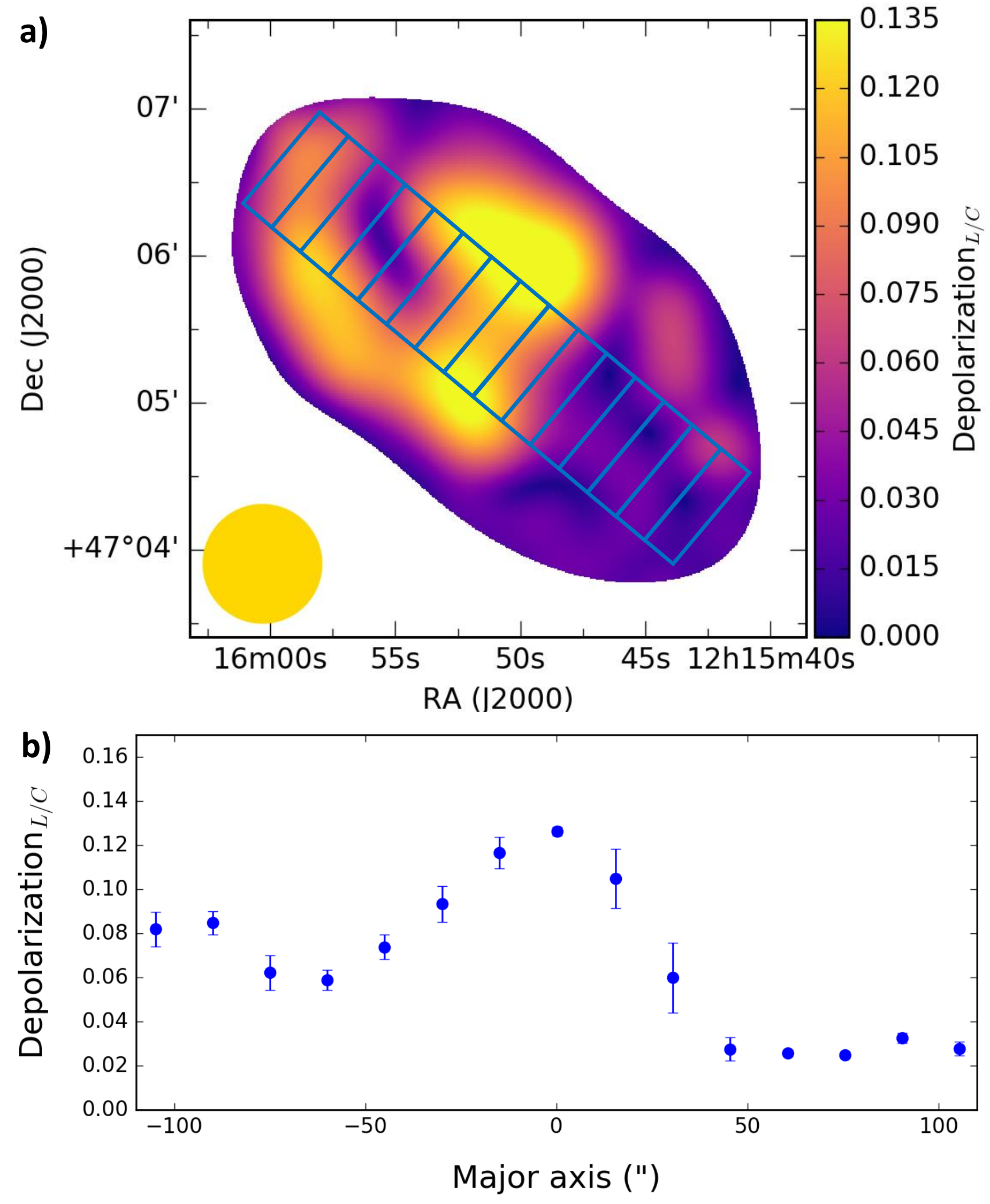}
	\caption{NGC~4217 depolarization of L-band/C-band. 
	{\bf a)} Map with clipping level of three sigma of the nonthermal intensity map at L-band. The beam size is 48". The strip along the disk is used for the profile below. The size of each of the 15 boxes is 15" x 48". {\bf b)} NGC~4217 depolarization profile between L-band and C-band within the strip along the disk. The x-axis value of -100" is at the northeastern (approaching) side of the galaxy.}	
	\label{fig:N4217_ratiodegreepol}
		\label{fig:N4217_ratiodegreepol_profile}
\end{figure}

To calculate the depolarization, first the nonthermal maps as well as the polarization map at C-band were all smoothed to 48" to match the L-band D-configuration polarization map, then they were all regridded to the same pixel scale. The nonthermal degree of polarization was calculated for both bands and then the ratio (depolarization) was calculated.
In Figure~\ref{fig:N4217_ratiodegreepol}a), the depolarization of NGC~4217 is shown, that is the ratio of the nonthermal degree of polarization between L-band and C-band. This is a measure for how much polarized emission is detected in L-band in comparison to C-band and therefore how much is depolarized in L-band. The smaller the value, the more depolarization is occurring. In the southwest part of the galaxy (right half of the major axis), smaller values are found. Thus the receding side is affected more by depolarization, whereas the approaching side of the galaxy is affected less strongly by depolarization and exhibits larger values.

In Figure~\ref{fig:N4217_ratiodegreepol_profile}b), we show the depolarization disk profile along the major axis using a box width of 48" and box height of 15". Negative major axis values correspond to the northeast (approaching) side of the galaxy. The profile was measured with NOD3. The depolarization is the lowest toward the center of the galaxy, reaching its highest value of $\sim$0.13. The depolarization increases (the value decreases) toward both sides of the major axis. In the northern disk (left), the depolarization is less strong in comparison to the southern disk (right) and decreases slightly toward the edge.


\subsection{Magnetic field strength via equipartition}
\label{sec:equipartition}

The maps of the magnetic field strength were determined with the revised equipartition formula by \citet[][their eq. 3]{beckkrause2005}:
\begin{equation}
\text{B}_{eq} = \left(\frac{4 \pi (2\upalpha + 1) (\text{K}_0 + 1) \text{I}_{nt} \text{E}_p^{1-2\upalpha} (\upnu/2c_1)^{\upalpha}}{(2\upalpha - 1) c_2(\upalpha) l c_4(i)}\right)^{1/(3+\upalpha)},
\end{equation}
with the nonthermal spectral index distribution $\upalpha$ defined to be positive, the number density ratio K$_0$ of protons and electrons, the nonthermal intensity map I$_{nt}$, the proton energy that corresponds to the proton rest mass E$_p$~=~938.28~MeV~=~1.5033~$\times$~10$^3$~erg, the path length $l$ through the source (in unit of pc) and the following constants
\begin{align}
c_1 &= 3 e/ (4\pi \text{m}_e^3 c^5) = 6.26428 \cdot 10^{18} \text{erg}^{-2} \text{s}^{-1} \text{G}^{-1},\\ \nonumber
c_2 &= \frac{1}{4} \ c_3 \ ((2 \upalpha + 1) + 7/3)/(2 \upalpha +2) \ \Gamma\left[(3 \upalpha + 1) / 6 \right]\\ 
		& \ \ \ \times \Gamma\left[(3 \upalpha + 5) / 6 \right] ,\\
c_3 &= \sqrt{3} e^3 / (4 \pi m_e c^2) = 1.86558 \cdot 10^{-23} \text{erg G}^{-1} \text{sr}^{-1},\\
c_4(i) &= \text{cos(incl)}^{\upalpha + 1}.
\label{eq:c}
\end{align}
If the field is completely turbulent and has an isotropic angle, $c_4$ becomes:
\begin{equation}
c_4(i) = (2/3)^{(\upalpha + 1)/2}.
\label{eq:c4}
\end{equation} 
The number density ratio of protons and electrons is chosen to be K$_0$~=~100, which is consistent with local Milky Way CR data \citep[see e.g.,][]{seta2019}. K$_0$ can be assumed to be constant in the energy range of a few GeV, where equipartition holds and losses are small or affect protons and electrons in the same way. The spectral indices of the injection spectrum of protons $\gamma_p$ and electrons $\gamma_e$ with $\upalpha$ = ($\gamma_e$ - 1)/2 are then the same.

The calculations were done pixel-by-pixel by taking the nonthermal intensity map and the resolved nonthermal spectral index map into account. The pixel values of the magnetic field strength B$_{eq}$ were then calculated according to the revised equipartition formula. To better represent the path lengths through an edge-on spiral galaxy, we did not use a constant value, but a spheroid model with path lengths varying from {\it l}~=~18\,kpc in the center to 6\,kpc toward the halo.

\begin{figure}[h]
	 	\includegraphics[width=0.5\textwidth]{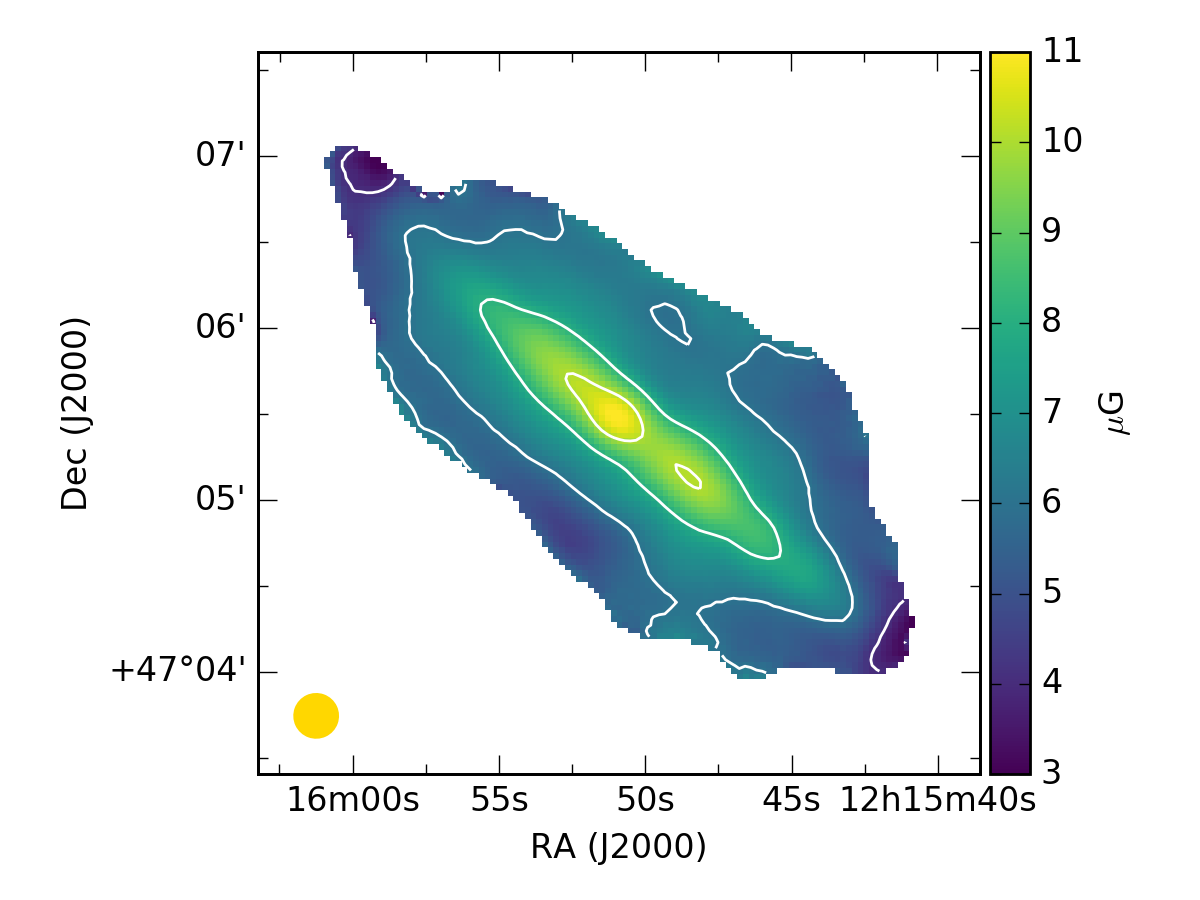}
	\caption{NGC~4217 magnetic field strength calculated pixel-by-pixel via the equipartition formula, using the nonthermal spectral index map and the intensity map with a beam size of 13.4".}
	\label{fig:N4217_bfeld_genau}
\end{figure}

The magnetic field strength map is shown in Fig.~\ref{fig:N4217_bfeld_genau}. 
The central region of the galaxy shows total magnetic field strengths of 11.0\,$\upmu$G in Fig.~\ref{fig:N4217_bfeld_genau}. The mean disk field strength is 9.0\,$\upmu$G. The halo magnetic field is roughly constant at a value of 7.4\,$\upmu$G. These values are in good agreement with the values of other spiral galaxies \citep{beck2019}.

\subsection{1D Model SPINNAKER}

The transport processes of CRs into the halo (perpendicular to the galaxy disc) are derived by using a 1D CR transport model with the program \textsc{spinnaker} by V. Heesen \citep{heesenetal2016, heesenetal2018}, similar to \citet{steinetal2019a, steinetal2019b}. Either advection with a constant advection speed $V$ or diffusion with a diffusion coefficient of $D = D_0$ E$^{\upmu}_{GeV}$ is modeled.

This model is applied to the total intensity profile of the pure synchrotron maps, which represents the synchrotron intensity emitted by the cosmic-ray electrons (CREs) as well as on the corresponding spectral index map of NGC~4217. With a box size of 200"~$\times$~6", the intensity value of two boxes at the same distance to the midplane (z) of both halo sides are averaged.

Apart from the choice between diffusion and advection, other important adjustable parameters are the injection index ($\gamma_0$), the magnetic field and the corresponding magnetic field scale height.
The magnetic field strength is described as a double exponential function with the superposition of a thin and a thick disk:
\begin{equation}
    B(z) = B_1 \exp(-z/h_{\rm B1}) + (B_0-B_1) \exp(-z/h_{\rm B2}),
\end{equation}
where $B_0$ is the magnetic field strength in total, $B_1$ is the thin disk magnetic field strength, and $h_{\rm B1}$ and $h_{\rm B2}$ are the scale heights of the thin and thick disk, respectively.

We applied the model via a manual tuning of all parameters based on $\chi^2$. Beforehand, we constrained several parameters within a certain parameter range, which were B$_0$ in the range of 9 to 12\,$\upmu$G, B$_1$ $<$ B$_0$, $\upgamma_0$ in the range of 2.3 to 3 to result in spectral indices between -0.65 and -1, and the scale heights of the magnetic field should follow $h_{\rm B1}$ $<$ $h_{\rm B2}$.

\begin{table}
 \captionof{table}{NGC~4217 \textsc{spinnaker} best-fitting parameters.}  
\label{tab:N4217spinnaker}
\centering
\begin{tabular}{lc}
\hline \hline
 Parameter &\\
\hline
B$_0$ ($\upmu$G)  & $11$   \\
B$_1$ ($\upmu$G) &  $6$  \\
\hline
\multicolumn{2}{c}{Model (ii) -- Advection (constant speed)} \\
$\upgamma_0$ & $2.7$  \\
h$_{\rm B1}$ (kpc)        &  $2.5$  \\
h$_{\rm B2}$ (kpc)        &  $3.4$  \\
$V_0$ (km\,s$^{-1}$) & 350 \\
$\upchi_T^2$      &    $0.3$  \\
\hline
\multicolumn{2}{c}{Model (i) -- Diffusion} \\
$\upgamma_0$ & $2.9$ \\
h$_{\rm B1}$ (kpc)        &  $1.5$  \\
h$_{\rm B2}$ (kpc)        &  $11.4$  \\
D (10$^{28}$\,cm$^2$\,s$^{-1}$)      &  $5.4$ \\
$\upchi_T^2$      &    $1.2$ \\
\hline
\end{tabular}
\end{table}

\begin{figure}
	\centering
		\includegraphics[width=0.5\textwidth]{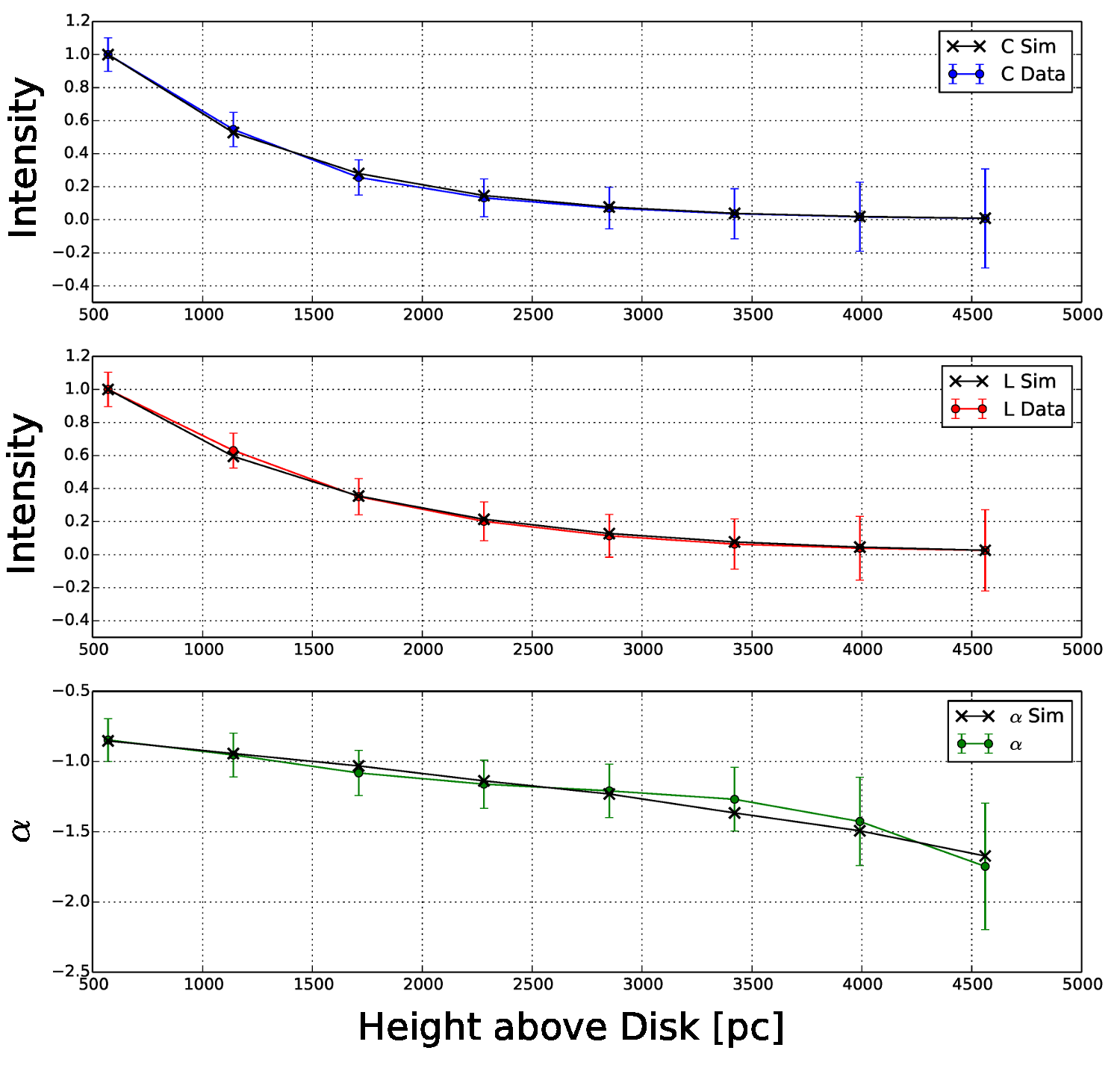}
	\caption[NGC~4217 cosmic-ray transport model with advection]{Cosmic-ray transport model for NGC~4217 with \textsc{spinnaker}, which is in agreement with advection. The errors are represented by a weighted standard deviation.}
	\label{fig:N4217-spinteractive-adv}
\end{figure}

The result using advection as the main transport process is displayed in Fig.~\ref{fig:N4217-spinteractive-adv}. The C-band data are shown in the first panel, the L-band data in the second, and the spectral index calculated from the corresponding data points of the C-band and L-band data is shown in the third panel. The total $\upchi_T^2$~=~0.3 suggests a good accordance with the advection model, while the diffusion model is significantly worse with $\upchi_T^2$~=~1.2 (Table~\ref{tab:N4217spinnaker}). The resulting advection speed is 350~$\pm$~20\,km~s$^{-1}$. The magnetic field strength for the thin component of 11\,$\upmu$G is chosen to be comparable to the mean magnetic field strength from equipartition (Section~\ref{sec:equipartition}). The nonthermal spectral index ($\upalpha)$ can be determined with \textsc{spinnaker} from $\upgamma_0$ = 2.7 via $\upalpha$~=~(1-$\upgamma_0$)/2. This results in $\upalpha$~=~$-$0.85, which is in good agreement with the mean value of the spectral index map (Fig.~\ref{fig:N4217_nonthermal_SpI}) 
of $\upalpha$~=~$-$0.8. 

\begin{figure}
	\centering
		\includegraphics[width=0.5\textwidth]{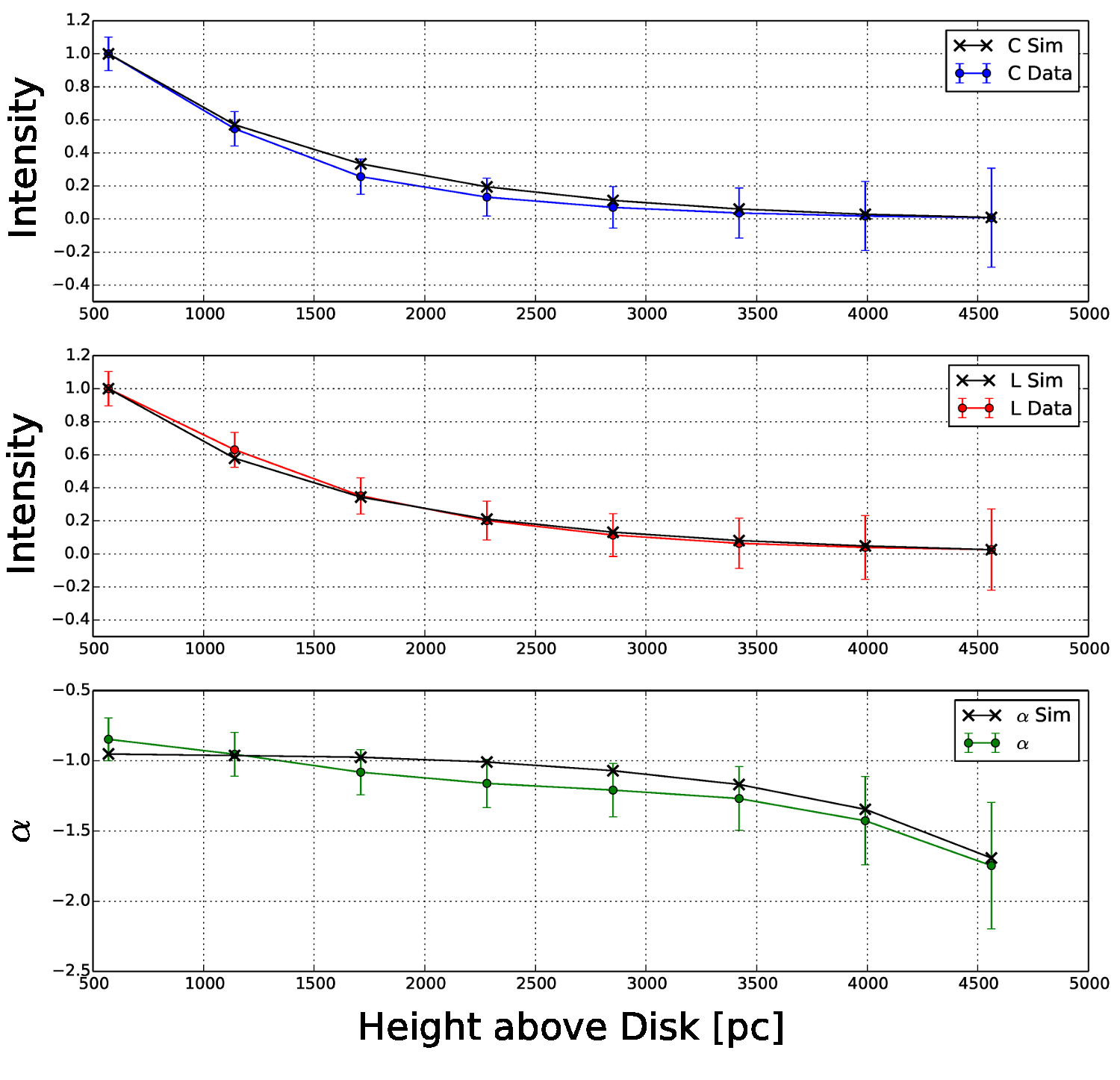}
	\caption[NGC~4217 cosmic-ray transport model with diffusion]{cosmic-ray transport model for NGC~4217 with \textsc{spinnaker} with diffusion for comparison. The errors are represented by a weighted standard deviation.}
	\label{fig:N4217-spinteractive-dif}
\end{figure}

For comparison, the results achieved with diffusion are presented in Fig.~\ref{fig:N4217-spinteractive-dif}. The difference of the fit quality of the three profiles $\upchi_T^2$ in comparison to advection is not very large.  Whereas the C- and L-band data are fit equally well in both cases, a less good fit to the nonthermal spectral index with diffusion is reached, which is also easily visible by eye. Furthermore, the nonthermal spectral index is less comparable with the measured value, $\upgamma_0$ = 2.9 leading to $\upalpha$~=~$-$0.95. In summary, the result gives a slightly better match for advection.

\section{Discussion}
\label{sec:discussion}

\subsection{Cosmic-ray transport in NGC~4217}
\label{sec:discussion_transport}
In the following paragraphs we discuss the different indicators for the transport mechanism in this galaxy as well as the general appearance of the radio halo.

\subsubsection{Total intensity distribution}
The radio continuum halo of NGC~4217 is dumbbell-shaped, which can be interpreted as a synchrotron-loss dominated halo, where the CREs lose their energy mainly due to synchrotron losses and are not able to escape from the galaxy’s gravitational potential. The smaller extent of observed radio emission from the halo above the central region of the galaxy is thereby attributed to the larger magnetic fields in this region, which leads to stronger losses. 
The scale height fits (Section~\ref{sec:scaleheights}) show that two-component exponential fits at all three frequencies are best approximating the radio total intensity profile. This indicates that advection may be the dominating CRE transport \citep{heesenetal2016}.

\subsubsection{Halo scale height}
Following \citet{steinetal2019b} in determining the frequency dependence of the halo scale height ($\bar{z_0} \propto \upnu^x$), we investigate the type of CRE propagation that is dominating. In the case of dominating synchrotron losses \citep{mulcahy2018,krauseetal2018,steinetal2019a} an exponent $x=-0.25$ is expected for diffusive transport and $x=-0.5$ for advective transport. This is using the assumption that the CREs emit most of their energy within the galaxy halo (\emph{electron calorimeter}), which might not be fully valid for the LOFAR 150~MHz data.

Calculating the exponent with the mean halo scale heights from Table~\ref{tab:N4217scaleheights}, the result is $x = -0.68 \pm 0.15$ between 6\,GHz and 1.5\,GHz. The derived radio scale heights are affected by the contribution from thermal emission. But we do not expect a significant increase of this exponent if we consider only the synchrotron scale height. As \citet{schmidtetal2019} show for NGC~4565, exponential scale heights of total radio and synchrotron emission seem to be the same within the errors at C-band. In conclusion, this exponent could only increase slightly, if we consider the synchrotron halo scale height for C-band. This gives a clear preference for advection as the dominating CRE transport mechanism.

\subsubsection{Integrated flux densities of the disk to halo}
Following \citet{steinetal2019b}, we determined the integrated flux densities of the disk and halo as well as their ratios at the two CHANG-ES frequencies and the LOFAR frequency (see Table~\ref{tab:N4217scaleheights}). The disk to halo ratio is 0.75 at C-band (6\,GHz), 0.62 at L-band (1.5\,GHz) and 0.22 at the LOFAR frequency (0.15\,GHz).

In order to compare the results of NGC~4217 to other galaxies, we additionally show the values for NGC~4013 and NGC~4666 from \citet{steinetal2019b} in Table~\ref{tab:N4013N4666N4217amp}. For NGC~4666, no LOFAR data were available. NGC~4013 has a relatively low star formation rate and its radio halo is diffusion dominated \citep{steinetal2019b}. NGC~4666 is a superwind galaxy with a high-velocity advective wind originating from a high star formation rate across the entire galaxy \citep{steinetal2019a}. Therefore we can compare three galaxies with different properties and different dominating transport processes.

\begin{table}
\captionof{table}{Ratios of the mean amplitudes $\bar{w}$\,(disk/halo) and ratios of the mean scale heights $\bar{z}$\,(disk/halo) at three frequencies, derived from Gaussian fits for NGC~4013 \citep{steinetal2019b}, from exponential fits for NGC~4666 \citep{steinetal2019b}, and for NGC~4217 from this work (Table~\ref{tab:N4217scaleheights}). For each galaxy, the last row (Flux dens. (disk/halo)) represents the ratio of flux densities between disk to halo, calculated with $\bar{w} \times \bar{z}$.}
\label{tab:N4013N4666N4217amp}
\centering
\begin{tabular}{lccc}
\hline \hline
 &  C-band & L-band & LOFAR \\
 & 6\,GHz & 1.5\,GHz & 0.15\,GHz\\
\hline
NGC 4013 &&&\\
$\bar{w}$\,(disk/halo) &  9.4~$\pm$~2.8 & 4.5~$\pm$~2.4 & 1.9$\pm$~0.9\\
$\bar{z}$\,(disk/halo) &  0.18~$\pm$~0.06 & 0.18~$\pm$~0.02 & 0.15~$\pm$~0.02 \\
Flux dens. (disk/halo) &  1.7~$\pm$~0.4  & 0.8~$\pm$~0.2 & 0.3~$\pm$~0.1 \\
\hline
NGC 4666 &&&\\
$\bar{w}$\,(disk/halo) & 1.38~$\pm$~0.33 & 0.97~$\pm$~0.13 &-  \\
$\bar{z}$\,(disk/halo) & 0.26~$\pm$~0.10 & 0.34~$\pm$~0.06 & - \\
Flux dens. (disk/halo) & 0.36~$\pm$~0.16 & 0.33~$\pm$~0.07 &- \\
\hline
NGC 4217 &&&\\
$\bar{w}$\,(disk/halo)& 3.2~$\pm$~0.3 & 1.9~$\pm$~0.2 & 0.70~$\pm$~0.09 \\
$\bar{z}$\,(disk/halo) & 0.24~$\pm$~0.02 & 0.32~$\pm$~0.03 & 0.32~$\pm$~0.09 \\
Flux dens. (disk/halo) & 0.75~$\pm$~0.10 & 0.62~$\pm$~0.08 & 0.22~$\pm$~0.07 \\
\hline
\end{tabular}
\end{table}

\begin{figure}
	\centering
		\includegraphics[width=0.5\textwidth]{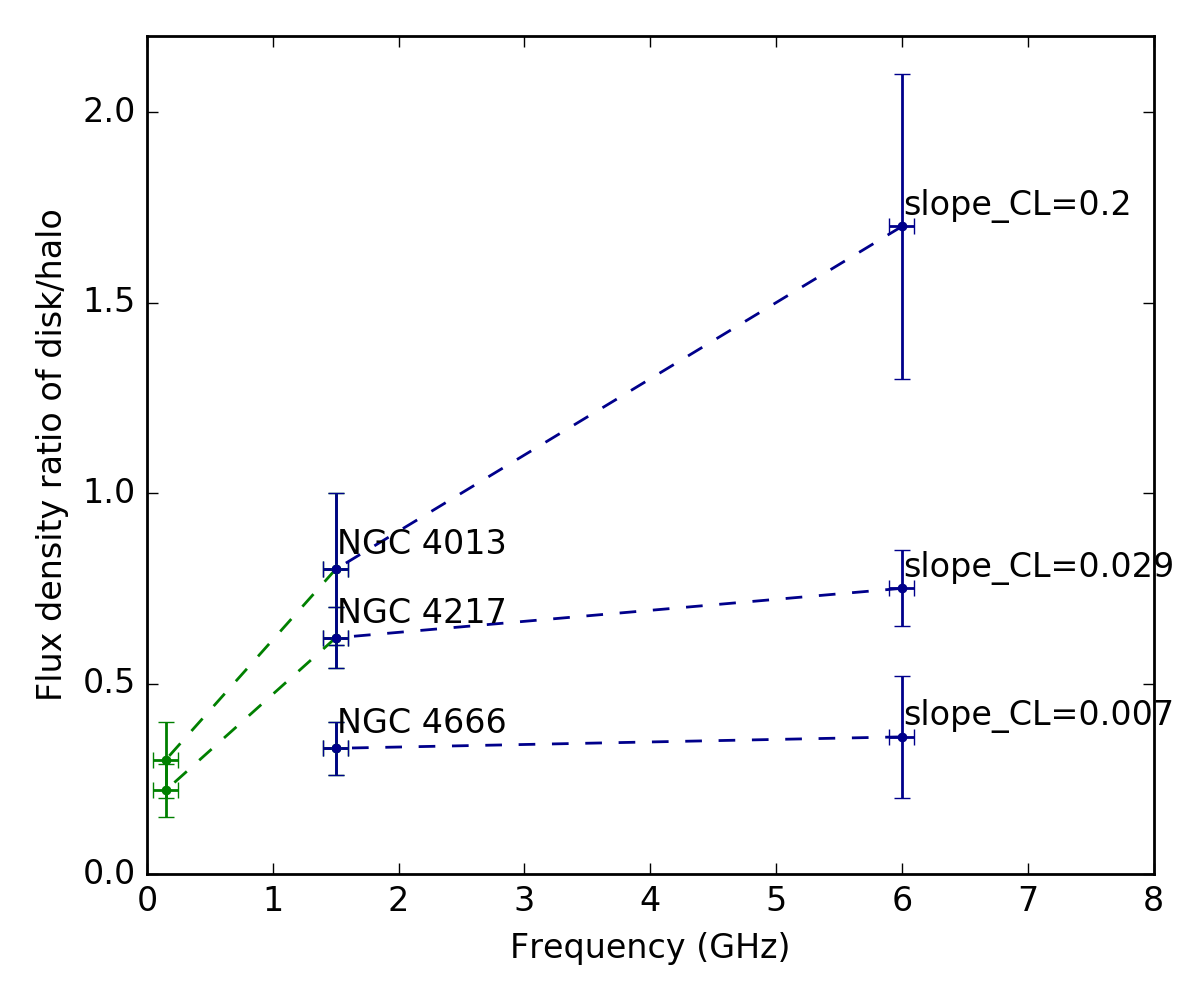}
	\caption{Integrated flux density ratios of the disk and halo at different frequencies for the three CHANG-ES galaxies NGC 4217 (this work), NGC~4013, and NGC~4666 from \citet{steinetal2019b}. The green data points are from LOFAR, the blue data points are from the CHANG-ES frequencies; slope\_CL is the slope between the flux density ratios of the disk and halo at the two CHANG-ES frequencies.}
	\label{fig:PhD_gal_fluxdensratio}
\end{figure}

According to Table~\ref{tab:N4013N4666N4217amp}, the ratios of the mean scale heights ($\bar{z}$) between disk and halo do not show a frequency dependence; the variations are minor for all three galaxies. In contrast, the ratios of the mean amplitudes ($\bar{w}$) between disk and halo show a strong frequency dependence. The product of these two ratios is equivalent to the flux density ratio between disk and halo and results in a frequency dependence. NGC~4217 exhibits a stronger frequency dependence in comparison to the superwind galaxy NGC~4666 and a less strong frequency dependence in comparison to NGC~4013.

To present this visually, the three frequencies are plotted against the disk to halo flux density ratios (Table~\ref{tab:N4013N4666N4217amp}, row Flux dens.) for all three galaxies. The result is presented in Fig.~\ref{fig:PhD_gal_fluxdensratio}. The green data points are those from LOFAR. The dark blue data points are from C- and L-band. 
As visible in Fig.~\ref{fig:PhD_gal_fluxdensratio}, for NGC~4217, this slope is different between the LOFAR/L-band points in comparison to the L-band/C-band points. To check the LOFAR data point, we investigate the ratio of the disk and halo with Table~\ref{tab:N4217scaleheights} separately. We find a ratio of the halo flux density between L-band and LOFAR of 5.9, which means that the halo at LOFAR frequencies has 5.9 times more flux density than the L-band halo. This results in a spectral index of 0.77. The ratio of the disk flux density between L-band and LOFAR is 2.1, which gives a spectral index of 0.32. This is quite small and could hint at thermal absorption occurring in the disk. 

\begin{figure}
\centering
	\includegraphics[width=0.5\textwidth]{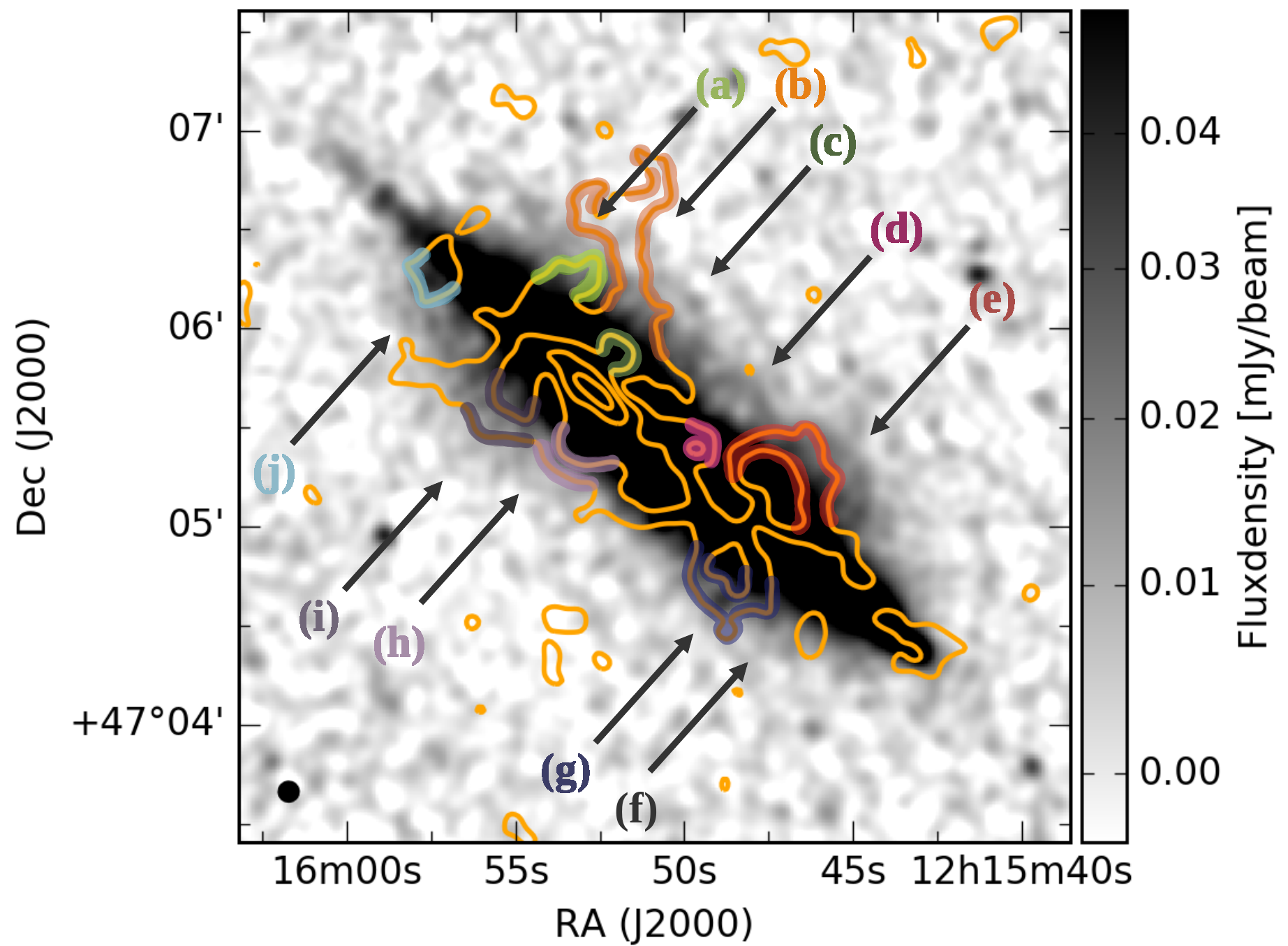}
	\caption{NGC~4217 total intensity image of C-band C-configuration with logarithmic scaling (same as Fig.~\ref{fig:N4217-bw}). Multi-colored contours are overlaid from polarization (Stokes Q and U, see Fig.~\ref{fig:N4217_Ccomb_pol7}) at 3, 6, 9, 12, 15 $\upsigma$ levels with a  $\upsigma$ of 4.8\,$\upmu$Jy/beam. The arrows indicate the regions discussed in the text and are labeled similarly to Fig.~\ref{fig:N4217-bw}. The different colored contours indicate a connection of the polarized emission with the structures in total intensity (arrows). We note that structures (c) and (d) only coincide with polarized emission near the disk.}
	\label{fig:N4217-bw_pol}
\end{figure}

The dark blue lines between the C- and L-band data points have the slope alpha\_CL, added in the plot at the location of the C-Band data point. The slope, alpha\_CL, is the smallest for NGC~4666, where the absolute flux density ratio between disk and halo stays nearly the same. This is probably caused by the galaxy-wide wind, which is the dominating transport of the CREs. This affects both frequencies in the same way. In NGC~4013, we find the highest inclination of alpha\_CL~=~0.2, where the absolute flux density is highly frequency dependent. In contrast, this galaxy is diffusion dominated. The fraction of the disk to the total flux density decreases toward lower frequencies, as we trace older CREs, which have moved further out in the halo. NGC~4217 seems to be in between these two extreme cases. As we have seen with the other indicators, this hints at advection being the dominant transport in this galaxy, but with diffusion also playing a role.

\subsubsection{SPINNAKER}
The 1D transport model found slightly better matches for advection than for diffusion as the main transport process. The constant advection speed found with this model is 350\,km~s$^{-1}$, which is comparable to that of NGC~4666 \citep{steinetal2019a}. Furthermore, the advection speed is comparable to the escape velocity of the galaxy, calculated following \citet{miskolczietal2019}, with a range of  v$_{\text{esc}}$~$\approx$~280~--~560\,km~s$^{-1}$, varying the outer radius of the dark matter halo R$_{\text{max}}$ between 13~--~20\,kpc and the radial distance r between 1~--~13\,kpc. Indeed, \citet{heesenetal2018} found a correlation between the advection speed and the escape velocity in several edge-on spiral galaxies. As the 1D model gives reasonable results also for diffusion, we conclude that an additional contribution of diffusion as the transport process of the CRs is also in accordance with the current data.

\subsection{Loops, shells and a symmetric superbubble structure}
\label{sec:discussion_loops}
The total intensity distribution of NGC~4217 at C-band shows loops, shells and a supberbubble structure throughout the galaxy. We discuss ten of these in detail. Note that most of the structures are above or around 3$\upsigma$. We consider structures (a), (c), (d) and (e) convincing loop structures (3$\upsigma$-detections). The other structures show extended emission or filamentary structures with some of them having faint loops rather being at the 2$\upsigma$-level  (Compare to appendix Fig.~\ref{fig:N4217-bw_pol_3sig} to see the total intensity image above 3$\upsigma$). When we compare them to H$\upalpha$ emission and an optical image in Fig.~\ref{fig:N4217-bubble_threeparts}, we find a strong correlation with H$\upalpha$. A few loop and shell structures coincide with dust filaments in the optical, reaching far into the halo.

\subsubsection{Comparison to polarized intensity}
To compare these structures of the total intensity to the polarized emission, Fig.~\ref{fig:N4217-bw_pol} shows an overlay. The background image in gray scale is the total intensity at C-band from the C-configuration only, with polarized intensity at C-band shown in multi-colored contours, from Fig.~\ref{fig:N4217_Ccomb_pol7}.

In general, polarization features can be associated with features in the total intensity maps. We colorized each polarized emission feature corresponding to the total intensity structure and its letter with the same color. Some loop and shell structures in total intensity show extents in polarized emission. For example, structures (a) and (b) are almost cospatial with the most extended halo feature in polarized emission. Note that structure (b) is the most extended structure in total intensity. Structure (c) has a peak of polarized intensity closer to the disk, similar to the bubble-like structure (d). The most prominent super-bubble feature (e), can be associated with a significant polarized feature that also resembles a bubble. This NW polarized feature creates an almost symmetric structure with a SE counterpart (below the disk) that in turn is associated with structure (g) in total intensity. In total intensity, (g) and (f) are visible as two different structures, whereas in polarization, only (g) appears. Note, that at location (g) there is an outflow structure in total intensity closer to the disk (see Fig.~\ref{fig:N4217-bw}, insert), which coincides with the bubble structure in polarized intensity. The faintest extended structure (f) is the only one which is not apparent in polarized emission, while the small structure (h) and the elongated structure (i) can be associated with bumps in the outer polarization contour. Toward larger radii to the northeast, the polarization distribution ends. The total intensity shell (j) seems associated with an additional patch of polarized emission that is not connected to the rest of the polarized emission.

\subsubsection{Origin and nature of the loops, shells and superbubble structure}
The loops, shells, and superbubble structure in NGC~4217 seem to be more prominent above the plane. Even if the galaxy is nearly edge-on, the halo side above the disk is oriented toward us and thus we see more structures, while these are probably partly hidden by the disk on the halo side below the disk.

The structures we observe in NGC~4217 are not comparable with AGN-driven, wind-blown bubbles like in the center of NGC~3079 \citep{lietal2019} and also not comparable with superbubbles (around 100\,pc in diameter) found within disks of external galaxies like the LMC \citep{dehortaetal2014}. The structures we find in NGC 4217 are distributed over the entire galaxy and have diameters of several kpc; it is an interesting possibility that they may follow the spiral structure. In summary, we observe dust filaments in the optical B-band image, seemingly following some of the loops and shells. The H$\upalpha$ emission is either extended smoothly within some of the structures (e, g, and i) or the two locations of the individual loop structures, where they are connected to the disk, are slightly extended (a, b, c, d, h). Especially the superbubble structure (e) is prominent in all investigated emissions, with the counterpart below the disk mainly being prominent in polarization. At that location, the radio continuum emission shows a second total intensity peak in the southwest part of the disk. This is also observable in the near-infrared J-band of 2MASS \citep{2MASS2006}. Furthermore, \citet{thomsonetal2004} found prominent OB associations at that location. Thus, we conclude a major star formation event, happening not too long ago, to be the origin of these symmetric superbubbles. 

Scale-invariant models of the global magnetic field often show both small and large-scale magnetic field lines looping over the spiral arms (e.g., \citet[][Figs. 6, 7]{henriksen2017} and \citet[][Fig. 3]{woodfindenetal2019}). These loops can be several kpc in size and are perhaps best thought of as "arcs" that are arched spaces over the spiral arms. The spaces would be filled with magnetic fields that can support ionized gas, similar to what is observed in the solar corona, which possibly explains our observations. The central "finger" in the observed structures (see the simplified cartoon of the loop and bubble-like morphologies in Fig.~\ref{fig:bubble-model}) 
might be a region of strong magnetic field looping
very close to the spiral arm. Other models of bubble structures also have such finger-like features in the center of the bubble \citep[e.g.,][]{maclowetal1989}.

\subsection{Polarized intensity peak on the approaching side}
\label{sec:discussion_pol}
\subsubsection{Observations}
The polarized emission at C-band is distributed over large parts of the regions emitting in total radio continuum. Nevertheless, less polarized intensity is observed on the receding side, with some holes in the distribution. In comparison, a peak of the polarized intensity is detected on the approaching side. The polarization at L-band was only observed in the central parts and on the approaching galaxy side. The receding side seems to be completely depolarized - not even RM-synthesis could recover the signal from that side of the galaxy. One reason for that is the effect of Faraday rotation, which is very strong at L-band, causing wavelength-dependent rotation of the polarization angle and depolarization. As a result, we only see a Faraday screen at L-band, which is a layer in the halo located closer to us.

The depolarization map between L- and C-band (Fig.~\ref{fig:N4217_ratiodegreepol}) shows that the northeastern side of the galaxy is less strongly depolarized compared to the southwestern (receding) side. According to the depolarization profile of the disk (Fig.~\ref{fig:N4217_ratiodegreepol_profile}), the depolarization is smallest in the center of the galaxy, increases toward larger distances from the center along the major axis, and is stronger on the receding side. 

In general, when comparing the receding and approaching side of the disk (without the halo) at C-band, we find about 20\% more polarized emission on the approaching side. At L-band, about 70\% more polarized emission is observed on the approaching side of the disk. This trend was also seen in NGC~4666 \citep{steinetal2019a}. Indeed, Krause et al. (in prep.) show that the majority of the CHANG-ES galaxies have stronger polarization on the approaching side in C-band.

\subsubsection{General explanations and a simplified model}
Considering the disk only, it is likely that, also at C-band, the observed polarized emission is not "Faraday-thin", which means we do not see radiation from the entire galaxy, but just from the part that is closer to us. Polarized emission from the part further away is depolarized due to different depolarization effects, such as the wavelength-dependent internal Faraday depolarization \citep{sokoloffetal1998}.

\begin{figure}[h!]
	\centering
		\includegraphics[width=0.5\textwidth]{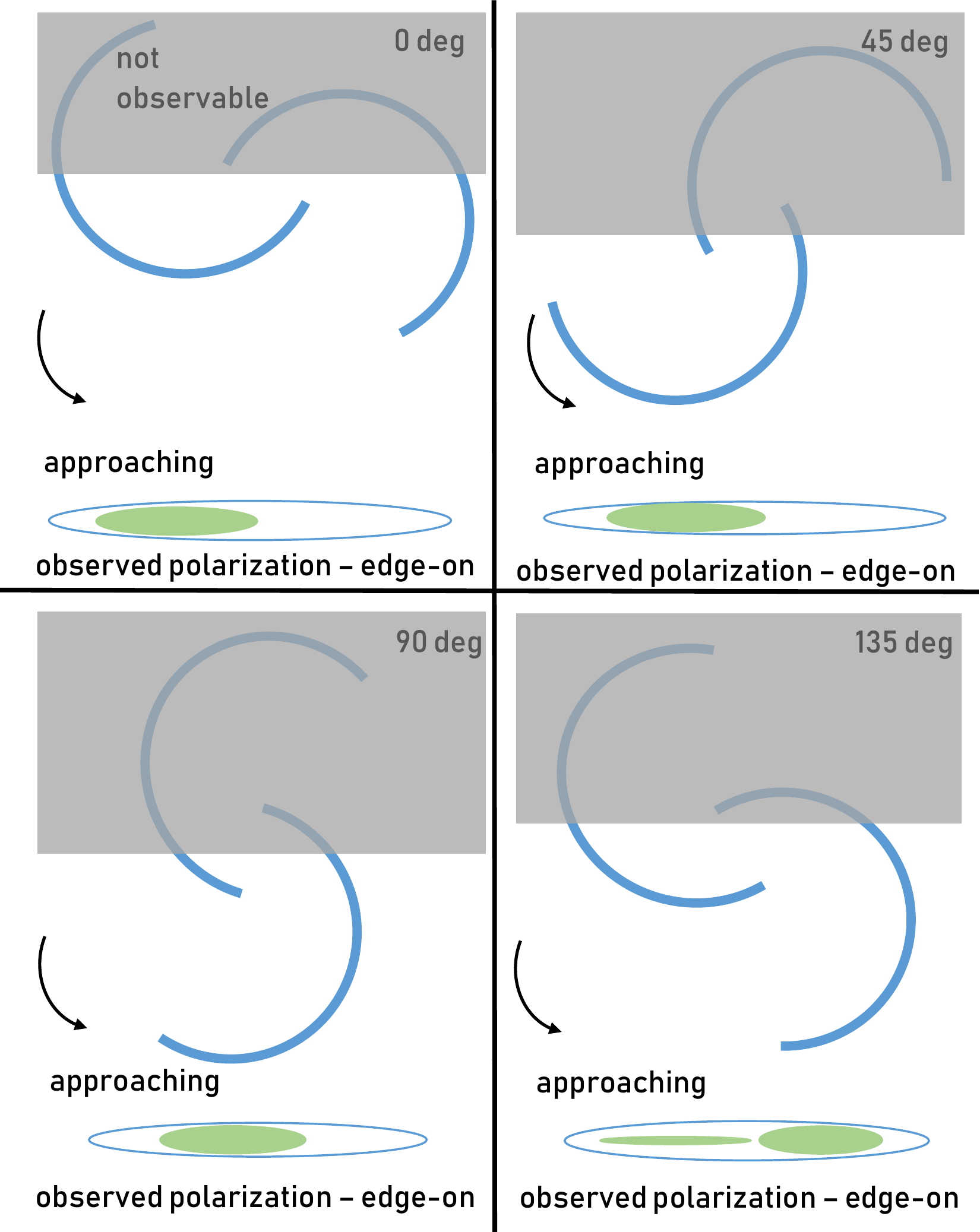}
	\caption{Simplified two arm spiral galaxy model (face-on) for four different viewing angles to explain why the emission is stronger on the approaching side. Trailing spiral arms are assumed. Furthermore, we assume the part of the galaxy that is further away from us (gray shaded) cannot be observed due to Faraday depolarization. The observed extent of the galaxy seen edge-on with the predicted polarized emission is presented in green, which is the observed polarization originating from the spiral arm closer to the observe, with the magnetic field more perpendicular to the line of sight.}
	\label{fig:model_depol}
\end{figure}

Considering possible geometries of the galaxy with respect to us, and using the fact that spiral arms are generally trailing, it is most likely that the inner (concave) side of the receding spiral arm points toward us (Fig.~\ref{fig:model_depol}). Also, the outer (convex) side of the trailing spiral arm on the approaching side points toward us. With this in mind, we propose two scenarios:

1) One reason to the asymmetry could be the emission process of synchrotron radiation. We assume that the magnetic field of the disk follows the spiral structure of the galaxy. Observations show either polarization close to the star forming spiral arms or magnetic arms between the star forming spiral arms, with the field lines running approximately parallel to the spiral arms \citep[see e.g.,][]{beck2016}. Polarized emission is emitted perpendicular to the magnetic field lines. Thus, mainly the magnetic field perpendicular to the line of sight contributes to the polarized emission that we observe. On the approaching side, closer to us, the field is perpendicular to the line of sight and we observe more polarized emission. Less polarized emission can reach us on the receding side, as we see mostly magnetic field lines parallel to the line of sight. 

2) Another reason might be that there is a difference between the inner spiral arm, mostly visible on the receding side, and the outer spiral arm, mostly visible on the approaching side. We argue that there is more turbulence and hence depolarization at the inner (concave) side of the spiral arm in comparison to the outer (convex) side of the spiral arm and that, additionally, the magnetic field is preferably located on the outer side of the spiral arm. Therefore, the polarized intensity is preferably observed on the convex side. This is elaborated in the following: 

We assume galactic matter and magnetic fields to be located inside of the corotation radius, where the rotation velocity of the matter and the spiral pattern are similar. Inside of the corotation radius, gas and magnetic fields move from the inner (concave) side into the spiral arms (upstream) and move out on the outer (convex) side. The physical properties of the infalling gas on the concave side is likely to result in a high Reynolds number which leads to turbulence. Turbulence in the infalling gas on the concave side of the arm causes Faraday depolarization of the polarized emission on its way to the observer through a galaxy in an edge-on view. This turbulent infalling gas, after crossing the arm to the convex side, may contribute to a stronger magnetic field both in the arm and on the convex edge, due to large-scale dynamo action, with a time delay between star formation and magnetic field amplification. This can explain an increased Faraday depolarization trailing the spiral arm. A recent paper by \citet{shanahanetal2019} shows that there are large RM values on the inner side of a spiral arm of the Milky Way, going along with that argument.

As a summary of the model, we present a simplified model of the disk of a two-arm spiral galaxy in Fig.~\ref{fig:model_depol}, to explain the observations with four different viewing angle geometries ($0\degr$, $45\degr$, $90\degr$, $135\degr$). We assume trailing spiral arms. The predicted polarized emission is shown in green below each model galaxy. The four panels represent half of a galaxy rotation, which is symmetric to the second half of a rotation by adding 180$\degr$. For most viewing angles, we see predominantly polarized emission (shown in green) from the approaching side. The model galaxy shows more polarized emission from the receding side or from the central parts at rotation angles between roughly 100$\degr$ and 145$\degr$, and correspondingly between roughly 280$\degr$ and 325$\degr$, thus a quarter of the rotation angle possibilities. This leads to the prediction that about 75\% of edge-on spiral galaxies will manifest the observable polarization on the approaching side. From the above model, we conclude the geometry of NGC~4217 to be likely between 0$\degr$ and 90$\degr$ or correspondingly 180$\degr$ and 270$\degr$ (compare with Fig.~\ref{fig:model_depol}).\\

\noindent
Possible caveats:

In case of more than two spiral arms the general explanations still holds, but the effect of the difference in polarized emission from the two sides (approaching and receding) would be weakened. If the spiral arms are more wound up, which means the pitch angles of the spiral arms are smaller, the asymmetry can still be explained, but we would also observe polarization from the receding side.

Furthermore, shock compression of the magnetic field by density waves and hence enhanced anisotropic turbulent fields (causing strong polarized intensity) occurs at the inner (concave) edge of the spiral arm, which may seem contradictory to our proposed model. Nevertheless, for the face-on spiral galaxy M51, shock compression only occurs in the inner spiral arms. At larger galactic radii, where density waves are weaker, polarized emission in M51 is measured mostly at the outer (convex) side \citep[see Fig.~5 in][]{patrikeevetal2006}. Thus, density waves (if any) are expected to affect the distribution of polarized emission mainly in the inner part of a galaxy.

In the above explanations, only the galaxy disk was considered. Obviously, in the big  picture, both the halo and the disk have magnetic fields and polarized emission which contribute. We have assumed here that the depolarization and difference of the two galaxy sides are dominated by the disk field.


\section{Summary and conclusions}
\label{sec:summary}

In this paper, we analyze the radio continuum data of the galaxy NGC~4217 from CHANG-ES VLA observations at 6\,GHz and 1.5\,GHz with supplemental LOFAR data at 150\,MHz as well as the hot gas with Chandra X-ray data. 

\begin{itemize}
\item The total radio intensity distribution of NGC~4217 at C-band shows several apparent filaments, loops, shell and bubble-like structures as well as a superbubble-like structure. Most of them coincide with H$\upalpha$ emission extensions or disk footprints as well as polarized emission features, and a few of them are associated with dust filaments reaching far into the halo, as seen in a combined SDSS u- and g-band optical "B-band" image. 

\item The most prominent super-bubble-like structure has almost symmetric halo components on each side of the disk and is observed in polarized intensity, total intensity and optical images (B-band, H$\upalpha$). Within the disk at this off-center SW location resides a peak of C-band radio continuum emission that is second in intensity only to the peak at the galaxy center. This off-center peak is cospatial with prominent OB complexes \citep{thomsonetal2004}. Thus, massive star formation is probably occurring locally in the disk, forming a wind-blown superbubble \citep[][and references therein]{kimetal2017} via the kinetic energy input of tens of SN explosions and stellar winds. The magnetic field orientations of the symmetric superbubble-like structure are oriented almost perpendicular to the major axis of the galaxy. This region does not seem to influence the entire galaxy but dominates in most of the southwestern part.

\item The CHANG-ES and LOFAR data indicate that the halo is synchrotron-loss dominated, with an advective wind with a velocity similar to or slightly less than the escape velocity. Specific observations that favor this scenario are as follows: (1) The radio continuum halo of NGC~4217 is dumbbell-shaped, which can be interpreted as a synchrotron-loss dominated halo. (2) NGC~4217 shows a well defined two-component exponential total radio intensity profile with the disk [halo] scale heights of 0.24 [1.0]\,kpc, 0.42 [1.3]\,kpc and 0.46 [1.4]\,kpc at C-band, L-band, and LOFAR, respectively, which points to advection \citep{heesenetal2016}. (3) The exponent of the frequency dependence of the scale heights at C- and L-band suggests a synchrotron-loss dominated halo with advection. (4) The total flux density ratio of the disk and halo hint at advection being the dominant process and diffusion being less important in this galaxy. (5) The 1D transport model \textsc{spinnaker} is in slightly better agreement with advection than with diffusion as the main transport process. The modeled advection speed is 350\,km~s$^{-1}$, which is similar to the escape velocity of the galaxy.

\item At C-band, the disk flux density contributes nearly as much as the halo to the total flux density of the galaxy, at L-band the fraction of the disk contribution decreases, and at the LOFAR frequency the disk only contributes 18\% to the total flux density.

\item The CHANG-ES observations of NGC~4217 at C-band show a large radio halo with magnetic field orientations from polarized intensity suggesting an X-shaped and large-scale regular magnetic field, which means the magnetic field orientations of the polarized emission regions within the mid-plane are oriented plane-parallel and become vertical in the halo of the galaxy. We present this by a color composite image of the magnetic flow lines. At L-band, polarized emission is found in the northeastern half of the galaxy in the D-configuration data. RM-synthesis could not reconstruct the polarized emission from the southwestern (receding) side of the galaxy. The magnetic field strength of the disk is 9\,$\upmu$G. 

\item With supplemental Chandra data, we show similarities between the diffuse hot gas of NGC~4217 and the diffuse polarized emission. 

\item The results from RM-synthesis indicate that the radial component of the disk magnetic field is oriented inward. Furthermore, a helical outflow in the northwestern part of the galaxy was found in the RM map, which is extended nearly 7\,kpc into the halo. 

\item At C-band, about 20\% more polarized emission in the disk is observed on the approaching side, at L-band about 70\% more polarized emission in the disk is observed on the approaching side. This suggests that Faraday depolarization could also be occurring at C-band. With our introduced disk model we are able to explain the observations of higher polarized intensities on the approaching side of the galaxy and predict that roughly 75\% of edge-on spiral galaxies will show more polarization on the approaching side.
\end{itemize}

In this paper, we reveal galaxy-wide outflows of the multi-phase material of the nearby edge-on star forming galaxy NGC~4217. These are observed in radio continuum and optical wavelengths and are probably induced by stellar winds and/or the combined action of several supernova explosions. Deeper observations are recommended to show more details of these structures.

By studying the RM map of NGC~4217 from RM-synthesis at C-band, we reveal a helical magnetic field structure in the outflow of the northwestern halo with a scale that is in accordance with theoretical predictions. Further studies of magnetic field structures in spiral galaxies could lead to a more comprehensive picture of helical fields and their scales. 
By analyzing a sub-sample of three spiral galaxies of the CHANG-ES sample, general conclusions on the transport processes of the CR particles into the halo are drawn. Galaxies with weak star formation reveal diffusive transport that leads to thin radio halos, while advective transport in galaxies with moderate star formation leads to medium-sized radio halos, and superwind galaxies form large boxy radio halos. Further investigations with a larger sample could quantify this correlation.

The general behavior of magnetic fields in spiral galaxies is considered by inferring a model to explain the observation of weaker polarized emission on the receding side of NGC~4217 and the majority of edge-on galaxies in the CHANG-ES sample \citep{krauseetal2020}. Our model roughly explains the observed percentage of galaxies showing this behavior. A larger sample of edge-on galaxies could clarify the applicability of the model. 

\begin{acknowledgements}
We thank the anonymous referee for help in improving this work. This research is kindly supported and funded by the Hans-B\"ockler Foundation. This research was also supported and funded by the DFG Research Unit 1254 "Magnetisation of Interstellar and Intergalactic Media: The Prospects of Low-Frequency Radio Observations", has made use of the NASA's Astrophysics Data System Bibliographic Services, and the NASA/IPAC Extragalactic Database (NED) which is operated by the Jet Propulsion Laboratory, California Institute of Technology, under contract with the National Aeronautics and Space Administration. \\
The National Radio Astronomy Observatory is a facility of the National Science Foundation operated under cooperative agreement by Associated Universities, Inc.\\
Funding for SDSS-III has been provided by the Alfred P. Sloan Foundation, the Participating Institutions, the National Science Foundation, and the U.S. Department of Energy Office of Science (http://www.sdss3.org). 
SDSS-III is managed by the Astrophysical Research Consortium for the Participating Institutions of the SDSS-III Collaboration including the University of Arizona, the Brazilian Participation Group, Brookhaven National Laboratory, Carnegie Mellon University, University of Florida, the French Participation Group, the German Participation Group, Harvard University, the Instituto de Astrofisica de Canarias, the Michigan State/Notre Dame/JINA Participation Group, Johns Hopkins University, Lawrence Berkeley National Laboratory, Max Planck Institute for Astrophysics, Max Planck Institute for Extraterrestrial Physics, New Mexico State University, New York University, Ohio State University, Pennsylvania State University, University of Portsmouth, Princeton University, the Spanish Participation Group, University of Tokyo, University of Utah, Vanderbilt University, University of Virginia, University of Washington, and Yale University. \\
This publication makes use of data products from the Wide-field Infrared Survey Explorer, which is a joint project of the University of California, Los Angeles, and the Jet Propulsion Laboratory/California Institute of Technology, funded by the National Aeronautics and Space Administration.
\end{acknowledgements}

\bibliography{Bibliography}
\bibliographystyle{aa}

\begin{appendix}
\section{Additional figures}
We show here an additional figure to inspect the bubble-like features throughout the galaxy without the marked locations. 

\begin{figure}
	\includegraphics[width=0.75\textwidth]{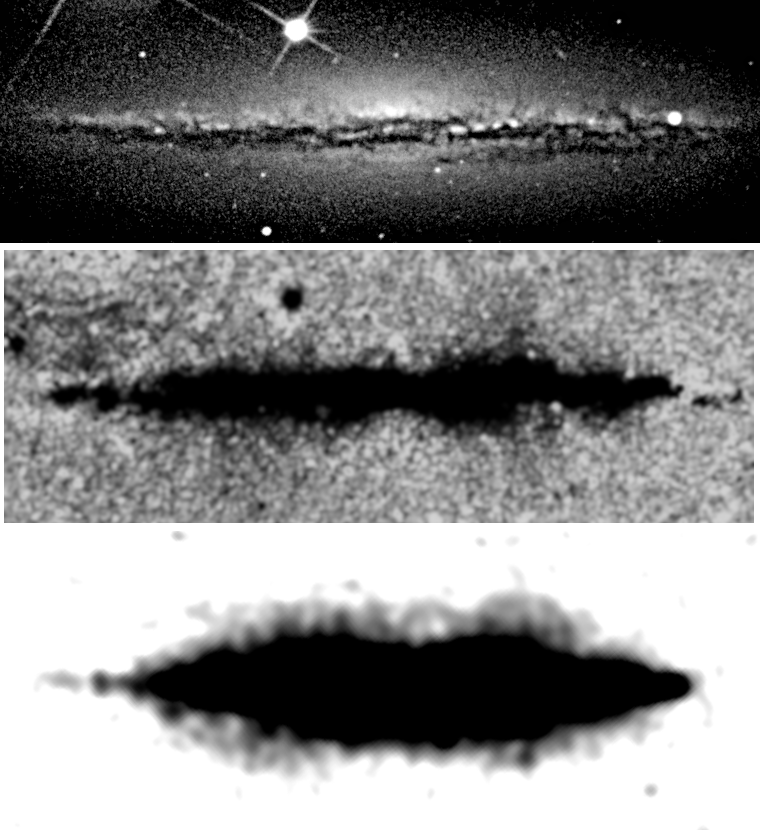}
	\caption{NGC~4217 rotated by $40\degr$, same scale. Top: Artificial optical B-band image created by combining the u and g filters from SDSS, sharpened by unsharp masked filter of GIMP. Middle: H$\upalpha$ image \citep{rand1996}, with edge detection filter applied using GIMP. Bottom: C-configuration C-band radio map intensities shown above 3$\upsigma$-level (9$\upmu$Jy.)}
	\label{fig:N4217-bband_halpha}
	\label{fig:N4217-bw_pol_3sig}
\end{figure}

\end{appendix}
\end{document}